\documentclass[11pt]{article}

\newcommand{\slpart}{\mbox{$\partial \hspace{-0.50em}/$}}

\newcommand{\sln}{\mbox{$n \hspace{-0.50em}/$}}
\newcommand{\slP}{\mbox{$P \hspace{-0.65em}/$}}
\newcommand{\slQ}{\mbox{$Q \hspace{-0.65em}/$}}
\newcommand{\slq}{\mbox{$q \hspace{-0.45em}/$}}

\usepackage{amssymb}
\usepackage{amsmath}
\usepackage{amsfonts}
\usepackage{ulem}
\usepackage{axodraw}
\usepackage{simplewick}
\usepackage[toc,page]{appendix}
\usepackage{hyperref}

\title{Pion-Nucleon Scattering in Kadyshevsky Formalism: II Baryon Exchange Sector}
\author{J.W.Wagenaar and T.A.Rijken}

\begin{document}
\allowdisplaybreaks \maketitle

\abstract{In this paper, which is the second part in a series of
two, we construct tree level baryon exchange and resonance
amplitudes for $\pi N$/$MB$-scattering in the framework of the
Kadyshevsky formalism. We use this formalism to formally implement
absolute pair suppression, where we make use of the method of
Takahashi and Umezawa. The resulting amplitudes are Lorentz
invariant and causal. We continue studying the frame dependence of
the Kadyshevsky integral equation using the method of Gross and
Jackiw. The invariant amplitudes, including those for meson
exchange, are linked to the phase-shifts using the partial wave
basis.}

\section{Introduction}\label{bexchres}

In the previous paper, referred to as paper I \cite{JWT1}, we have
given a motivation for constructing a pion-nucleon ($\pi N$)
scattering, or more generally a meson-baryon ($MB$) scattering,
model. We have given the main ingredients of the model and,
besides others, the (theoretical) results for meson exchange
processes.

In this paper, referred to as paper II, we present the results in
the baryon sector. We construct tree level amplitudes for baryon
exchange and resonance or, to put it in other words, $u$- and
$s$-channel baryon exchange diagrams in the Kadyshevsky formalism
\cite{Kad64,Kad67,Kad68,Kad70}.

The Kadyshevsky formalism is equivalent to Feynman formalism,
since it can be derived using the same S-matrix formula. The main
features for exploiting the Kadyshevsky formalism is that all
particles are on the mass-shell at the cost of an extra quasi
particle, which carries four momentum only. A three dimensional
Lippmann-Schwinger type of integral equation comes about
naturally, without any approximations as for instance in
\cite{henk1} and \cite{Ver76}. Especially at second order, this
formalism provides a covariant, though frame dependent
\footnote{By frame dependent we mean: dependent on a vector
$n^\mu$.}, separation of positive and negative energy
contributions. In this way it is a natural basis for implementing
pair suppression, which may also be interesting for relativistic
many body theories.

In \cite{henk1} pair suppression is assumed by considering
positive states in the integral equation only. Here, we implement
pair suppression formally, and to our knowledge for the first
time, in a covariant and frame independent way. This is done by
using a method based on the Takahashi-Umezawa (TU) method
\cite{Tak53a,Tak53b,Ume56}, see also paper I. In paper I we
studied the $n$-dependence of the integral equation using the
method of Gross and Jackiw (GJ) \cite{Gross69}. This we will
continue here.

In section \ref{pairsupp} we start with introducing the concept of
pair suppression. After discussing how it can be implemented
formally we apply it to $\pi N$ system. The amplitudes are
calculated in section \ref{smelements}. In section \ref{pwe} we
use the helicity basis and make a partial wave expansion to
introduce the phase-shifts. We show how the amplitudes are related
to these phase-shifts. This is done for the entire model.

\section{Pair Suppression Formalism}\label{pairsupp}

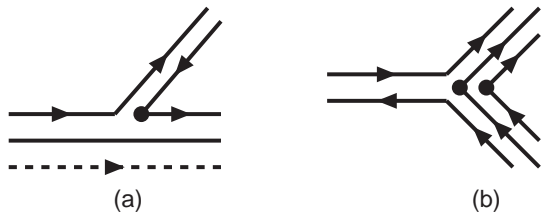
\begin{figure}[hbt]
\begin{center}
\begin{picture}(200,75)(0,0)
 \SetPFont{Helvetica}{9}
 \SetScale{1.0} \SetWidth{1.5}

 \DashArrowLine(0,15)(80,15){3}
 \Line(0,25)(80,25)
 \ArrowLine(0,35)(40,35)
 \ArrowLine(40,35)(75,75)
 \ArrowLine(80,70)(50,35)
 \ArrowLine(50,35)(80,35)
 \Vertex(50,35){3}

 \PText(45,0)(0)[b]{(a)}

 \ArrowLine(120,50)(165,50)
 \ArrowLine(165,40)(120,40)
 \ArrowLine(165,50)(190,75)
 \ArrowLine(190,15)(165,40)
 \ArrowLine(170,45)(200,75)
 \ArrowLine(200,15)(170,45)
 \ArrowLine(180,45)(200,65)
 \ArrowLine(200,25)(180,45)
 \Vertex(170,45){3}
 \Vertex(180,45){3}

 \PText(180,0)(0)[b]{(b)}

\end{picture}
\end{center}
 \caption{\sl (a) $MMM$ ($MBB$) vertex and (b) $M B\bar{B}$ vertex}
 \label{fig:pairsupp}
\end{figure}

To understand the idea of pair suppression at low energy, picture
a general meson-baryon (MB) vertex in terms of their constituent
quarks (see figure \ref{fig:pairsupp}). As stated in \cite{SN78}
every time a quark - anti-quark ($q\bar{q}$) pair is created from
the vacuum the vertex is damped. This idea is supported by
\cite{witten} who's author considers a vertex creating a baryon -
anti-baryon ($B\bar{B}$) pair in a large $N$, $SU(N)$ theory
\footnote{In a $SU(N)$ theory a baryon is represented as a $q^N$
state, whereas a meson is always a $q\bar{q}$ state, independent
of $N$.}. Such a vertex is comparable to figure
\ref{fig:pairsupp}(b), but now $N-1$ pairs need to be created. It
is claimed in \cite{witten} that such vertices are indeed
suppressed. Although it is questionable whether $N=3$ is really
large, we assume that pair suppression holds for $SU_{(F)}(3)$
theories at low energy.

Now, one could imagine that this principle should also apply for
the creation of a meson - anti-meson ($M\bar{M}$) pair and
therefore pair suppression should be implemented in the meson
exchange sector (paper I). For the reason why we have not done
this one should look again at figure \ref{fig:pairsupp} and
consider the large $N$, $SU(N)$ theory again. For the creation of
a $M\bar{M}$ pair at the vertex only one extra $q\bar{q}$ pair
needs to be created instead of the $N-1$ pairs in the $B\bar{B}$
case and is therefore much likelier to happen. Going back to the
real $SU_{(F)}(3)$ the difference is only one $q\bar{q}$ pair,
nevertheless we assume that a $M\bar{M}$ pair creation is not
suppressed.

Also from physical point of view it is nonsense to imply pair
suppression in the meson sector. In order to see this one has to
realize that an anti-meson is also a meson. So, assuming pair
suppression in the meson sector means that a triple meson ($MMM$)
vertex is suppressed, which makes it impossible to consider meson
exchange in meson-baryon scattering as we did in paper I. From
figure \ref{fig:pairsupp}(a) we see that the $MMM$ vertex is of
the same order (in number of $q\bar{q}$ creations, as compared to
figure \ref{fig:pairsupp}(b)) as the meson-baryon-baryon ($MBB$)
vertex in $SU_{(F)}(3)$. So, suppressing the $MMM$ vertex means
that we should also suppress the $MBB$ vertex and no description
of $MB$-scattering in terms of $MB$ vertices is possible at all!

This does not mean, however, that there's no pair suppression what
so ever in the meson sector. As can be seen from the amplitudes in
paper I we only considered $MBB$ vertices, whereas in principle
also $MB\bar{B}$ vertices could have been included. The latter
vertices are not considered using the argument of pair suppression
as discussed above. We will come back to this later.

Since we suppressed the  $MB\bar{B}$ vertex it means that pair
suppression should also be active in the Vector Meson Dominance
(VMD) \cite{sakurai69} model describing nucleon Compton scattering
($\gamma N\rightarrow\gamma N$). From electron Compton scattering
it is well-know that the Thomson limit is exclusively due to the
negative energy electron states (see for instance section 3-9 of
\cite{sakurai67}). However, since the nucleon is composite it may
well be that the negative energy contribution is produced by only
one of the constituents \cite{brod} and it is not necessary to
create an entire anti-baryon.

The suppression of negative energy states may harm the causality
and Lorentz invariance condition. Therefore, the question may
raise whether it is possible to include pair suppression and still
maintain causality and Lorentz invariance. The following example
shows that it should in principle be possible: Imagine an
infinitely dense medium where all anti-nucleon states are filled,
i.e. the Fermi energy of the anti-nucleons $\bar{p}_F= \infty$,
and that for nucleons $p_F=0$. An example would be an anti-neutron
star of infinite density. Then, in such a medium pair production
in $\pi N$-scattering is Pauli-blocked, because all anti-nucleon
states are filled. Denoting the ground-state by $|\Omega\rangle$,
one has, see e.g. \cite{Bow73},
\begin{eqnarray*}
S_F(x-y) &=& -i \langle
\Omega|T[\psi(x)\bar{\psi}(y)]|\Omega\rangle\ ,
\end{eqnarray*}
which gives in momentum space \cite{Bow73}
\begin{eqnarray*}
S_F(p;p_F,\bar{p}_F) &=&\frac{ \mbox{$p \hspace{-0.55em}/$}
+M}{2E_p} \left\{
 \frac{1-n_F(p)}{p_0-E_p+i\varepsilon} + \frac{n_F(p)}{p_0-E_p-i\varepsilon}
\right.\nonumber\\ && \hspace{1cm} \left.
-\frac{1-\bar{n}_F(p)}{p_0+E_p-i\varepsilon}
 - \frac{\bar{n}_F(p)}{p_0+E_p+i\varepsilon} \right\}\ .
\end{eqnarray*}
At zero temperature $T=0$ the non-interacting fermion functions
$n_F, \bar{n}_F$ are defined by
\begin{eqnarray*}
 n_F
&=&
 \left\{
 \begin{array}{c}
    1,\ |{\bf p}|<p_F \\ 0,\ |{\bf p}| > p_F
 \end{array}
 \right.  \ \ ,\ \
 \bar{n}_F =
 \left\{
 \begin{array}{c}
    1,\ |{\bf p}|< \bar{p}_F \\ 0,\ |{\bf p}| > \bar{p}_F
 \end{array}
 \right.\ .
\end{eqnarray*}
In the medium sketched above, clearly $n_F(p)=0$ and
$\bar{n}_F(p)=1$, which leads to a propagator
$S_{ret}(p;0,\infty)$. This propagator is causal and Lorentz
invariant.

The above (academic) example may perhaps convince a sceptical
reader that a perfect relativistic model with 'absolute pair
suppression' is feasible indeed.

As far as our results are concerned we refer to section
\ref{smelements}, where we will see that intermediate baryon
states are represented by retarded (-like) propagators, which have
the nice feature to be causal and $n$-independent. We, therefore,
have a theory that is relativistic and yet it does contain
(absolute) pair suppression.

\subsection{Equations of Motion}

Consider a Lagrangian containing not only the free fermion part,
but also a (simple) coupling between fermions and a scalar
\begin{eqnarray}
 \mathcal{L}
&=&
 \mathcal{L}_{free}+\mathcal{L}_I\nonumber\\*
&=&
 \bar{\psi}\left(\frac{i}{2}\overrightarrow{\slpart}-\frac{i}{2}\overleftarrow{\slpart}-M\right)\psi
 +g\,\bar{\psi}\Gamma\psi\cdot\phi\label{Lag}
\end{eqnarray}
The Euler-Lagrange equation for the fermion part reads
\begin{eqnarray}
 \left(i\slpart-M\right)\psi=-g\Gamma\psi\cdot\phi\label{EulerLag}
\end{eqnarray}
In order to incorporate pair suppression we pose that the
transitions between positive and negative energy fermion states
vanish in the interaction part of \eqref{Lag}, i.e.
$\overline{\psi^{(+)}}\Gamma\psi^{(-)}=\overline{\psi^{(-)}}\Gamma\psi^{(+)}=0$.
So, we impose {\it absolute} pair suppression. From now on, when
we speak of pair suppression we mean absolute pair suppression,
unless it is mentioned otherwise. Of course it is in principle
possible to allow for some pair production. This can be done for
instance by not eliminating the terms
$\overline{\psi^{(+)}}\Gamma\psi^{(-)}$ and
$\overline{\psi^{(-)}}\Gamma\psi^{(+)}$ in \eqref{Lag}, but
allowing them with some small coupling $g'\ll g$. This, however,
makes the situation much more complicated and is not worked out
here.

Since half of the term on the rhs of \eqref{EulerLag} finds its
origin in such vanished terms, it is reduced by a factor 2 by the
pair suppression condition.

Making the split up $\psi=\psi^{(+)}+\psi^{(-)}$, which is
invariant under orthochronous Lorentz transformations, in
\eqref{EulerLag} we assume both parts are independent, so that we
have
\begin{subequations}
\begin{eqnarray}
 \left(i\slpart-M\right)\psi^{(+)}&=&-\frac{g}{2}\,\Gamma\psi^{(+)}\cdot\phi\ ,\label{EulerLaga}\\
 \left(i\slpart-M\right)\psi^{(-)}&=&-\frac{g}{2}\,\Gamma\psi^{(-)}\cdot\phi\ .\label{EulerLagb}
\end{eqnarray}
\end{subequations}

One might wonder why we did not consider independent positive and
negative energy fields from the start in \eqref{Lag}. Although
this would not cause any trouble in the interaction part
($\mathcal{L}_I$) it will in the free part. The quantum condition
in such a situation would be
$\left\{\psi^{(\pm)}(x),\pi^{(\pm)}(y)\right\}=i\delta^3(x-y)$.
This is in conflict with the important relations between the
positive and negative energy components
\begin{eqnarray}
 \left\{\psi^{(+)}(x),\overline{\psi^{(+)}}(y)\right\}&=&\left(i\slpart+M\right)\Delta^{+}(x-y)\ ,\nonumber\\
 \left\{\psi^{(-)}(x),\overline{\psi^{(-)}}(y)\right\}&=&-\left(i\slpart+M\right)\Delta^{-}(x-y)\ ,\label{complusmin}
\end{eqnarray}
which we do need. Therefore we don't make the split up in the
Lagrangian, but in the equations of motion.

The assumption that both parts $\psi^{(+)}$ and $\psi^{(-)}$ are
independent means that besides the anti-commutation relations in
\eqref{complusmin} all others are zero.

In order to incorporate pair suppression in the meson sector (see
paper I) the only thing to do is to exclude the transitions
$\overline{\psi^{(+)}}\Gamma\psi^{(-)}$ and
$\overline{\psi^{(-)}}\Gamma\psi^{(+)}$ in the interaction
Lagrangians. By doing so, only $u$ and $\bar{u}$ spinors will
contribute. Therefore, only these spinors are present in the
results for meson exchange (paper I).

For baryon exchange and resonance diagrams the implications for
pair suppression are less trivial. We, therefore, discuss how pair
suppression can be implemented in these situation in the following
subsections.

\subsection{Takahashi Umezawa Scheme for Pair
Suppression}\label{TUpairsupp}

In order to obtain the interaction Hamiltonian in case of pair
suppression we set up the theory very similar to the TU scheme
\cite{Tak53a,Tak53b,Ume56} introduced and applied in paper I.
Since we only make the split-up in the fermion fields, the scalar
fields are unaffected and therefore not included in this
subsection.

We start with defining the currents
\begin{eqnarray}
 \mbox{\boldmath $j$}_{\psi^{(\pm)},a}(x)
&=&
 \left(-\frac{\partial\mathcal{L}_I}{\partial\mbox{\boldmath $\psi$}^{(\pm)}(x)},
 -\frac{\partial\mathcal{L}_I}{\partial(\partial_\mu \mbox{\boldmath $\psi$}^{(\pm)})(x)}\right)
 \ .\label{currents}
\end{eqnarray}
Solutions to the equations of motion resulting from a general
(interaction) Lagrangian are Yang-Feldman (YF) \cite{Yang50} type
of equations
\begin{eqnarray}
 \mbox{\boldmath $\psi$}^{(\pm)}(x)
&=&
 \psi^{(\pm)}(x)
 +\frac{1}{2}\int d^4y\ D_a(y)\left(i\slpart+M\right)\theta[n(x-y)]
 \nonumber\\
&&
 \phantom{\psi^{(\pm)}(x)+\frac{1}{2}\int}\times
 \Delta(x-y)\cdot\mbox{\boldmath $j$}_{\psi^{(\pm)},a}(y)
 \ .\label{heisfields}
\end{eqnarray}
Here, we have chosen to use the retarded Green functions again,
this, in order to be close to the treatment of paper I.

Furthermore, we introduce the auxiliary fields
\begin{eqnarray}
 \psi^{(\pm)}(x,\sigma)
&=&
 \psi^{(\pm)}(x)
 \mp i\int_{-\infty}^{\sigma} d^4yD_a(y)\left(i\slpart+M\right)
 \Delta^{\pm}(x-y)\cdot\mbox{\boldmath $j$}_{\psi^{(\pm)},a}(y)\
 .\label{auxsig}
\end{eqnarray}
Combining these two equations (\eqref{heisfields} and
\eqref{auxsig}) we get
\begin{eqnarray}
 \mbox{\boldmath $\psi$}^{(\pm)}(x)
&=&
 \psi^{(\pm)}(x/\sigma)
 +\frac{1}{4}\int d^4y\left[D_a(y)\left(i\slpart+M\right),\epsilon(x-y)\vphantom{\frac{A}{A}}\right]
 \Delta(x-y)\cdot\mbox{\boldmath $j$}_{\psi^{(\pm)},a}(y)\nonumber\\
&&
 \pm\frac{i}{2}\int d^4y\ \theta[n(x-y)]D_a(y)\left(i\slpart+M\right)
 \Delta^{(1)}(x-y)\cdot\mbox{\boldmath $j$}_{\psi^{(\pm)},a}(y)
 \ .\label{TUpairheis}
\end{eqnarray}
The factor $1/2$ in \eqref{heisfields} is essential. This becomes
clear when we decompose $\Delta^{\pm}(x-y)=\frac{\pm
i}{2}\,\Delta(x-y)+\frac{1}{2}\,\Delta^{(1)}(x-y)$ in
\eqref{auxsig}. The first part ($\Delta$) combines with
\eqref{heisfields} to the second term on the rhs of
\eqref{TUpairheis} and the second part ($\Delta^{(1)}$) gives a
new contribution to $\mbox{\boldmath $\psi$}^{(\pm)}$ as compared
to $\mbox{\boldmath $\psi$}$ in the original treatment. We see
that if we add $\mbox{\boldmath $\psi$}^{(+)}$ and
$\mbox{\boldmath $\psi$}^{(-)}$ we get back the $\mbox{\boldmath
$\psi$}$ in the original treatment, again. This makes the factor
$1/2$ difference in the first part of \eqref{TUpairheis} easier to
understand.

Similar to the treatment in appendix C of paper I, it can be shown
that $\psi^{(\pm)}(x)$ and $\psi^{(\pm)}(x,\sigma)$ satisfy the
same commutation relation and that the unitary operator connecting
the two is related to the S-matrix. Following similar steps the
defining equation for the interaction Hamiltonian is
\begin{eqnarray}
 \left[\psi^{(\pm)}(x),\mathcal{H}_I(y;n)\vphantom{\frac{A}{A}}\right]
&=&
 U^{-1}[\sigma]\left[D_a(y)(\pm)\left(i\slpart+M\right)\Delta^{\pm}(x-y)\cdot\mbox{\boldmath
 $j$}_{\psi^{(\pm)},a}(y)\right]U[\sigma]\ ,\label{TUpairhint}
\end{eqnarray}

Having discussed the formalism to implement pair suppression, now,
we're going to apply it.

\subsection{(Pseudo) Scalar Coupling}\label{pscoupling}

In the (pseudo) scalar sector of the theory including pair
suppression we start with the following interaction Lagrangian
\begin{equation}
 \mathcal{L}_I
=
 g\,\overline{\mbox{\boldmath $\psi$}^{(+)}}\Gamma\mbox{\boldmath $\psi$}^{(+)}\cdot\mbox{\boldmath $\phi$}
 +g\,\overline{\mbox{\boldmath $\psi$}^{(-)}}\Gamma\mbox{\boldmath $\psi$}^{(-)}\cdot\mbox{\boldmath $\phi$}\ ,
 \label{psc1}
\end{equation}
\footnote{We note that this interaction Lagrangian \eqref{psc1} is
charge invariant.} where $\Gamma=1$ or $\Gamma=i\gamma^5$. We will
not use the specific forms for $\Gamma$ until the discussion of
the amplitudes in section \ref{smelements}. This, in order to be
as general as possible.

From \eqref{psc1} we deduce the currents according to
\eqref{currents}
\begin{eqnarray}
 \mbox{\boldmath $j$}_{\psi^{(\pm)},a}
&=&
 \left(-g\,\Gamma\mbox{\boldmath $\psi$}^{(\pm)}\cdot\mbox{\boldmath $\phi$},0\right)\ ,\nonumber\\
 \mbox{\boldmath $j$}_{\phi,a}
&=&
 \left(-g\,\overline{\mbox{\boldmath $\psi$}^{(+)}}\Gamma\mbox{\boldmath $\psi$}^{(+)}
 -g\,\overline{\mbox{\boldmath $\psi$}^{(-)}}\Gamma\mbox{\boldmath
 $\psi$}^{(-)},0\right)\ .\label{psc2}
\end{eqnarray}
The fields in the H.R. can be expressed in terms of fields in the
I.R. using \eqref{TUpairheis}
\begin{subequations}
\begin{eqnarray}
 \mbox{\boldmath $\psi$}^{(\pm)}(x)
&=&
 \psi^{(\pm)}(x/\sigma)
 \mp\frac{ig}{2}\int d^4y\ \theta[n(x-y)]\left(i\slpart+M\right)\Delta^{(1)}(x-y)
 \nonumber\\
&&
 \phantom{\psi^{(\pm)}(x/\sigma)\mp\frac{ig}{2}\int}\times
 \Gamma\psi^{(\pm)}(y)\cdot\phi(y)\ ,\label{psc3a}\\
 \mbox{\boldmath $\phi$}(x)
&=&
 \phi(x/\sigma)+\frac{1}{4}\int d^4y\left[D_a(y),\epsilon(x-y)\right]\Delta(x-y)
 \cdot\mbox{\boldmath $j$}_{\phi,a}(y)\nonumber\\
&=&
 \phi(x/\sigma)\ .\label{psc3b}
\end{eqnarray}
\end{subequations}
Equation \eqref{psc3a} was found by assuming that the coupling
constant is small and considering only contributions up to order
$g$, just as in paper I.

With the expressions \eqref{psc3a} and \eqref{psc3b} and the
definition of the commutator of the (fermion) fields with the
interaction Hamiltonian \eqref{TUpairhint} we get
\begin{eqnarray}
 \left[\psi^{(+)}(x),\mathcal{H}_I(y;n)\right]
&=&
 -g\left(i\slpart+M\right)\Delta^{+}(x-y)\Gamma\psi^{(+)}(y)\cdot\phi(y)\nonumber\\
&&
 +\frac{ig^2}{2}\left(i\slpart+M\right)\Delta^{+}(x-y)\int d^4z\,\Gamma\,\theta[n(y-z)]\nonumber\\
&&
 \phantom{+\,}\times
 \left(i\slpart_y+M\right)\Delta^{(1)}(y-z)\,\Gamma\,\psi^{(+)}(z)\cdot\phi(z)\phi(y)\
 , \nonumber\\
 \left[\psi^{(-)}(x),\mathcal{H}_I(y;n)\right]
&=&
 g\left(i\slpart+M\right)\Delta^{-}(x-y)\Gamma\psi^{(-)}(y)\cdot\phi(y)\nonumber\\*
&&
 +\frac{ig^2}{2}\left(i\slpart+M\right)\Delta^{-}(x-y)\int d^4z\,\Gamma\,\theta[n(y-z)]\nonumber\\
&&
 \phantom{+\,}\times
 \left(i\slpart_y+M\right)\Delta^{(1)}(y-z)\,\Gamma\,\psi^{(-)}(z)\cdot\phi(z)\phi(y)
 \ .\label{psc4}
\end{eqnarray}
Here, we have not included the commutator of the scalar field
$\phi$ with the interaction Hamiltonian, because \eqref{psc4}
already contains enough information to get the interaction
Hamiltonian
\begin{eqnarray}
 \mathcal{H}_I(x;n)
&=&
 -g\,\overline{\psi^{(+)}}\Gamma\psi^{(+)}\cdot\phi-g\,\overline{\psi^{(-)}}\Gamma\psi^{(-)}\cdot\phi\nonumber\\
&&
 +\frac{ig^2}{2}\int d^4y\left[\overline{\psi^{(+)}}\,\Gamma\,\phi\right]_x
 \theta[n(x-y)]\left(i\slpart_x+M\right)
 \Delta^{(1)}(x-y)\left[\Gamma\psi^{(+)}\phi\vphantom{\frac{a}{a}}\right]_y\nonumber\\
&&
 -\frac{ig^2}{2}\int d^4y\left[\overline{\psi^{(-)}}\,\Gamma\,\phi\right]_x
 \theta[n(x-y)]\left(i\slpart_x+M\right)
 \Delta^{(1)}(x-y)\left[\Gamma\psi^{(-)}\phi\vphantom{\frac{a}{a}}\right]_y
 \ .\label{psc5}
\end{eqnarray}
In \eqref{psc5} we see that the interaction Hamiltonian contains
terms proportional to $\Delta^{(1)}(x-y)$ which are of order
$O(g^2)$. These terms will be essential to get covariant and
$n$-independent S-matrix elements and amplitudes at order
$O(g^2)$.

If we would include external quasi fields in interaction
Lagrangian \eqref{psc1}, then the terms of order $g^2$ in the
interaction Hamiltonian \eqref{psc5} would be quartic in the quasi
field. Two quasi fields can be contracted
\begin{eqnarray}
 \bar{\chi}(x)\bcontraction{}{\chi}{(x)}{\bar{\chi}}\chi(x)\bar{\chi}(y)\chi(y)
 =\bar{\chi}(x)\theta[n(x-y)]\chi(y)\ .\label{psc5a}
\end{eqnarray}
So, the terms of order $g^2$ get an additional factor
$\theta[n(x-y)]$. However, since these terms already contain such
a factor, we make the identification
$\theta[n(x-y)]\theta[n(x-y)]\rightarrow\theta[n(x-y)]$.
Therefore, all relevant $\pi N$ terms in \eqref{psc5} are
quadratic in the external quasi field, just as we want. This
argument is valid for all couplings.

\subsection{(Pseudo) Vector Coupling}\label{pvcoupling}

Here, we repeat the steps of the previous subsection (section
\ref{pscoupling}) but now in the case of (pseudo) vector coupling.
The interaction Lagrangian reads
\begin{equation}
 \mathcal{L}_I
=
 \frac{f}{m_\pi}\,\overline{\mbox{\boldmath $\psi$}^{(+)}}
 \Gamma_\mu\mbox{\boldmath $\psi$}^{(+)}\cdot\partial^\mu\mbox{\boldmath $\phi$}
 +\frac{f}{m_\pi}\,\overline{\mbox{\boldmath $\psi$}^{(-)}}
 \Gamma_\mu\mbox{\boldmath $\psi$}^{(-)}\cdot\partial^\mu\mbox{\boldmath $\phi$}\ ,
 \label{pvc1}
\end{equation}
where $\Gamma_\mu=\gamma_\mu$ or $\Gamma_\mu=\gamma_5\gamma_\mu$.
From \eqref{pvc1} we deduce the currents
\begin{eqnarray}
 \mbox{\boldmath $j$}_{\psi^{(\pm)},a}
&=&
 \left(-\frac{f}{m_\pi}\,\Gamma_\mu\mbox{\boldmath $\psi$}^{(\pm)}\cdot\partial^\mu\mbox{\boldmath $\phi$},0\right)\ ,\nonumber\\
 \mbox{\boldmath $j$}_{\phi,a}
&=&
 \left(0,-\frac{f}{m_\pi}\,\overline{\mbox{\boldmath $\psi$}^{(+)}}\Gamma_\mu\mbox{\boldmath $\psi$}^{(+)}
 -\frac{f}{m_\pi}\,\overline{\mbox{\boldmath $\psi$}^{(-)}}\Gamma_\mu\mbox{\boldmath $\psi$}^{(-)}\right)\ .\label{pvc2}
\end{eqnarray}
The fields in the H.R. are expressed in terms of fields in the
I.R. as follows
\begin{subequations}
\begin{eqnarray}
 \mbox{\boldmath $\psi$}^{(\pm)}(x)
&=&
 \psi^{(\pm)}(x/\sigma)\mp\frac{if}{2m_\pi}\int
 d^4y\theta[n(x-y)]\left(i\slpart+M\right)\Delta^{(1)}(x-y)\nonumber\\*
&&
 \phantom{\psi^{(\pm)}(x/\sigma)\mp\frac{if}{2m_\pi}\int}\times
 \Gamma_\mu\psi^{(\pm)}(y)\cdot\partial^\mu\phi(y)\ ,\\
 \mbox{\boldmath $\phi$}(x)
&=&
 \phi(x/\sigma)\ ,\\
 \partial^\mu\mbox{\boldmath $\phi$}(x)
&=&
 \left[\partial^\mu\phi(x,\sigma)\right]_{x/\sigma}
 -\frac{f}{m_\pi}\ n^\mu\,\overline{\psi^{(+)}}(x)n\cdot\Gamma\psi^{(+)}(x)\nonumber\\
&&
 -\frac{f}{m_\pi}\
 n^\mu\,\overline{\psi^{(-)}}(x)n\cdot\Gamma\psi^{(-)}(x)\ .\label{pvc3}
\end{eqnarray}
\end{subequations}
The commutators of the different fields with the interaction
Hamiltonian are
\begin{eqnarray}
 \left[\psi^{(+)}(x),\mathcal{H}_I(y;n)\right]
&=&
 \frac{f}{m_\pi}
 \left(i\slpart+M\right)\Delta^{+}(x-y)\left[\vphantom{\frac{A}{A}}
 -\Gamma_\mu\psi^{(+)}\cdot\partial^\mu\phi\right.\nonumber\\
&&
 \left.+\frac{f}{m_\pi}\ n\cdot\Gamma\psi^{(+)}\,\overline{\psi^{(+)}}n\cdot\Gamma\psi^{(+)}
 +\frac{f}{m_\pi}\ n\cdot\Gamma\psi^{(+)}\,\overline{\psi^{(-)}}n\cdot\Gamma\psi^{(-)}\right]_y
 \nonumber\\
&&
 +\frac{if^2}{2m^2_\pi}\left(i\slpart+M\right)\Delta^{+}(x-y)\int d^4z\,\Gamma_\mu\,\theta[n(y-z)]\nonumber\\
&&
 \phantom{+\frac{if^2}{2m^2_\pi}}\times
 \left(i\slpart_y+M\right)\Delta^{(1)}(y-z)\,\Gamma_\nu\,\psi^{(+)}(z)\cdot\partial^\nu\phi(z)\partial^\mu\phi(y)
 \ ,\nonumber\\
 \left[\psi^{(-)}(x),\mathcal{H}_I(y;n)\right]
&=&
 -\frac{f}{m_\pi}
 \left(i\slpart+M\right)\Delta^{-}(x-y)\left[\vphantom{\frac{A}{A}}
 -\Gamma_\mu\psi^{(-)}\cdot\partial^\mu\phi\right.\nonumber\\
&&
 \left.+\frac{f}{m_\pi}\ n\cdot\Gamma\psi^{(-)}\,\overline{\psi^{(+)}}n\cdot\Gamma\psi^{(+)}
 +\frac{f}{m_\pi}\ n\cdot\Gamma\psi^{(-)}\,\overline{\psi^{(-)}}n\cdot\Gamma\psi^{(-)}\right]_y
 \nonumber\\
&&
 -\frac{if^2}{2m^2_\pi}\left(i\slpart+M\right)\Delta^{-}(x-y)\int d^4z\,\Gamma_\mu\,\theta[n(y-z)]\nonumber\\
&&
 \phantom{+\frac{if^2}{2m^2_\pi}}\times
 \left(i\slpart_y+M\right)\Delta^{(1)}(y-z)\,\Gamma_\nu\,\psi^{(-)}(z)\cdot\partial^\nu\phi(z)\partial^\mu\phi(y)
 \ ,\label{pvc4}
\end{eqnarray}
and from these equations we deduce the interaction Hamiltonian
\begin{eqnarray}
 \mathcal{H}_I(x;n)
&=&
 -\frac{f}{m_\pi}\,\overline{\psi^{(+)}}\Gamma_\mu\psi^{(+)}\cdot\partial^\mu\phi
 -\frac{f}{m_\pi}\,\overline{\psi^{(-)}}\Gamma_\mu\psi^{(-)}\cdot\partial^\mu\phi\nonumber\\
&&
 +\frac{f^2}{2m^2_\pi}\left[\overline{\psi^{(+)}}\,n\cdot\Gamma\,\psi^{(+)}\right]^2
 +\frac{f^2}{2m^2_\pi}\left[\overline{\psi^{(-)}}\,n\cdot\Gamma\,\psi^{(-)}\right]^2\nonumber\\
&&
 +\frac{f^2}{m^2_\pi}\left[\overline{\psi^{(+)}}\,n\cdot\Gamma\,\psi^{(+)}\right]
 \left[\overline{\psi^{(-)}}\,n\cdot\Gamma\,\psi^{(-)}\right]
 \nonumber\\
&&
 +\frac{if^2}{2m^2_\pi}\int d^4y\left[\overline{\psi^{(+)}}\Gamma_\mu\partial^\mu\phi\right]_x
 \theta[n(x-y)]\left(i\slpart+M\right)
 \nonumber\\
&&
 \phantom{+\frac{if^2}{2m^2_\pi}\int}\times
 \Delta^{(1)}(x-y)\left[\Gamma_\nu\psi^{(+)}\partial^\nu\phi\right]_y\nonumber\\
&&
 -\frac{if^2}{2m^2_\pi}\int d^4y\left[\overline{\psi^{(-)}}\Gamma_\mu\partial^\mu\phi\right]_x
 \theta[n(x-y)]\left(i\slpart+M\right)
 \nonumber\\
&&
 \phantom{+\frac{if^2}{2m^2_\pi}\int}\times
 \Delta^{(1)}(x-y)\left[\Gamma_\nu\psi^{(-)}\partial^\nu\phi\right]_y\ .\label{pvc5}
\end{eqnarray}
As in \eqref{psc5} there are also terms proportional to
$\Delta^{(1)}(x-y)$ quadratic in the coupling constant. Also,
\eqref{pvc5} contains contact terms, but they do not contribute to
$\pi N$-scattering.

\subsection{$\pi N\Delta_{33}$ Coupling}\label{deltacoupling}

At this point we deviated from \cite{henk1} as far as the
interaction Lagrangian is concerned. For the description of the
coupling of the $\Delta_{33}$, which is a spin-3/2 field, to $\pi
N$ we follow \cite{pasc,pasctim} by using the gauge invariant
interaction Lagrangian
\begin{eqnarray}
 \mathcal{L}_I
&=&
 g_{gi}\,\epsilon^{\mu\nu\alpha\beta}\left(\partial_{\mu}
 \overline{\mbox{\boldmath $\Psi$}^{(+)}_\nu}\right)\gamma_5\gamma_{\alpha}
 \mbox{\boldmath $\psi$}^{(+)}\left(\partial_{\beta}\mbox{\boldmath $\phi$}\right)
 +g_{gi}\,\epsilon^{\mu\nu\alpha\beta}\overline{\mbox{\boldmath $\psi$}^{(+)}}
 \gamma_5\gamma_{\alpha}\left(\partial_{\mu}\mbox{\boldmath $\Psi$}^{(+)}_{\nu}\right)
 \left(\partial_{\beta}\mbox{\boldmath $\phi$}\right)
 \nonumber\\
&&
 +g_{gi}\,\epsilon^{\mu\nu\alpha\beta}\left(\partial_{\mu}
 \overline{\mbox{\boldmath $\Psi$}^{(-)}_\nu}\right)\gamma_5\gamma_{\alpha}
 \mbox{\boldmath $\psi$}^{(-)}\left(\partial_{\beta}\mbox{\boldmath $\phi$}\right)
 +g_{gi}\,\epsilon^{\mu\nu\alpha\beta}\overline{\mbox{\boldmath $\psi$}^{(-)}}
 \gamma_5\gamma_{\alpha}\left(\partial_{\mu}\mbox{\boldmath $\Psi$}^{(-)}_{\nu}\right)
 \left(\partial_{\beta}\mbox{\boldmath $\phi$}\right)
 \ .\label{deltaex1}
\end{eqnarray}
Here, $\Psi_\mu$ represents the spin-3/2 $\Delta_{33}$ field. As
is mentioned in \cite{pasc} the $\Psi_\mu$ field does not only
contain spin-3/2 components but also spin-1/2 components. By using
the interaction Lagrangian as in \eqref{deltaex1} it is assured
that only the spin-3/2 components of the $\Delta_{33}$ field
couple.

From \eqref{deltaex1} we deduce the currents
\begin{eqnarray}
 \mbox{\boldmath $j$}_{\phi,a}(x)
&=&
 \left[0,\vphantom{\frac{A}{A}}
 -g_{gi}\,\epsilon^{\mu\nu\alpha\beta}\left(\partial_{\mu}
 \overline{\mbox{\boldmath $\Psi$}^{(+)}_\nu}\right)\gamma_5\gamma_{\alpha}\mbox{\boldmath $\psi$}^{(+)}
 -g_{gi}\,\epsilon^{\mu\nu\alpha\beta}\overline{\mbox{\boldmath $\psi$}^{(+)}}
 \gamma_5\gamma_{\alpha}\left(\partial_{\mu}\mbox{\boldmath $\Psi$}^{(+)}_{\nu}\right)
 \right.\nonumber\\
&&
 \left.\vphantom{\frac{A}{A}}\phantom{\left[\right.0,}
 -g_{gi}\,\epsilon^{\mu\nu\alpha\beta}\left(\partial_{\mu}
 \overline{\mbox{\boldmath $\Psi$}^{(-)}_\nu}\right)\gamma_5\gamma_{\alpha}\mbox{\boldmath $\psi$}^{(-)}
 -g_{gi}\,\epsilon^{\mu\nu\alpha\beta}\overline{\mbox{\boldmath $\psi$}^{(-)}}
 \gamma_5\gamma_{\alpha}\left(\partial_{\mu}\mbox{\boldmath $\Psi$}^{(-)}_{\nu}\right)\right]\nonumber\\
 \mbox{\boldmath $j$}_{\psi^{(\pm)},a}(x)
&=&
 \left[-g_{gi}\,\epsilon^{\mu\nu\alpha\beta}\gamma_5\gamma_{\alpha}\left(\partial_{\mu}\mbox{\boldmath $\Psi$}^{(+)}_{\nu}\right)
 \left(\partial_{\beta}\mbox{\boldmath $\phi$}\right),0\right]\nonumber\\
 \mbox{\boldmath $j$}_{\Psi^{(\pm)}_\nu,a}(x)
&=&
 \left[0,-g_{gi}\,\epsilon^{\mu\nu\alpha\beta}\gamma_5\gamma_{\alpha}
 \mbox{\boldmath $\psi$}^{(+)}\left(\partial_{\beta}\mbox{\boldmath
 $\phi$}\right)\right]\ .\label{deltaex2}
\end{eqnarray}
To avoid lengthy equations we express the commutators of the
various fields with the interaction Hamiltonian in terms of fields
in the H.R. \eqref{TUpairhint}
\begin{eqnarray}
 \left[\phi(x),\mathcal{H}_I(y;n)\vphantom{\frac{A}{A}}\right]
&=&
 U(\sigma)i\Delta(x-y)\overleftarrow{\partial_\beta^y}\left[
 -g_{gi}\,\epsilon^{\mu\nu\alpha\beta}\left(\partial_{\mu}
 \overline{\mbox{\boldmath $\Psi$}^{(+)}_\nu}\right)\gamma_5\gamma_{\alpha}\mbox{\boldmath $\psi$}^{(+)}
 \right.\nonumber\\
&&
 -g_{gi}\,\epsilon^{\mu\nu\alpha\beta}\overline{\mbox{\boldmath $\psi$}^{(+)}}
 \gamma_5\gamma_{\alpha}\left(\partial_{\mu}\mbox{\boldmath $\Psi$}^{(+)}_{\nu}\right)
 -g_{gi}\,\epsilon^{\mu\nu\alpha\beta}\left(\partial_{\mu}
 \overline{\mbox{\boldmath $\Psi$}^{(-)}_\nu}\right)\gamma_5\gamma_{\alpha}\mbox{\boldmath $\psi$}^{(-)}
 \nonumber\\
&&
 \left.-g_{gi}\,\epsilon^{\mu\nu\alpha\beta}\overline{\mbox{\boldmath $\psi$}^{(-)}}
 \gamma_5\gamma_{\alpha}\left(\partial_{\mu}\mbox{\boldmath $\Psi$}^{(-)}_{\nu}\right)
 \right]_yU^{-1}(\sigma)\ ,\nonumber
\end{eqnarray}
\begin{eqnarray}
 \left[\psi^{\pm}(x),\mathcal{H}_I(y;n)\vphantom{\frac{A}{A}}\right]
&=&
 U(\sigma)(\pm)\left(i\slpart_x+M\right)\Delta^{\pm}(x-y)\left[
 -g_{gi}\,\epsilon^{\mu\nu\alpha\beta}\gamma_5\gamma_{\alpha}
 \mbox{\boldmath $\psi$}^{(+)}\left(\partial_{\beta}\mbox{\boldmath $\phi$}\right)\right]_yU^{(-1)}(\sigma)\ ,\nonumber
\end{eqnarray}
\begin{eqnarray}
 \left[\Psi^{\pm}_\mu(x),\mathcal{H}_I(y;n)\vphantom{\frac{A}{A}}\right]
&=&
 U(\sigma)(\pm)\left(i\slpart_x+M_\Delta\right)(-)\nonumber\\*
&&
 \times\left(g_{\mu\nu}-\frac{1}{3}\,\gamma_\mu\gamma_\nu
 +\frac{2\partial_\mu\partial_\nu}{3M^2_\Delta}-\frac{1}{3M^2_\Delta}\left(\gamma_\mu i\partial_\nu-i\partial_\mu\gamma_\nu\right)
 \right)\Delta^{\pm}(x-y)\overleftarrow{\partial_\rho^y}\nonumber\\
&&
 \times\left(-g_{gi}\,\epsilon^{\mu\nu\alpha\beta}\gamma_5\gamma_{\alpha}
 \mbox{\boldmath $\psi$}^{(+)}\left(\partial_{\beta}\mbox{\boldmath $\phi$}\right)\right)_yU^{-1}(\sigma)\
 ,\label{deltaex3}
\end{eqnarray}
where the fields in the H.R. are expressed in terms of fields in
the I.R. using \eqref{TUpairheis}
\begin{eqnarray}
 \mbox{\boldmath $\psi$}^{(\pm)}(x)
&=&
 \psi^{(\pm)}(x/\sigma)\pm\frac{i}{2}\int d^4y\ \theta[n(x-y)](i\slpart+M)\Delta^{(1)}(x-y)\nonumber\\
&&
 \phantom{\psi^{(\pm)}(x/\sigma)\pm\frac{i}{2}\int}
 \times g_{gi}\,\epsilon^{\mu\nu\alpha\beta}\gamma_5\gamma_{\alpha}
 \left[\left(\partial_{\mu}\Psi^{(\pm)}_{\nu}\right)\left(\partial_{\beta}\phi\right)
 \vphantom{\frac{a}{a}}\right]_y\ ,\nonumber\\
 \partial_\rho\mbox{\boldmath $\phi$}(x)
&=&
 \left[\partial_\rho\phi(x,\sigma)\right]_{x/\sigma}\nonumber\\
&&
 -g_{gi}\,\epsilon^{\mu\nu\alpha\beta}n_\rho\left(\partial_{\mu}\overline{\Psi^{(+)}_{\nu}}\right)\gamma_5\gamma_{\alpha}\psi^{(+)}n_\beta
 -g_{gi}\,\epsilon^{\mu\nu\alpha\beta}n_\rho\overline{\psi^{(+)}}\gamma_5\gamma_{\alpha}\left(\partial_{\mu}\Psi^{(+)}_{\nu}\right)n_\beta
 \nonumber\\
&&
 -g_{gi}\,\epsilon^{\mu\nu\alpha\beta}n_\rho\left(\partial_{\mu}\overline{\Psi^{(-)}_{\nu}}\right)\gamma_5\gamma_{\alpha}\psi^{(-)}n_\beta
 -g_{gi}\,\epsilon^{\mu\nu\alpha\beta}n_\rho\overline{\psi^{(-)}}\gamma_5\gamma_{\alpha}\left(\partial_{\mu}\Psi^{(-)}_{\nu}\right)n_\beta
 \ ,\nonumber\\
 \partial_\rho\mbox{\boldmath $\Psi$}^{(\pm)}_\mu(x)
&=&
 \left[\partial_\rho\Psi^{(\pm)}_\mu(x,\sigma)\right]_{x/\sigma}\nonumber\\
&&
 +\frac{g_{gi}}{2}\left[\left(i\slpart_x+M_\Delta\right)n_\rho n_\gamma+\sln\left(i\partial_\rho n_\gamma+n_\rho i\partial_\gamma\right)
 -2\sln n_\rho n_\gamma n\cdot i\partial\vphantom{\frac{A}{A}}\right]\nonumber\\
&&
 \phantom{+\frac{g_{gi}}{2}}\times
 \left(g_{\mu\nu}-\frac{1}{3}\,\gamma_\mu\gamma_\nu\right)
 \epsilon^{\rho\nu\alpha\beta}\gamma_5\gamma_\alpha\psi^{(\pm)}\left(\partial_\beta\phi\right)\nonumber\\
&&
 \mp\frac{ig_{gi}}{2}\,\int d^4y\theta[n(x-y)]\left(i\slpart_x+M_\Delta\right)\left[g_{\mu\nu}-\frac{1}{3}\,\gamma_\mu\gamma_\nu\right]
 \nonumber\\
&&
 \phantom{\mp\frac{ig_{gi}}{2}\,\int}\times
 \partial_\rho\partial_\gamma\Delta^{(1)}(x-y)
 \left[\epsilon^{\rho\nu\alpha\beta}\gamma_5\gamma_\alpha\psi^{(\pm)}
 \left(\partial_\beta\phi\right)\right]_y\ .\label{deltaex4}
\end{eqnarray}
Here, we have already used that $\partial_\rho\mbox{\boldmath
$\Psi$}^{(\pm)}_\mu(x)$ always appears in combination with
$\epsilon^{\rho\mu\alpha\beta}$. Therefore, we have eliminated
terms that are symmetric in $\rho$ and $\mu$.

With these ingredients we can construct the interaction
Hamiltonian. Because it contains a lot of terms we only focus on
those terms that contribute to $\pi N$-scattering
\begin{eqnarray}
&&
 \mathcal{H}_I(x;n)=\nonumber\\
&=&
 -g_{gi}\,\epsilon^{\mu\nu\alpha\beta}\left(\partial_{\mu}\overline{\Psi^{(+)}_{\nu}}\right)\gamma_5\gamma_{\alpha}\psi^{(+)}
 \left(\partial_{\beta}\phi\right)
 -g_{gi}\,\epsilon^{\mu\nu\alpha\beta}\overline{\psi^{(+)}}\gamma_5\gamma_{\alpha}\left(\partial_{\mu}\Psi^{(+)}_{\nu}\right)
 \left(\partial_{\beta}\phi\right)
 \nonumber\\
&&
 -g_{gi}\,\epsilon^{\mu\nu\alpha\beta}\left(\partial_{\mu}\overline{\Psi^{(-)}_{\nu}}\right)\gamma_5\gamma_{\alpha}\psi^{(-)}
 \left(\partial_{\beta}\phi\right)
 -g_{gi}\,\epsilon^{\mu\nu\alpha\beta}\overline{\psi^{(-)}}\gamma_5\gamma_{\alpha}\left(\partial_{\mu}\Psi^{(-)}_{\nu}\right)
 \left(\partial_{\beta}\phi\right)\nonumber\\
&&
 -\frac{g^2_{gi}}{2}\,\epsilon^{\mu\nu\alpha\beta}\overline{\psi^{(+)}}\gamma_5\gamma_{\alpha}\left(\partial_{\beta}\phi\right)
 \left[\vphantom{\frac{A}{A}}\left(i\slpart_x+M_\Delta\right)n_\mu n_{\mu'}
 +\sln\left(i\partial_\mu n_{\mu'}+n_\mu i\partial_{\mu'}\right)\right.\nonumber\\
&&
 \phantom{-\frac{g^2_{gi}}{2}}\left.\vphantom{\frac{A}{A}}
 -2\sln n_\mu n_{\mu'} n\cdot i\partial\right]
 \left(g_{\nu\nu'}-\frac{1}{3}\,\gamma_\nu\gamma_{\nu'}\right)
 \epsilon^{\mu'\nu'\alpha'\beta'}\gamma_5\gamma_{\alpha'}\psi^{(+)}\left(\partial_{\beta'}\phi\right)\nonumber\\
&&
 -\frac{g^2_{gi}}{2}\,\epsilon^{\mu\nu\alpha\beta}\overline{\psi^{(-)}}\gamma_5\gamma_{\alpha}\left(\partial_{\beta}\phi\right)
 \left[\vphantom{\frac{A}{A}}\left(i\slpart_x+M_\Delta\right)n_\mu n_{\mu'}
 +\sln\left(i\partial_\mu n_{\mu'}+n_\mu i\partial_{\mu'}\right)\right.\nonumber\\
&&
 \phantom{-\frac{g^2_{gi}}{2}}\left.\vphantom{\frac{A}{A}}
 -2\sln n_\mu n_{\mu'} n\cdot i\partial\right]
 \left(g_{\nu\nu'}-\frac{1}{3}\,\gamma_\nu\gamma_{\nu'}\right)
 \epsilon^{\mu'\nu'\alpha'\beta'}\gamma_5\gamma_{\alpha'}\psi^{(-)}\left(\partial_{\beta'}\phi\right)\nonumber\\
&&
 +\frac{ig^2_{gi}}{2}\int d^4y
 \left[\epsilon^{\mu\nu\alpha\beta}\overline{\psi^{(+)}}\gamma_5\gamma_{\alpha}\left(\partial_{\beta}\phi\right)\right]_x
 \theta[n(x-y)]\left(i\slpart_x+M_\Delta\right)\nonumber\\
&&
 \phantom{+\frac{ig^2_{gi}}{2}}\times
 \left(g_{\nu\nu'}-\frac{1}{3}\,\gamma_\nu\gamma_{\nu'}\right)
 \partial_\mu\partial_{\mu'}\Delta^{(1)}(x-y)
 \left[\epsilon^{\mu'\nu'\alpha'\beta'}\gamma_5\gamma_{\alpha'}\psi^{(+)}\left(\partial_{\beta'}\phi\right)\right]_y\nonumber\\
&&
 -\frac{ig^2_{gi}}{2}\int d^4y
 \left[\epsilon^{\mu\nu\alpha\beta}\overline{\psi^{(-)}}\gamma_5\gamma_{\alpha}\left(\partial_{\beta}\phi\right)\right]_x
 \theta[n(x-y)]\left(i\slpart_x+M_\Delta\right)\nonumber\\
&&
 \phantom{+\frac{ig^2_{gi}}{2}}\times
 \left(g_{\nu\nu'}-\frac{1}{3}\,\gamma_\nu\gamma_{\nu'}\right)
 \partial_\mu\partial_{\mu'}\Delta^{(1)}(x-y)
 \left[\epsilon^{\mu'\nu'\alpha'\beta'}\gamma_5\gamma_{\alpha'}\psi^{(-)}\left(\partial_{\beta'}\phi\right)\right]_y
 \ .\label{deltaex5}
\end{eqnarray}

\section{S-Matrix Elements and Amplitudes}\label{smelements}

Since the Kadyshevsky rules as presented in appendix A of paper I
do not contain pair suppression, we're going to derive the
amplitudes from the S-matrix. The basic ingredients, namely the
interaction Hamiltonians, we have constructed in the previous
section (sections \ref{pscoupling}, \ref{pvcoupling} and
\ref{deltacoupling}) for different couplings. As in paper I we
also consider here equal initial and final states, i.e. $\pi N$
($MB$) scattering. For the results for general $MB$ initial and
final states we refer to appendix \ref{kadampinv}

\subsection{(Pseudo) Scalar Coupling}\label{smpsc}

For the pseudo scalar coupling case we collect all $g^2$
contributions to the S-matrix (see \eqref{psc5})
\begin{eqnarray}
 S^{(2)}
&=&
 (-i)^2\int d^4xd^4y\,\theta[n(x-y)]\mathcal{H}_I(x)\mathcal{H}_I(y)\nonumber\\
&=&
 -g^2\int d^4xd^4y\,\theta[n(x-y)]\left[\overline{\psi^{(+)}}\,\Gamma\phi\right]_x
 \left(i\slpart+M\right)
 \Delta^{+}(x-y)\left[\Gamma\psi^{(+)}\phi\vphantom{\frac{a}{a}}\right]_y\ ,\nonumber\\
 S^{(1)}
&=&
 (-i)\int d^4x\,\mathcal{H}_I(x)\nonumber\\
&=&
 \frac{g^2}{2}\int d^4xd^4y\left[\overline{\psi^{(+)}}\,\Gamma\phi\right]_x
 \theta[n(x-y)]\left(i\slpart_x+M\right)
 \Delta^{(1)}(x-y)\left[\Gamma\psi^{(+)}\phi\vphantom{\frac{a}{a}}\right]_y
 \ ,\label{smatrixps1}
\end{eqnarray}
which need to be added
\begin{eqnarray}
 S^{(2)}+S^{(1)}
&=&
 -\frac{ig^2}{2}\int d^4xd^4y\left[\overline{\psi^{(+)}}\,\Gamma\phi\right]_x
 \theta[n(x-y)]\left(i\slpart+M\right)
 \Delta(x-y)\left[\Gamma\psi^{(+)}\phi\vphantom{\frac{a}{a}}\right]_y
 \ .\quad\label{smatrixps2}
\end{eqnarray}
We see here that indeed the $\Delta^{(1)}(x-y)$ propagator in the
interaction Hamiltonian \eqref{psc5} is crucial, since it combines
with the $\Delta^{(+)}(x-y)$ propagator \eqref{smatrixps1} to form
a $\Delta(x-y)$ propagator \eqref{smatrixps2}. Together with the
$\theta[n(x-y)]$ in \eqref{smatrixps2} we recognize the causal
retarded (-like) character as we already mentioned in the section
\ref{pairsupp}. The S-matrix element is therefore covariant and if
we analyze its $n$-dependence using the GJ method \cite{Gross69}
as in paper I we would see that it is $n$-independent (for
vanishing external quasi momenta, of course).

Also we notice that the initial and final states are still
positive energy states. We started with a separation of positive
and negative energy states in section \ref{pairsupp} and after the
whole procedure this is still valid for the end-states. However,
we have to notice that inside an amplitude, negative energy
propagates via the $\Delta(x-y)$ propagator, but this is also the
case in our example of the infinite dense anti-nucleon star of
section \ref{pairsupp}. Moreover, in \cite{henk1} pair suppression
is assumed by only considering positive energy end-states, and
this is what we have achieved formally.

All the above observations are also valid in the case of (pseudo)
vector coupling and the $\pi N\Delta_{33}$ coupling of section
\ref{smpvc} and section \ref{smpnD}, respectively as we will see.

The last important observation is that in \eqref{smatrixps2} it
does not matter whether the derivative just acts on the
$\Delta(x-y)$ propagator or also on the $\theta[n(x-y)]$ function
\footnote{This is because $\delta(x^0-y^0)\Delta(x-y)=0$}.
Therefore, the $\bar{P}$-method of paper I can be applied,
although it is not really necessary. This situation is contrary to
ordinary baryon exchange, where the $\bar{P}$ method can only be
applied for the summed diagrams, as explained in paper I.\\

\begin{figure}[hbt]
\begin{center}
\begin{picture}(400,110)(0,0)
 \SetPFont{Helvetica}{9}
 \SetScale{1.0} \SetWidth{1.5}
 \DashArrowLine(50,90)(100,90){4}
 \ArrowLine(100,90)(150,90)
 \ArrowLine(50,20)(100,20)
 \DashArrowLine(100,20)(150,20){4}
 \Vertex(100,90){3}
 \Vertex(100,20){3}
 \ArrowLine(100,20)(100,90)

 \PText(40,90)(0)[b]{q}
 \PText(40,20)(0)[b]{p}
 \PText(160,90)(0)[b]{p'}
 \PText(160,20)(0)[b]{q'}
 \Text( 90,55)[]{$P$}


 \SetWidth{0.2}
 \ArrowLine(50,00)(100,20)
 \ArrowArc (65,55)(49.5,315,45)
 \ArrowLine(100,90)(150,110)
 \Text(40,00)[]{$\kappa$}
 \Text(160,110)[]{$\kappa'$}
 \Text(125,55)[]{$\kappa_1$}
 \PText(100,0)(0)[b]{(a)}

 \SetScale{1.0} \SetWidth{1.5}
 \DashArrowLine(220,90)(270,55){4}
 \DashArrowLine(320,55)(370,90){4}
 \ArrowLine(220,20)(270,55)
 \ArrowLine(320,55)(370,20)
 \Vertex(270,55){3}
 \Vertex(320,55){3}
 \ArrowLine(270,55)(320,55)

 \PText(210,90)(0)[b]{q}
 \PText(210,20)(0)[b]{p}
 \PText(380,90)(0)[b]{q'}
 \PText(380,20)(0)[b]{p'}
 \Text(295,65)[]{$P$}


 \SetWidth{0.2}
 \ArrowLine(220,115)(270,55)
 \ArrowArc (295,80)(35.5,225,315)
 \ArrowLine(320,55)(390,115)
 \Text(210,115)[]{$\kappa$}
 \Text(380,115)[]{$\kappa'$}
 \Text(295,40)[]{$\kappa_1$}
 \PText(290,00)(0)[b]{(b)}

\end{picture}
\end{center}
  \caption{\sl Baryon exchange (a) and resonance (b) diagrams}\label{fig:barexres}
\end{figure}
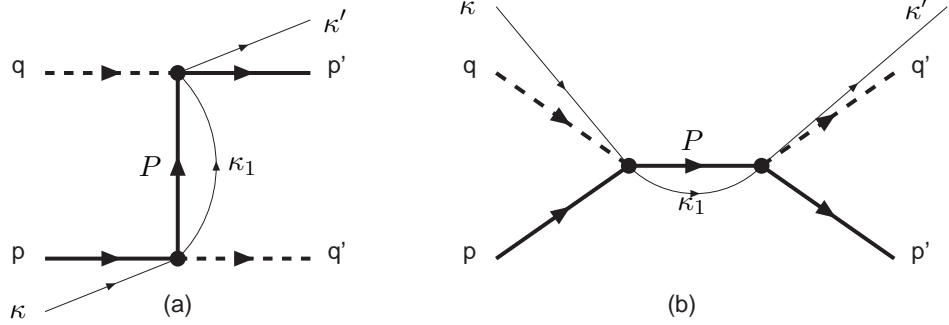

\noindent The summed S-matrix elements \eqref{smatrixps2} lead to
baryon exchange and resonance Kadyshevsky diagrams, which are
exposed in figure \ref{fig:barexres}. We're going to treat them
separately.

\noindent The amplitude for the (pseudo) scalar baryon exchange
and resonance resulting from the S-matrix in \eqref{smatrixps2}
are
\begin{eqnarray}
 M_{\kappa'\kappa}(u)
&=&
 \frac{g^2}{2}\,\int\frac{d\kappa_1}{\kappa_1+i\varepsilon}\,\bar{u}(p's')\left[
 \vphantom{\frac{A}{A}}\Gamma
 \left(\slP_u+M_B\right)\Gamma\right]u(ps)\Delta(P_u)
 \ ,\nonumber\\
 M_{\kappa'\kappa}(s)
&=&
 \frac{g^2}{2}\,\int\frac{d\kappa_1}{\kappa_1+i\varepsilon}\,\bar{u}(p's')\left[
 \vphantom{\frac{A}{A}}\Gamma
 \left(\slP_s+M_B\right)\Gamma\right]u(ps)\Delta(P_s)\
 .\qquad\qquad\label{smatrixps3}
\end{eqnarray}
Here $P_i=\Delta_i+n\bar{\kappa}-n\kappa_1$ and
$\Delta(P_i)=\epsilon(P_i^0)\delta(P_i^2-M_B^2)$ ($i=u,s$). The
$\Delta_i$ stand for
\begin{eqnarray}
 \Delta_u&=&\frac{1}{2}\left(p'+p-q'-q\right)\ ,\nonumber\\*
 \Delta_s&=&\frac{1}{2}\left(p'+p+q'+q\right)\ .\label{smatrixps4}
\end{eqnarray}
After expanding the $\delta(P_i^2-M_B^2)$-function the $\kappa_1$
integral can be performed
\begin{eqnarray}
 \delta(P_i^2-M_B^2)
&=&
 \frac{1}{|\kappa_1^{+}-\kappa_1^{-}|}
 \left(\delta(\kappa_1-\kappa_1^{+})+\delta(\kappa_1-\kappa_1^{-})\right)\ ,\nonumber\\
 \kappa_1^{\pm}
&=&
 \Delta_i\cdot n+\bar{\kappa}\pm A_i\ .\label{smatrixps5}
\end{eqnarray}
The $\epsilon(P^0_i)$ selects both solutions with a relative minus
sign. This yields for the amplitudes
\begin{eqnarray}
 M^{S}_{\kappa'\kappa}(u)
&=&
 \frac{g_{S}^2}{2}\ \bar{u}(p's')\left[M+M_B-\slQ+\bar{\kappa}\sln\right]u(ps)\
 \frac{1}{\left(\Delta_u\cdot n+\bar{\kappa}\right)^2-A^2_u+i\varepsilon}\ ,\nonumber\\
 M^{PS}_{\kappa'\kappa}(u)
&=&
 \frac{g_{PS}^2}{2}\ \bar{u}(p's')\left[M-M_B-\slQ+\bar{\kappa}\sln\right]u(ps)\
 \frac{1}{\left(\Delta_u\cdot n+\bar{\kappa}\right)^2-A^2_u+i\varepsilon}\ ,\nonumber\\
 M^{S}_{\kappa'\kappa}(s)
&=&
 \frac{g_{S}^2}{2}\ \bar{u}(p's')\left[M+M_B+\slQ+\bar{\kappa}\sln\right]u(ps)\
 \frac{1}{\left(\Delta_s\cdot n+\bar{\kappa}\right)^2-A^2_s+i\varepsilon}\ ,\nonumber\\
 M^{PS}_{\kappa'\kappa}(s)
&=&
 \frac{g_{PS}^2}{2}\ \bar{u}(p's')\left[M-M_B+\slQ+\bar{\kappa}\sln\right]u(ps)\
 \frac{1}{\left(\Delta_s\cdot n+\bar{\kappa}\right)^2-A^2_s+i\varepsilon}
 \ ,\label{smatrixps7}
\end{eqnarray}
where $S$ and $PS$ stand for {\it scalar} and {\it pseudo scalar},
respectively. Taking the limit of $\kappa'=\kappa=0$ in
\eqref{smatrixps7} we get
\begin{eqnarray}
 M^S_{00}(u)
&=&
 \frac{g_{S}^2}{2}\ \bar{u}(p's')\left[M+M_B-\slQ\right]u(ps)\,\frac{1}{u-M_B^2+i\varepsilon}
 \ ,\nonumber\\
 M^{PS}_{00}(u)
&=&
 \frac{g_{PS}^2}{2}\ \bar{u}(p's')\left[M-M_B-\slQ\right]u(ps)\,\frac{1}{u-M_B^2+i\varepsilon}
 \ ,\nonumber\\
 M^S_{00}(s)
&=&
 \frac{g_{S}^2}{2}\ \bar{u}(p's')\left[M+M_B+\slQ\right]u(ps)\,\frac{1}{s-M_B^2+i\varepsilon}\ ,\nonumber\\
 M^{PS}_{00}(s)
&=&
 \frac{g_{PS}^2}{2}\ \bar{u}(p's')\left[M-M_B+\slQ\right]u(ps)\,\frac{1}{s-M_B^2+i\varepsilon}\ ,\label{smatrixps8}
\end{eqnarray}
which is a factor $1/2$ of the result in \cite{henk1}. This factor
is because of the fact that we only took the positive energy
contribution. This difference can easily be intercepted by
considering an interaction Lagrangian as in \eqref{psc1} scaled by
a factor of $\sqrt{2}$ and eventually identifying $g/\sqrt{2}$ as
the physical coupling constant. We stress here that although we
have included absolute pair suppression formally, we still get a
factor $1/2$ of the usual Feynman expression.

In paper I we studied the $n$-dependence of the (approximation of
the) Kadyshevsky integral equation using the GJ method
\begin{eqnarray}
 M_{00} &=& M_{00}^{irr} + \int d\kappa\ M_{0\kappa}^{irr}\ G'_\kappa\ M_{\kappa0}\
 ,\nonumber\\*
 P^{\alpha\beta}\frac{\partial}{\partial n^\beta} M_{00}
&=&
 P^{\alpha\beta}\frac{\partial M_{00}^{irr}}{\partial n^\beta}
 +P^{\alpha\beta}\int d\kappa\left[\frac{\partial M_{0\kappa}^{irr}}{\partial n^\beta}\
 G'_\kappa\ M_{\kappa0} + M_{0\kappa}^{irr}\ G'_\kappa\
 \frac{\partial M_{\kappa0}}{\partial n^\beta}\right]\
 .\label{KIE12}
\end{eqnarray}
Important was that the integrand in the second line of
\eqref{KIE12} is of the form
\begin{equation}
 \int d\kappa\ \kappa\ h(\kappa) G'_\kappa\ ,\label{KIE13a}
\end{equation}
and in some cases a phenomenological "form factor" is needed
\begin{eqnarray}
 F(\kappa)=\left(\frac{\Lambda^2_\kappa}
 {\Lambda^2_\kappa-\kappa^2-i\epsilon(\kappa)\varepsilon}\right)^{N_\kappa}\
 .\label{KIE15}
\end{eqnarray}
For the details we refer to paper I. Whether \eqref{KIE13a}
applies and \eqref{KIE15} is necessary we need to check for every
exchange and resonance process.

In order to do so in the case of (P)S baryon exchange or resonance
we take a closer look at the denominators in \eqref{smatrixps7}
\begin{eqnarray}
 \left(\Delta_i\cdot n+\bar{\kappa}\right)^2-A^2_s
 =\Delta_i^2-M_B^2+2\Delta_i\cdot n\bar{\kappa}+\bar{\kappa}^2\
 .\label{smatrixps7a}
\end{eqnarray}
From this we conclude that all $n$-dependent terms in
\eqref{smatrixps7} are proportional to $\bar{\kappa}$, therefore
differentiating \eqref{smatrixps7} with respect to $n^\alpha$ will
yield a result linear proportional to $\kappa$. If we would only
consider (P)S baryon exchange or resonance in the Kadyshevsky
integral equation, then we indeed would have a situation as in
\eqref{KIE13a}. Looking at the powers of $\kappa,\kappa'$ in
\eqref{smatrixps7} we see that $h(\kappa)$ in \eqref{KIE13a} will
be of the order $O(\frac{1}{\kappa^2})$ and the phenomenological
"form factor" \eqref{KIE15} would not be necessary.

\subsection{(Pseudo) Vector Coupling}\label{smpvc}

The $g^2$ contributions of (pseudo) vector coupling in the second
and first order of the S-matrix are
\begin{eqnarray}
 S^{(2)}
&=&
 (-i)^2\int d^4xd^4y\,\theta[n(x-y)]\mathcal{H}_I(x)\mathcal{H}_I(y)\nonumber\\
&=&
 -\frac{f^2}{m_\pi^2}\,\int d^4xd^4y\,\theta[n(x-y)]
 \left[\overline{\psi^{(+)}}\Gamma_\mu\left(\partial^\mu\phi\right)\right]_x
 \left(i\slpart+M\right)
 \nonumber\\
&&
 \phantom{-\frac{f^2}{m_\pi^2}\,\int}\times
 \Delta^{+}(x-y)\left[\Gamma_\nu\psi^{(+)}\left(\partial^\nu\phi\right)\right]_y
 \ ,\nonumber\\
 S^{(1)}
&=&
 (-i)\int d^4x\,\mathcal{H}_I(x)\nonumber\\
&=&
 \frac{f^2}{2m_\pi^2}\,\int d^4xd^4y\,
 \left[\overline{\psi^{(+)}}\Gamma_\mu\left(\partial^\mu\phi\right)\right]_x
 \theta[n(x-y)]\left(i\slpart+M\right)
 \nonumber\\*
&&
 \phantom{\frac{f^2}{2m_\pi^2}\,\int}\times
 \Delta^{(1)}(x-y)\left[\Gamma_\nu\psi^{(+)}\left(\partial^\nu\phi\right)\right]_y
 \ .\label{smatrixpv1}
\end{eqnarray}
Adding the two together
\begin{eqnarray}
 S^{(2)}+S^{(1)}
&=&
 -\frac{if^2}{2m_\pi^2}\,\int d^4xd^4y\,\theta[n(x-y)]
 \left[\overline{\psi^{(+)}}\Gamma_\mu\left(\partial^\mu\phi\right)\right]_x
 \left(i\slpart+M\right)\nonumber\\
&&
 \phantom{-\frac{f^2}{m_\pi^2}\,\int}\times
 \Delta(x-y)\left[\Gamma_\nu\psi^{(+)}\left(\partial^\nu\phi\right)\right]_y\
 ,\label{smatrixpv2}
\end{eqnarray}
leads again to a covariant, $n$-independent result
($\kappa'=\kappa=0$). See the text below \eqref{smatrixps2} about
this issue and other important observations.

The two Kadyshevsky diagrams resulting from \eqref{smatrixpv2} are
the same as shown in figure \ref{fig:barexres}. The amplitudes
that go with them, in case of (pseudo) vector coupling, are
\begin{eqnarray}
 M_{\kappa'\kappa}(u)
&=&
 \frac{f^2}{2m_\pi^2}\,\int\frac{d\kappa_1}{\kappa_1+i\varepsilon}\,\bar{u}(p's')\left[
 \vphantom{\frac{A}{A}}\left(\Gamma\cdot q\right)
 \left(\slP_u+M_B\right)\left(\Gamma\cdot q'\right)\right]u(ps)\Delta(P_u)
 \ ,\nonumber\\
 M_{\kappa'\kappa}(s)
&=&
 \frac{f^2}{2m_\pi^2}\,\int\frac{d\kappa_1}{\kappa_1+i\varepsilon}\,\bar{u}(p's')\left[
 \vphantom{\frac{A}{A}}\left(\Gamma\cdot q'\right)
 \left(\slP_s+M_B\right)\left(\Gamma\cdot q\right)\right]u(ps)\Delta(P_s)
 \ ,\label{smatrixpv3}
\end{eqnarray}
where $P_i$ and $\Delta(P_i)$ are defined below
\eqref{smatrixps3}. As far as the $\kappa_1$ integration is
concerned we take similar steps as in \eqref{smatrixps5}.

After some (Dirac) algebra the amplitudes in \eqref{smatrixpv3}
become
\begin{eqnarray}
 M^{V}_{\kappa'\kappa}(u)
&=&
 \frac{f_{V}^2}{2m_\pi^2}\,\bar{u}(p's')\,\left[-\left(M-M_B\right)
 \left(-M^2+\frac{1}{2}\left(u_{p'q}+u_{pq'}\right)+2M\slQ\right.\right.\nonumber\\
&&
 \left.
 -\frac{1}{2}\left(\kappa'-\kappa\right)\left(p'-p\right)\cdot n
 +\frac{1}{2}\left(\kappa'-\kappa\right)\left[\sln,\slQ\right]
 -\frac{1}{2}\left(\kappa'-\kappa\right)^2\right)\nonumber\\
&&
 -\frac{1}{2}\left(u_{pq'}-M^2\right)\left(\slQ+\frac{1}{2}\left(\kappa'-\kappa\right)\sln\right)
 \nonumber\\
&&
 -\frac{1}{2}\left(u_{p'q}-M^2\right)\left(\slQ-\frac{1}{2}\left(\kappa'-\kappa\right)\sln\right)
 \nonumber\\
&&
 \left.+\bar{\kappa}\left(-(p'-p)\cdot n\ \slQ\,+2Q\cdot n\ \slQ
 +M^2\sln -\frac{\sln}{2}\left(u_{p'q}+u_{pq'}\right)\right)\right]u(p)\nonumber\\
&&
 \times\frac{1}{\left(\Delta_u\cdot n+\bar{\kappa}\right)^2-A^2_u+i\varepsilon}\
 ,\nonumber\\
 M^{PV}_{\kappa'\kappa}(u)
&=&
 \frac{f_{PV}^2}{2m_\pi^2}\,\bar{u}(p's')\,\left[-\left(M+M_B\right)
 \left(-M^2+\frac{1}{2}\left(u_{p'q}+u_{pq'}\right)+2M\slQ\right.\right.\nonumber\\
&&
 \left.
 -\frac{1}{2}\left(\kappa'-\kappa\right)\left(p'-p\right)\cdot n
 +\frac{1}{2}\left(\kappa'-\kappa\right)\left[\sln,\slQ\right]
 -\frac{1}{2}\left(\kappa'-\kappa\right)^2\right)\nonumber\\
&&
 -\frac{1}{2}\left(u_{pq'}-M^2\right)\left(\slQ+\frac{1}{2}\left(\kappa'-\kappa\right)\sln\right)
 \nonumber\\
&&
 -\frac{1}{2}\left(u_{p'q}-M^2\right)\left(\slQ-\frac{1}{2}\left(\kappa'-\kappa\right)\sln\right)
 \nonumber\\
&&
 \left.+\bar{\kappa}\left(-(p'-p)\cdot n\ \slQ\,+2Q\cdot n\ \slQ
 +M^2\sln
 -\frac{\sln}{2}\left(u_{p'q}+u_{pq'}\right)\right)\right]u(p)\nonumber\\*
&&
 \times\frac{1}{\left(\Delta_u\cdot n+\bar{\kappa}\right)^2-A^2_u+i\varepsilon}\
 ,\nonumber\\
 M^{V}_{\kappa'\kappa}(s)
&=&
 \frac{f_{V}^2}{2m_\pi^2}\,\bar{u}(p's')\,\left[-\left(M-M_B\right)
 \left(-M^2+\frac{1}{2}\left(s_{p'q'}+s_{pq}\right)-2M\slQ\right.\right.\nonumber\\
&&
 \left.
 -\frac{1}{2}\left(\kappa'-\kappa\right)\left(p'-p\right)\cdot n
 -\frac{1}{2}\left(\kappa'-\kappa\right)\left[\sln,\slQ\right]
 -\frac{1}{2}\left(\kappa'-\kappa\right)^2\right)\nonumber\\
&&
 +\frac{1}{2}\left(s_{p'q'}-M^2\right)\left(\slQ+\frac{1}{2}\left(\kappa'-\kappa\right)\sln\right)
 \nonumber\\
&&
 +\frac{1}{2}\left(s_{pq}-M^2\right)\left(\slQ-\frac{1}{2}\left(\kappa'-\kappa\right)\sln\right)
 \nonumber\\
&&
 \left.
 +\bar{\kappa}\left((p'-p)\cdot n\ \slQ\,+2Q\cdot n\ \slQ+M^2\,\sln\
 -\frac{\sln}{2}\left(s_{p'q'}+s_{pq}\right)\right)\right]u(p)\nonumber\\
&&
 \times\frac{1}{\left(\Delta_s\cdot n+\bar{\kappa}\right)^2-A^2_s+i\varepsilon}\ ,\nonumber\\
 M^{PV}_{\kappa'\kappa}(s)
&=&
 \frac{f_{PV}^2}{2m_\pi^2}\,\bar{u}(p's')\,\left[-\left(M+M_B\right)
 \left(-M^2+\frac{1}{2}\left(s_{p'q'}+s_{pq}\right)-2M\slQ\right.\right.\nonumber\\
&&
 \left.
 -\frac{1}{2}\left(\kappa'-\kappa\right)\left(p'-p\right)\cdot n
 -\frac{1}{2}\left(\kappa'-\kappa\right)\left[\sln,\slQ\right]
 -\frac{1}{2}\left(\kappa'-\kappa\right)^2\right)\nonumber\\
&&
 +\frac{1}{2}\left(s_{p'q'}-M^2\right)\left(\slQ+\frac{1}{2}\left(\kappa'-\kappa\right)\sln\right)
 \nonumber\\
&&
 +\frac{1}{2}\left(s_{pq}-M^2\right)\left(\slQ-\frac{1}{2}\left(\kappa'-\kappa\right)\sln\right)
 \nonumber\\
&&
 \left.
 +\bar{\kappa}\left((p'-p)\cdot n\ \slQ\,+2Q\cdot n\ \slQ+M^2\,\sln\
 -\frac{\sln}{2}\left(s_{p'q'}+s_{pq}\right)\right)\right]u(p)\nonumber\\
&&
 \times\frac{1}{\left(\Delta_s\cdot n+\bar{\kappa}\right)^2-A^2_s+i\varepsilon}\ .\label{smatrixpv5}
\end{eqnarray}
Here, (P)V stands for {\it (pseudo) vector}. Taking the limit
$\kappa'=\kappa=0$
\begin{eqnarray}
 M^{V}_{00}(u)
&=&
 \frac{f_{V}^2}{2m_\pi^2}\,\bar{u}(p's')\left[\vphantom{\frac{A}{A}}
 -\left(M-M_B\right)\left(-M^2+u+2M\slQ\right)-\left(u-M^2\right)\slQ
 \right]u(p)\nonumber\\
&&
 \times\frac{1}{u-M_B^2+i\varepsilon}\ ,\nonumber\\
 M^{PV}_{00}(u)
&=&
 \frac{f_{PV}^2}{2m_\pi^2}\,\bar{u}(p's')\left[\vphantom{\frac{A}{A}}
 -\left(M+M_B\right)\left(-M^2+u+2M\slQ\right)-\left(u-M^2\right)\slQ \right]u(p)
 \nonumber\\
&&
 \times\frac{1}{u-M_B^2+i\varepsilon}\ ,\nonumber\\
 M^{V}_{00}(s)
&=&
 \frac{f_{V}^2}{2m_\pi^2}\,\bar{u}(p's')\left[\vphantom{\frac{A}{A}}
 -\left(M-M_B\right)\left(-M^2+s-2M\slQ\right)+\left(s-M^2\right)\slQ\right]u(p)
 \nonumber\\*
&&
 \times\frac{1}{s-M_B^2+i\varepsilon}\ ,\nonumber\\
 M^{PV}_{00}(s)
&=&
 \frac{f_{PV}^2}{2m_\pi^2}\,\bar{u}(p's')\left[\vphantom{\frac{A}{A}}
 -\left(M+M_B\right)\left(-M^2+s-2M\slQ\right)+\left(s-M^2\right)\slQ\right]u(p)
 \nonumber\\
&&
 \times\frac{1}{s-M_B^2+i\varepsilon}\ ,\label{smatrixpv6}
\end{eqnarray}
where we, again, get factor $1/2$ from the result in \cite{henk1}
for the same reason as mentioned in section \ref{smpsc}.

Studying the $n$-dependence of the amplitudes \eqref{smatrixpv5}
in light of the $n$-dependence of the Kadyshevsky integral
equation as before (section \ref{smpsc}), we see that, again, all
$n$-dependent terms in \eqref{smatrixpv5} are linear proportional
to either $\kappa$ or $\kappa'$. Therefore, when we would only
consider (P)V baryon exchange or resonance in the Kadyshevsky
integral equation, we would, again, find ourself in a similar
situation as in \eqref{KIE13a}, when studying the $n$-dependence.
However, looking at the powers of $\kappa$ and $\kappa'$ in
\eqref{smatrixpv5} we notice that the function $h(\kappa)$ in
\eqref{KIE13a} is of higher order then $O(\frac{1}{\kappa^2})$.
Therefore, the phenomenological "form factor" \eqref{KIE15} would
be necessary.

\subsection{$\pi N\Delta_{33}$ Coupling}\label{smpnD}

As far as the $\pi N\Delta_{33}$ coupling is concerned we find the
following $g_{gi}^2$ contribution in the second and first order of
the S-matrix from \eqref{deltaex5}
\begin{eqnarray}
 S^{(2)}
&=&
 (-i)^2\int d^4xd^4y\,\theta[n(x-y)]\mathcal{H}_I(x)\mathcal{H}_I(y)\nonumber\\
&=&
 -g_{gi}^2\int d^4xd^4y\,\theta[n(x-y)]
 \left[\epsilon^{\mu\nu\alpha\beta}\overline{\psi^{(+)}}\gamma_5\gamma_\alpha\partial_\beta\phi\right]_x
 \partial_\mu^x\partial_{\mu'}^y\left(i\slpart+M_\Delta\right)
 \nonumber\\
&&
 \phantom{-g_{gi}^2\int}\times
 \Lambda_{\nu\nu'}\Delta^{+}(x-y)\left[\epsilon^{\mu'\nu'\alpha'\beta'}
 \gamma_5\gamma_{\alpha'}\psi^{(+)}\partial_{\beta'}\phi\right]_y
 \ ,\nonumber\\
 S^{(1)}
&=&
 (-i)\int d^4x\mathcal{H}_I(x)\nonumber\\
&=&
 \frac{g_{gi}^2}{2}\ \int d^4xd^4y
 \left[\epsilon^{\mu\nu\alpha\beta}\overline{\psi^{(+)}}\gamma_5\gamma_\alpha\partial_\beta\phi\right]_x
 \theta[n(x-y)]\partial_\mu\partial_{\mu'}\left(i\slpart+M_\Delta\right)
 \nonumber\\
&&
 \phantom{-g_{gi}^2\int}\times
 \left(g_{\nu\nu'}-\frac{1}{3}\,\gamma_\nu\gamma_{\nu'}\right)
 \Delta^{(1)}(x-y)\left[\epsilon^{\mu'\nu'\alpha'\beta'}
 \gamma_5\gamma_{\alpha'}\psi^{(+)}\partial_{\beta'}\phi\right]_y\nonumber\\
&&
 +\frac{ig_{gi}^2}{2}\,
 \left[\epsilon^{\mu\nu\alpha\beta}\overline{\psi^{(+)}}\gamma_5\gamma_\alpha\partial_\beta\phi\right]
 \left[\vphantom{\frac{A}{A}}\left(i\slpart+M_\Delta\right)n_\mu n_{\mu'}
 +\sln\left(n_\mu i\partial_{\mu'}+i\partial_\mu n_{\mu'}\right)\right.\nonumber\\
&&
 \phantom{+\frac{ig_{gi}^2}{2}\,\left[\right.}\left.\vphantom{\frac{A}{A}}
 -2\sln n_{\mu}n_{\mu'}n\cdot i\partial\right]
 \left(g_{\nu\nu'}-\frac{1}{3}\,\gamma_\nu\gamma_{\nu'}\right)
 \left[\epsilon^{\mu'\nu'\alpha'\beta'}\gamma_5\gamma_{\alpha'}\psi^{(+)}\partial_{\beta'}\phi\right]
 \ ,\label{smatrixpnD1}
\end{eqnarray}
where
\begin{eqnarray}
 \Lambda_{\mu\nu}
&=&
 -\left[g_{\mu\nu}-\frac{1}{3}\,\gamma_\mu\gamma_\nu+\frac{2\partial_\mu\partial_\nu}{3M^2}
 -\frac{1}{3M_\Delta}\left(\gamma_\mu i\partial_\nu-\gamma_\nu
 i\partial_\mu\right)\right]\ .\label{smatrixpnD2}
\end{eqnarray}
Because of the anti-symmetric property of the epsilon tensor all
derivative terms in \eqref{smatrixpnD2} do not contribute.

Upon addition of the two contributions in \eqref{smatrixpnD1} we
find
\begin{eqnarray}
 S^{(2)}+S^{(1)}
&=&
 -\frac{ig_{gi}^2}{2}\,\int d^4xd^4y
 \left[\epsilon^{\mu\nu\alpha\beta}\overline{\psi^{(+)}}\gamma_5\gamma_\alpha\partial_\beta\phi\right]_x
 \partial_\mu\partial_{\mu'}\left(i\slpart+M_\Delta\right)\left(g_{\nu\nu'}-\frac{1}{3}\,\gamma_\nu\gamma_{\nu'}\right)
 \nonumber\\
&&
 \phantom{-\frac{ig_{gi}^2}{2}\ \int}\times
 \theta[n(x-y)]\Delta(x-y)
 \left[\epsilon^{\mu'\nu'\alpha'\beta'}\gamma_5\gamma_{\alpha'}\psi^{(+)}\partial_{\beta'}\phi\right]_y\
 .\label{smatrixpnD3}
\end{eqnarray}
Again, we have a similar situation for the S-matrix element as in
section \ref{smpsc}. Therefore, we refer for the discussion of
\eqref{smatrixpnD3} to the text below \eqref{smatrixps2}.

A difference of this S-matrix element as compared of those of the
forgoing subsections (sections \ref{smpsc} and \ref{smpvc}) is
that the derivatives do not only act on the $\Delta(x-y)$
propagator in \eqref{smatrixpnD3}, but also on the
$\theta[n(x-y)]$. Therefore, the $\bar{P}$ method of paper I can
be applied. Of course this is obvious since this method was
introduced in order to incorporate terms like the second term on
the rhs of $S^{(1)}$ in \eqref{smatrixpnD1}.\\

\noindent As in the previous subsections (sections \ref{smpsc} and
\ref{smpvc}) two amplitudes arise from this S-matrix:
$\Delta_{33}$ exchange and resonance, whose the Kadyshevsky
diagrams are shown in figure \ref{fig:barexres}. The amplitudes
are
\begin{eqnarray}
 M_{\kappa'\kappa}(u)
&=&
 -\frac{g_{gi}^2}{2}\int\frac{d\kappa_1}{\kappa_1+i\varepsilon}\ \epsilon^{\mu\nu\alpha\beta}
 \bar{u}(p's')\gamma_\alpha\gamma_5 q_\beta\left(\bar{P}_u\right)_{\mu}\left(\bar{P}_u\right)_{\mu'}
 \left(\bar{\slP}_u+M_\Delta\right)\nonumber\\
&&
 \phantom{ -\frac{g_{gi}^2}{2}\int}\times
 \left(g_{\nu\nu'}-\frac{1}{3}\,\gamma_\nu\gamma_{\nu'}\right)
 \Delta(P_u)\,\epsilon^{\mu'\nu'\alpha'\beta'}\gamma_{\alpha'}\gamma_5 q'_{\beta'}
 u(ps)\ ,\nonumber\\
 M_{\kappa'\kappa}(s)
&=&
 -\frac{g_{gi}^2}{2}\int\frac{d\kappa_1}{\kappa_1+i\varepsilon}\ \epsilon^{\mu\nu\alpha\beta}
 \bar{u}(p's')\gamma_\alpha\gamma_5 q'_\beta\left(\bar{P}_s\right)_{\mu}\left(\bar{P}_s\right)_{\mu'}
 \left(\bar{\slP}_s+M_\Delta\right)\nonumber\\
&&
 \phantom{ -\frac{g_{gi}^2}{2}\int}\times
 \left(g_{\nu\nu'}-\frac{1}{3}\,\gamma_\nu\gamma_{\nu'}\right)
 \Delta(P_s)\,\epsilon^{\mu'\nu'\alpha'\beta'}\gamma_{\alpha'}\gamma_5
 q_{\beta'}u(ps)\ ,\qquad\quad\label{smatrixpnD4}
\end{eqnarray}
where $\bar{P}_i=P_i+n\kappa_1$, $i=u,s$ . $P_i$ and $\Delta(P_i)$
are as before.

Performing the $\kappa_1$ integral is in this situation even
simpler then in the previous cases (section \ref{smpsc} and
\ref{smpvc}). As can be seen from \eqref{smatrixps5} the
$\Delta(P_i)$ in \eqref{smatrixpnD4} selects two solutions for
$\kappa_1$ (with a relative minus sign, due to $\epsilon(P^0_i)$),
which only need to be applied to the quasi scalar propagator
$1/(\kappa_1+i\varepsilon)$. This, because the $\bar{P}_i$ is
$\kappa_1$-independent. Contracting all the indices in
\eqref{smatrixpnD4} the amplitudes become
\begin{eqnarray}
 M_{\kappa'\kappa}(u)
&=&
 -\frac{g_{gi}^2}{2}\,\bar{u}(p's')\left[\left(\bar{\slP}_u+M_\Delta\right)\left(
 \bar{P}_u^2\left(q'\cdot q\right)
 -\frac{1}{3}\,\bar{P}_u^2\slq\slq'
 -\frac{1}{3}\,\bar{\slP}_u\slq\left(\bar{P}_u\cdot q'\right)
 \right.\right.\nonumber\\
&&
 \left.\left.\phantom{-\frac{g_{gi}^2}{2}\,\bar{u}(p's')[}
 +\frac{1}{3}\,\bar{\slP}_u\slq'\left(\bar{P}_u\cdot q\right)
 -\frac{2}{3}\,\left(\bar{P}_u\cdot q'\right)\left(\bar{P}_u\cdot q\right)\right)\right]u(ps)\nonumber\\
&&
 \times\frac{1}{\left(\Delta_u\cdot n+\bar{\kappa}\right)^2-A^2_u+i\varepsilon}\
 ,\nonumber\\
 M_{\kappa'\kappa}(s)
&=&
 -\frac{g_{gi}^2}{2}\,\bar{u}(p's')\left[\left(\bar{\slP}_s+M_\Delta\right)\left(
 \bar{P}_s^2\left(q'\cdot q\right)
 -\frac{1}{3}\,\bar{P}_s^2\slq'\slq
 -\frac{1}{3}\,\bar{\slP}_s\slq'\left(\bar{P}_s\cdot q\right)
 \right.\right.\nonumber\\
&&
 \left.\left.\phantom{-\frac{g_{gi}^2}{2}\,\bar{u}(p's')[}
 +\frac{1}{3}\,\bar{\slP}_s\slq\left(\bar{P}_s\cdot q'\right)
 -\frac{2}{3}\,\left(\bar{P}_s\cdot q'\right)\left(\bar{P}_s\cdot q\right)\right)\right]u(ps)\nonumber\\
&&
 \times\frac{1}{\left(\Delta_s\cdot n+\bar{\kappa}\right)^2-A^2_s+i\varepsilon}\
 ,\label{smatrixpnD6}
\end{eqnarray}
which leads, after some (Dirac) algebra, to
\begin{eqnarray}
 M_{\kappa'\kappa}(u)
&=&
 -\frac{g_{gi}^2}{2}\,\bar{u}(p's')\left[\vphantom{\frac{A}{A}}
 \frac{1}{2}\,\bar{P}_u^2\,\left(M+M_\Delta-\slQ+\bar{\kappa}\sln\right)(2m^2-t_{q'q})\right.\nonumber\\
&&
 -\frac{1}{3}\,\bar{P}_u^2\,
 \left(\vphantom{\frac{A}{A}}\left(M+M_\Delta\right)\slq\slq'
 +\frac{1}{2}\,\left(u_{pq'}-M^2\right)\slq
 \right.\nonumber\\
&&
 \phantom{-\frac{1}{3}\,\bar{P}_u^2\,\left(\right.}\left.\vphantom{\frac{A}{A}}
 +\frac{1}{2}\,\left(s_{pq}+t_{q'q}-M^2-4m^2\right)\slq'
 +\bar{\kappa}\sln\slq\slq'\right)\nonumber\\
&&
 -\frac{1}{12}\left(\bar{P}_u^2\,\slq
 +\frac{M_\Delta}{2}\left(s_{pq}-M^2-2m^2\right)
 -\frac{M_\Delta}{2}\,\slq'\slq+M_\Delta\bar{\kappa}\sln\slq\right)
 \left(\vphantom{\frac{A}{A}}-4m^2\right.\nonumber\\
&&
 \phantom{-\frac{1}{12}\left(\right.}\left.
 +s_{p'q'}-u_{pq'}+t_{q'q}-2\bar{\kappa}(p'-p)\cdot n
 +4\bar{\kappa}n\cdot Q-\left(\kappa'^2-\kappa^2\right)\vphantom{\frac{A}{A}}\right)\nonumber\\
&&
 +\frac{1}{12}\left(\bar{P}_u^2\,\slq'
 +\frac{M_\Delta}{2}\left(M^2-u_{pq'}\right)
 -\frac{M_\Delta}{2}\,\slq\slq'+M_\Delta\bar{\kappa}\sln\slq'\right)
 \left(\vphantom{\frac{A}{A}}-4m^2
 \right.\nonumber\\
&&
 \phantom{+\frac{1}{12}\left(\right.}\left.
 +s_{pq}-u_{p'q}+t_{q'q}+2\bar{\kappa}(p'-p)\cdot n
 +4\bar{\kappa}n\cdot Q+\left(\kappa'^2-\kappa^2\right)\vphantom{\frac{A}{A}}\right)\nonumber\\
&&
 -\frac{1}{24}\left(M+M_\Delta-\slQ +\bar{\kappa}\sln\vphantom{\frac{A}{A}}\right)
 \left(\vphantom{\frac{A}{A}}-4m^2+s_{p'q'}-u_{pq'}+t_{q'q}\right.\nonumber\\
&&
 \phantom{-\frac{1}{24}\left(\right.}\left.
 -2\bar{\kappa}(p'-p)\cdot n
 +4\bar{\kappa}n\cdot Q-\left(\kappa'^2-\kappa^2\right)
 \vphantom{\frac{A}{A}}\right)\left(\vphantom{\frac{A}{A}}-4m^2+s_{pq}
 \right.\nonumber\\
&&
 \phantom{-\frac{1}{24}\left(\right.}\left.\left.
 -u_{p'q}+t_{q'q}+2\bar{\kappa}(p'-p)\cdot n
 +4\bar{\kappa}n\cdot Q+\left(\kappa'^2-\kappa^2\right)
 \vphantom{\frac{A}{A}}\right)\right]u(ps)\nonumber\\
&&
 \times\frac{1}{\left(\Delta_u\cdot n+\bar{\kappa}\right)^2-A^2_u+i\varepsilon}\
 ,\nonumber
\end{eqnarray}
\begin{eqnarray}
 M_{\kappa'\kappa}(s)
&=&
 -\frac{g_{gi}^2}{2}\,\bar{u}(p's')\left[\vphantom{\frac{A}{A}}
 \frac{1}{2}\,\bar{P}_s^2\,\left(M+M_\Delta+\slQ+\bar{\kappa}\sln\right)(2m^2-t_{q'q})
 \right.\nonumber\\
&&
 -\frac{1}{3}\,\bar{P}_s^2\,
 \left(\vphantom{\frac{A}{A}}\left(M+M_\Delta\right)\slq'\slq
 -\frac{1}{2}\,\left(s_{pq}-M^2\right)\slq'\right.\nonumber\\
&&
 \phantom{-\frac{1}{3}\,\bar{P}_s^2\,\left(\right.}\left.\vphantom{\frac{A}{A}}
 -\frac{1}{2}\,\left(u_{pq'}+t_{q'q}-M^2-4m^2\right)\slq
 +\bar{\kappa}\sln\slq'\slq\right)\nonumber\\
&&
 -\frac{1}{12}\left(
 \bar{P}_s^2\slq'
 +\frac{M_\Delta}{2}\left(M^2+2m^2-u_{pq'}\right)
 +\frac{M_\Delta}{2}\,\slq\slq'+M_\Delta\bar{\kappa}\sln\slq'\right)
 \left(\vphantom{\frac{A}{A}}4m^2\right.\nonumber\\*
&&
 \phantom{-\frac{1}{12}\left(\right.}\left.
 +s_{pq}-u_{p'q}-t_{q'q}+2\bar{\kappa}(p'-p)\cdot n
 +4\bar{\kappa}n\cdot Q+\left(\kappa'^2-\kappa^2\right)\vphantom{\frac{A}{A}}\right)\nonumber\\
&&
 +\frac{1}{12}\left(\bar{P}_s^2\slq
 +\frac{M_\Delta}{2}\left(s_{pq}-M^2\right)
 +\frac{M_\Delta}{2}\,\slq'\slq+M_\Delta\bar{\kappa}\sln\slq\right)
 \left(\vphantom{\frac{A}{A}}4m^2\right.\nonumber\\
&&
 \phantom{+\frac{1}{12}\left(\right.}\left.
 +s_{p'q'}-u_{pq'}-t_{q'q}-2\bar{\kappa}(p'-p)\cdot n
 +4\bar{\kappa}n\cdot Q-\left(\kappa'^2-\kappa^2\right)\vphantom{\frac{A}{A}}\right)\nonumber\\
&&
 -\frac{1}{24}\left(M+M_\Delta+\slQ+\bar{\kappa}\sln\vphantom{\frac{A}{A}}\right)
 \left(\vphantom{\frac{A}{A}}4m^2+s_{p'q'}-u_{pq'}-t_{q'q}\right.\nonumber\\
&&
 \phantom{-\frac{1}{24}\left(\right.}\left.\vphantom{\frac{A}{A}}
 -2\bar{\kappa}(p'-p)\cdot n
 +4\bar{\kappa}n\cdot Q-\left(\kappa'^2-\kappa^2\right)\right)
 \left(\vphantom{\frac{A}{A}}4m^2+s_{pq}\right.\nonumber\\
&&
 \phantom{-\frac{1}{24}\left(\right.}\left.\left.\vphantom{\frac{A}{A}}
 -u_{p'q}-t_{q'q}+2\bar{\kappa}(p'-p)\cdot n
 +4\bar{\kappa}n\cdot Q+\left(\kappa'^2-\kappa^2\right)\right)\right]u(ps)\nonumber\\
&&
 \times\frac{1}{\left(\Delta_s\cdot n+\bar{\kappa}\right)^2-A^2_s+i\varepsilon}\
 ,\label{smatrixpnD7}
\end{eqnarray}
where
\begin{eqnarray}
 \bar{P}_u^2
&=&
 \frac{1}{2}\left(u_{p'q}+u_{pq'}\right)-\frac{1}{4}(\kappa'-\kappa)^2+2\bar{\kappa}\Delta_u\cdot
 n+\bar{\kappa}^2\ ,\nonumber\\
 \bar{P}_s^2
&=&
 \frac{1}{2}\left(s_{p'q'}+s_{pq}\right)-\frac{1}{4}(\kappa'-\kappa)^2+2\bar{\kappa}\Delta_s\cdot
 n+\bar{\kappa}^2\ ,\label{smatrixpnD7a}
\end{eqnarray}
and
\begin{eqnarray}
 \slq'
&=&
 \slQ-\frac{1}{2}\,\sln(\kappa'-\kappa)\ ,\nonumber\\
 \slq
&=&
 \slQ+\frac{1}{2}\,\sln(\kappa'-\kappa)\ ,\nonumber\\
 \slq'\slq
&=&
 -2M\slQ+\frac{1}{2}\,\left(s_{p'q'}+s_{pq}\right)
 -M^2-\frac{1}{2}\,(\kappa'-\kappa)(p'-p)\cdot n
 \nonumber\\
&&
 +\frac{1}{2}\,(\kappa'-\kappa)\left[\slQ,\sln\right]
 -\frac{1}{2}\,(\kappa'-\kappa)^2\ ,\nonumber\\
 \slq\slq'
&=&
 2M\slQ+\frac{1}{2}\,\left(u_{p'q}+u_{pq'}\right)
 -M^2-\frac{1}{2}\,(\kappa'-\kappa)(p'-p)\cdot n
 \nonumber\\
&&
 -\frac{1}{2}\,(\kappa'-\kappa)\left[\slQ,\sln\right]
 -\frac{1}{2}\,(\kappa'-\kappa)^2\ ,\nonumber\\
 \sln\slq'
&=&
 M\sln-(n\cdot p')
 -\frac{1}{2}\,\left[\slQ,\sln\right]+n\cdot Q
 -\frac{1}{2}\,(\kappa'-\kappa)\ ,\nonumber\\
 \sln\slq
&=&
 -M\sln+(n\cdot p')
 -\frac{1}{2}\,\left[\slQ,\sln\right]+n\cdot Q
 +\frac{1}{2}\,(\kappa'-\kappa)\ ,\nonumber\\
 \sln\slq'\slq
&=&
 -M^2\sln+\frac{1}{2}\,\left(s_{p'q'}+s_{pq}\right)\sln
 -\frac{1}{2}\,(\kappa'-\kappa)n\cdot(p'-p)\sln
 \nonumber\\
&&
 +\left(\kappa'-\kappa\right)\left(n\cdot Q\right)\sln
 -(\kappa'-\kappa)\slQ-2n\cdot(p'-p)\slQ
 -\frac{1}{2}\,(\kappa'-\kappa)^2\sln\ ,\nonumber\\
 \sln\slq\slq'
&=&
 -M^2\sln+\frac{1}{2}\,\left(u_{p'q}+u_{pq'}\right)\sln
 -\frac{1}{2}\,(\kappa'-\kappa)n\cdot(p'-p)\sln
 \nonumber\\*
&&
 -\left(\kappa'-\kappa\right)\left(n\cdot Q\right)\sln
 +(\kappa'-\kappa)\slQ+2n\cdot(p'-p)\slQ
 -\frac{1}{2}\,(\kappa'-\kappa)^2\sln
 \ .\label{smatrixpnD8}
\end{eqnarray}
Taking the limit $\kappa'=\kappa=0$ yields
\begin{eqnarray}
 M_{00}(u)
&=&
 -\frac{g_{gi}^2}{2}\,\bar{u}(p's')\left[\vphantom{\frac{A}{A}}
 \frac{u}{2}\left(M+M_\Delta-\slQ\right)(2m^2-t)
 \right.\nonumber\\
&&
 \phantom{-\frac{g_{gi}^2}{2}\,\bar{u}(p's')[}
 -\frac{u}{3}\left(\vphantom{\frac{A}{A}}\left(M+M_\Delta\right)
 \left(2M\slQ+u-M^2\right)
 -m^2\slQ\right)\nonumber\\
&&
 \phantom{-\frac{g_{gi}^2}{2}\,\bar{u}(p's')[}
 -\frac{1}{6}\left(\vphantom{\frac{A}{A}}
 u\slQ+M_\Delta\left(M\slQ-m^2\right)\right)
 \left(\vphantom{\frac{A}{A}}M^2-m^2-u\right)\nonumber\\
&&
 \phantom{-\frac{g_{gi}^2}{2}\,\bar{u}(p's')[}
 +\frac{1}{6}\left(\vphantom{\frac{A}{A}}
 u\slQ+M_\Delta\left(M^2-u-M\slQ\right)\right)
 \left(\vphantom{\frac{A}{A}}M^2-m^2-u\right)\nonumber\\
&&
 \phantom{-\frac{g_{gi}^2}{2}\,\bar{u}(p's')[} \left.
 -\frac{1}{6}\left(M+M_\Delta-\slQ\vphantom{\frac{A}{A}}\right)
 \left(\vphantom{\frac{A}{A}}M^2-m^2-u\vphantom{\frac{A}{A}}\right)^2\right]u(ps)
 \nonumber\\
&&
 \times\frac{1}{u-M_\Delta^2+i\varepsilon}\ ,\nonumber\\
 M_{00}(s)
&=&
 -\frac{g_{gi}^2}{2}\,\bar{u}(p's')\left[\vphantom{\frac{A}{A}}
 \frac{s}{2}\left(M+M_\Delta+\slQ\right)(2m^2-t)\right.\nonumber\\
&&
 \phantom{-\frac{g_{gi}^2}{2}\,\bar{u}(p's')\left[\right.}
 -\frac{s}{3}\left(\vphantom{\frac{A}{A}}\left(M+M_\Delta\right)\left(-2M\slQ+s-M^2\right)
 +m^2\slQ\right)\nonumber\\
&&
 \phantom{-\frac{g_{gi}^2}{2}\,\bar{u}(p's')\left[\right.}
 -\frac{1}{6}\left(s\slQ +M_\Delta\left(M\slQ+m^2\right)\vphantom{\frac{A}{A}}\right)
 \left(\vphantom{\frac{A}{A}}s-M^2+m^2\right)\nonumber\\
&&
 \phantom{-\frac{g_{gi}^2}{2}\,\bar{u}(p's')\left[\right.}
 +\frac{1}{6}\left(s\slQ+M_\Delta\left(s-M^2-M\slQ\right)\vphantom{\frac{A}{A}}\right)
 \left(\vphantom{\frac{A}{A}}s-M^2+m^2\right)\nonumber\\
&&
 \phantom{-\frac{g_{gi}^2}{2}\,\bar{u}(p's')\left[\right.}
 -\frac{1}{6}\left.\left(M+M_\Delta+\slQ\vphantom{\frac{A}{A}}\right)
 \left(\vphantom{\frac{A}{A}}s-M^2+m^2\right)^2\right]u(ps)
 \nonumber\\
&&
 \times\frac{1}{s-M_\Delta^2+i\varepsilon}\ .\label{smatrixpnD9}
\end{eqnarray}
Considering only the $\Delta_{33}$ exchange and resonance in the
Kadyshevsky integral equation and study its $n$-dependence, we see
from \eqref{smatrixpnD7} and \eqref{smatrixpnD8} that we have a
similar situation as in the previous subsection (section
\ref{smpvc}): all $n$-dependent terms in \eqref{smatrixpnD7} and
\eqref{smatrixpnD8} are either proportional to $\kappa$ or to
$\kappa'$ and therefore \eqref{KIE13a} applies. The function
$h(\kappa)$ is such that the phenomenological "form factor"
\eqref{KIE15} is necessary.

\section{Invariants and Partial Wave Expansion}\label{pwe}

In elastic scattering processes important (indirect) observables
are the phase-shifts. In this section we introduce the
phase-shifts by introducing the partial wave expansion, which is
particularly convenient for solving the Kadyshevsky integral
equation . By also using the helicity basis we're able to link the
amplitudes obtained in paper I and the previous section (section
\ref{smelements}) to the phase-shifts.

\subsection{Amplitudes and Invariants}

Following the standard procedure, see e.g. \cite{brans73}, the
most general form of the parity-conserving amplitude describing
$\pi N$-scattering in Kadyshevsky formalism is
\begin{eqnarray}
  M_{\kappa'\kappa}
&=&
 \bar{u}(p's')\left[ \vphantom{\frac{A}{A}} A + B\slQ
 + A'\sln + B'\left[\sln,\slQ\right]\right] u(ps)\ , \label{pwe2}
\end{eqnarray}
where the invariants $A,B,A'$ and $B'$ are functions of the
Mandelstam variables and of $\kappa$ and $\kappa'$. The
contribution of the invariants to the various exchange processes
is given in appendix \ref{kadampinv}.

In proceeding we don't keep $n^\mu$ general, but choose it to be
\cite{Kad67,Kad70}
\begin{equation}
 n^\mu = \frac{(p+q)^\mu}{\sqrt{s_{pq}}} = \frac{(p'+q')^\mu}{\sqrt{s_{p'q'}}}\ .\label{pwe3}
\end{equation}
With this choice, $n^\mu$ is not an independent variable anymore
and the number of invariants is reduced to two. This is made
explicit as follows
\begin{eqnarray}
 \bar{u}(p's')\left[\sln\right]u(ps)
&=&
 \frac{1}{\sqrt{s_{p'q'}}+\sqrt{s_{pq}}}\bar{u}(p's')\left[M_f+M_i+2\slQ\right]u(ps)
 \ ,\nonumber\\
 \bar{u}(p's')\left[\left[\sln,\slQ\right]\vphantom{\frac{A}{A}}\right]u(ps)
&=&
 0\ .\label{pwe4}
\end{eqnarray}
As a result of the choice \eqref{pwe3} the invariants $A$ and $B$
in \eqref{pwe2} receive contributions from the invariant $A'$. We,
therefore, redefine the amplitude
\begin{eqnarray}
  M_{\kappa'\kappa}
&=&
 \bar{u}(p's')\left[ \vphantom{\frac{A}{A}} A'' + B''\slQ\right] u(ps)
 \ ,\nonumber\\
 A''
&=&
 A+\frac{1}{\sqrt{s_{p'q'}}+\sqrt{s_{pq}}}\left(M_f+M_i\right)A'
 \ ,\nonumber\\
 B''
&=&
 B+\frac{2}{\sqrt{s_{p'q'}}+\sqrt{s_{pq}}}\,A'\ .\label{pwe5}
\end{eqnarray}

Besides the invariants $A''$ and $B''$, we also introduce the
invariants $F$ and $G$ very similar to \cite{Pil67} \footnote{The
difference is a normalization factor.}
\begin{eqnarray}
 M_{\kappa'\kappa}
&=&
 \chi^\dagger(s')\left[F+G\left(\mbox{\boldmath $\sigma$}\cdot{\bf \hat{p}'}\right)
 \left(\mbox{\boldmath $\sigma$}\cdot{\bf \hat{p}}\right)\vphantom{\frac{A}{A}}
 \right]\chi(s)\ ,\label{pwe6}
\end{eqnarray}
since we will use the helicity basis. Here, $\chi(s)$ is a
helicity state vector. In \cite{henk1} this expansion was used in
combination with the expansion of the amplitude in Pauli spinor
space. The connection between the two are also given there.

The relation between the invariants $A'',B''$ and $F,G$ is given
by
\begin{eqnarray}
 F
&=&
 \sqrt{(E'+M_f)(E+M_i)}\ \left\{\vphantom{\frac{A}{A}}
 A'' + \frac{1}{2}\left[ (W'-M_f)+(W-M_i)\right]\ B''\right\}\ ,\nonumber \\
&&
 \nonumber \\
 G
&=&
 \sqrt{(E'-M_f)(E-M_i)}\ \left\{\vphantom{\frac{A}{A}}
 -A'' + \frac{1}{2}\left[ (W'+M_f)+(W+M_i)\right]\ B''\right\}\ .\label{pwe8}
\end{eqnarray}

\subsection{Helicity Amplitudes and Partial Waves}\label{helamp}

In this subsection we want to link the invariants $A''$ and $B''$
to experimental observable phase-shifts. This is done by using the
helicity basis and the partial wave expansion. The procedure is
based on \cite{Jac59} and similar to \cite{Ver76}.

The helicity amplitude in terms of the invariants $F$ and $G$ (see
\eqref{pwe6}) is
\begin{eqnarray}
  M_{\kappa'\kappa}(\lambda_f,\lambda_i)
&=&
 C_{\lambda_f,\lambda_i}(\theta,\phi)
 \left[\vphantom{\frac{A}{A}} F +4\lambda_f\lambda_i G\right]\ ,
 \label{pwe9}
\end{eqnarray}
where
\begin{eqnarray}
 C_{\lambda_f,\lambda_i}(\theta,\phi)
&=&
 \chi^\dagger_{\lambda_f}(\hat{\bf p}')\cdot
 \chi_{\lambda_i}(\hat{\bf p}) = D^{1/2*}_{\lambda_i\lambda_f}(\phi,\theta,-\phi)\ .\label{pwe10}
\end{eqnarray}
Here, $D^J_{mm'}(\alpha,\beta,\gamma)$ are the Wigner D-matrices
\cite{Jac59} and the angles $\theta$ and $\phi$ are defined as the
polar angles of the CM-momentum ${\bf p}'$ in a coordinate system
that has ${\bf p}$ along the positive z-axis. In the following we
take as the scattering plane the xz-plane, i.e. $\phi=0$.
Furthermore, we introduce the functions $f_{1,2}$ by
\begin{equation}
 F = \frac{f_1}{4\pi}\ \ ,\ \ G =  \frac{f_2}{4\pi}\ .\label{pwe11}
\end{equation}
Then, with these settings the helicity amplitude \eqref{pwe9} is
\begin{eqnarray}
  M_{\kappa'\kappa}(\lambda_f,\lambda_i)
&=&
 \frac{1}{4\pi}\,d^{1/2}_{\lambda_i\lambda_f}(\theta)
 \left(\vphantom{\frac{A}{A}} f_1 + 4\lambda_f\lambda_i f_2\right)\ ,
 \label{pwe12}
\end{eqnarray}
Next, we make the partial wave expansion of the helicity
amplitudes in the CM-frame very similar to \cite{Pil67}
\footnote{The difference is again a normalization factor. We use
the same normalization as \cite{henk1} and \cite{Ver76}.}
\begin{eqnarray}
 M_{\kappa'\kappa}(\lambda_f\lambda_i)
&=&
 (4\pi)^{-1}\sum_J (2J+1) M^J_{\kappa'\kappa}(\lambda_f\lambda_i)\
 D^{J*}_{\lambda_i,\lambda_f}(\phi,\theta,-\phi)\ ,\nonumber\\
&=&
 (4\pi)^{-1}e^{i(\lambda_i-\lambda_f)\phi}\sum_J (2J+1)
 M^J_{\kappa'\kappa}(\lambda_f\lambda_i)\ d^{J}_{\lambda_i,\lambda_f}(\theta)\ ,
 \quad\label{pwe14}
\end{eqnarray}
Using the partial wave expansion as in \eqref{pwe14} we obtain the
Kadyshevsky integral equation (paper I) in the partial wave basis.
Here, we just show the result; for the details we refer to
\cite{Ver76}
\begin{eqnarray}
 M^J_{00}(\lambda_f\lambda_i)
&=&
 M^{irr\ J}_{00}(\lambda_f\lambda_i)
 +\sum_{\lambda_n} \int_0^\infty k_n^2dk_n\
 M^{irr\ J}_{0\kappa}(\lambda_f\lambda_n)
 \nonumber\\
&&
 \times G'_\kappa\left(W_n;W\right)\
 M^J_{\kappa0}(\lambda_n\lambda_i)\ . \label{pwe16}
\end{eqnarray}
As mentioned in paper I, the $\kappa$-label is fixed after
integration.

Because of the summation over the intermediate helicity states the
partial wave Kadyshevsky integral equation \eqref{pwe16} is a
coupled integral equation. It can be decoupled using the
combinations $f_{(J-1/2)+}$ and $f_{(J+1/2)-}$ defined by
\begin{equation}
 \left(\begin{array}{c}
        f_{L+} \\
        f_{(L+1)-}
       \end{array} \right) =
 \left(\begin{array}{cc}
        +1 & +1 \\
        +1 & -1
       \end{array} \right)\
 \left(\begin{array}{c}
        M^J(+1/2\ 1/2) \\
        M^J(-1/2\ 1/2)
       \end{array} \right)\ ,
 \label{pwe18}
\end{equation}
here we introduced $L \equiv J-1/2$ \footnote{The labels $L+$ and
$(L+1)-$ in \eqref{pwe18} and their relation to total angular
momentum $J$ come from parity arguments as is best explained in
\cite{Jac59}.}.

In \eqref{pwe18} and in the following we omit the subscript $00$
for the final amplitudes where $\kappa$ and $\kappa'$ are put to
zero.

A similar expansion as \eqref{pwe18} holds for $M^{irr\
J}_{\kappa'\kappa}(\lambda_f\lambda_i)$ and what one gets is
\begin{eqnarray}
 f_{L\pm}(W',W)
&=&
 f^{irr}_{L\pm}(W',W) +
 \sum_{\lambda_n} \int_0^\infty k_n^2dk_n\ f^{irr}_{L\pm}(W',W_n)
 \nonumber\\
&&
 \times G\left(W_n;W\right)\ f_{L\pm}(W_n,W)\ .\label{pwe19}
\end{eqnarray}

The two-particle unitarity relation for the partial-wave helicity
states reads \cite{Pil67}
\begin{equation}
 i\left[ M^J(\lambda_f\lambda_i)- M^{J*}(\lambda_i\lambda_f)\right] =
 2 \sum_{\lambda_n} k\ M^{J*}(\lambda_f\lambda_n) M^J(\lambda_i\lambda_n)\ ,
 \label{pwe20}
\end{equation}
In a similar manner as for the partial wave Kadyshevsky integral
equation \eqref{pwe16}, also the unitarity relation (\ref{pwe20})
decouples for the combinations \eqref{pwe18}. One gets
\begin{equation}
 Im f_{L\pm}(W) = k\ f_{L\pm}^*(W) f_{L\pm}(W)\ ,\label{pwe21}
\end{equation}
which allows for the introduction of the elastic phase-shifts
\begin{equation}
 f_{L\pm}(W) = \frac{1}{k}\ e^{i\delta_{L\pm}(W)} \sin \delta_{L\pm}(W)\ .
 \label{pwe22}
\end{equation}
From \eqref{pwe22} we see that once we have found the invariants
$f_{L\pm}(W)$ by solving the partial wave Kadyshevsky integral
equation \eqref{pwe16} we can determine the phase-shifts. The
relation between the invariants $f_{L\pm}(W)$ and the invariants
$f_{1,2}$ is
\begin{eqnarray}
 f_{L\pm}
&=&
 \frac{1}{2}\int_{-1}^{+1}dx\ \left[P_{L}(x) f_1+ P_{L \pm 1}(x) f_2\right] \nonumber\\
&=&
 f_{1,L} + f_{2,L \pm 1}\ , \label{pwe27}
\end{eqnarray}
where $x=cos\,\theta$.

\subsection{Partial Wave Projection}

Via the equations \eqref{pwe27}, \eqref{pwe11} and (\ref{pwe8}),
the partial waves $f_{L\pm}$ can be traced back to the partial
wave projection of the invariant amplitudes $A''$ and $B''$, which
means that we are looking for the partial wave projections of the
invariants $A,B,A',B'$.

Before doing so we include form factors in the same way as in
\cite{henk1}. As mentioned there, they are needed to regulate the
high energy behavior and to take into account the extended size of
the mesons and baryons. We take them to be
\begin{eqnarray}
 F(\Lambda)
&=&
 e^{\frac{-\left({\bf k}_f-{\bf k}_i\right)^2}{\Lambda^2}}\qquad\text{for $t$-channel}
 \ ,\nonumber\\
 F(\Lambda)
&=&
 e^{\frac{-\left({\bf k}_f^2+{\bf k}^2_i\right)}{\Lambda^2}}\qquad\text{for $u,s$-channel}
 \ .\label{pwe28}
\end{eqnarray}

The partial wave projection includes an integration over
$cos\,\theta=x$. We, therefore, investigate the $x$-dependence of
the invariants. Main concern is the propagators. We want to write
them in the form $1/(z\pm x)$, which is especially difficult for
the propagators in the t-channel, because of the square root in
$A_t$. We therefore use the identity
\begin{eqnarray}
 \frac{1}{\omega(\omega+a)}
&=&
 \frac{1}{\omega^2-a^2}+\frac{2a}{\pi}\,\int_0^\infty\frac{d\lambda}{\lambda^2+a^2}
 \left[\frac{1}{\omega^2+\lambda^2}-\frac{1}{\omega^2-a^2}\right]
 \ ,\quad\label{pwe29}
\end{eqnarray}
which holds for $\omega,a\in\mathbb{R}$. With this identity we
write the propagators as
\begin{eqnarray}
 \frac{1}{2A_t}\ \frac{1}{\Delta_t\cdot n+\bar{\kappa}-A_t+i\varepsilon}
&=&
 -\frac{1}{2p'p}\left[\frac{1}{2}+\frac{\Delta_t\cdot n+\bar{\kappa}}{\pi}\,
 \int\frac{d\lambda}{f_\lambda(\bar{\kappa})}\right]
 \frac{1}{z_t(\bar{\kappa})-x}\nonumber\\
&&
 +\frac{1}{2p'p}\,\frac{\Delta_t\cdot n+\bar{\kappa}}{\pi}\,
 \int\frac{d\lambda}{f_\lambda(\bar{\kappa})}
 \,\frac{1}{z_{t,\lambda}-x}\ ,\nonumber\\
 \frac{1}{2A_t}\ \frac{1}{-\Delta_t\cdot n+\bar{\kappa}-A_t+i\varepsilon}
&=&
 -\frac{1}{2p'p}\left[\frac{1}{2}-\frac{\Delta_t\cdot n-\bar{\kappa}}{\pi}\,
 \int\frac{d\lambda}{f_\lambda(-\bar{\kappa})}\right]
 \frac{1}{z_t(-\bar{\kappa})-x}\nonumber\\
&&
 -\frac{1}{2p'p}\,\frac{\Delta_t\cdot n-\bar{\kappa}}{\pi}\,
 \int\frac{d\lambda}{f_\lambda(-\bar{\kappa})}
 \,\frac{1}{z_{t,\lambda}-x}\ ,\nonumber\\
 \frac{1}{\left(\bar{\kappa}+\Delta_u\cdot n\right)^2-A_u^2}
&=&
 -\frac{1}{2p'p}\,\frac{1}{z_u(\bar{\kappa})+x}\ ,\label{pwe30}
\end{eqnarray}
where $p'p=|{\bf p}'||{\bf p}|$ and
\begin{eqnarray}
 f_\lambda(\bar{\kappa})
&=&
 \lambda^2+\left(\Delta_t\cdot n\right)^2+\bar{\kappa}^2+2\bar{\kappa}\Delta_t\cdot n\ ,
 \nonumber\\
 z_i(\bar{\kappa})
&=&
 \frac{1}{2p'p}\left[p'+p+M^2-\bar{\kappa}^2-2\bar{\kappa}\Delta_i^0-(\Delta_i^0)^2\right]\ ,\nonumber\\
 z_{t,\lambda}
&=&
 \frac{1}{2p'p}\left[p'+p+M^2+\lambda^2\right]\ .\label{pwe31}
\end{eqnarray}
The invariants are expanded in polynomials of $x$, like
\begin{eqnarray}
 j^{\pm}(t)
&=&
 \left[X^j(\pm)+xY^j(\pm)\right]D^{(1)}(\pm\Delta_t,n,\bar{\kappa})\nonumber\\*
&=&
 \frac{1}{2p'p}\left[\left(X_1^j(\pm)+xY_1^j(\pm)\vphantom{\frac{A}{A}}\right)\frac{F(\Lambda_t)}{z_t(\pm\bar{\kappa})-x}
 +\left(X_2^j(\pm)+xY_2^j(\pm)\vphantom{\frac{A}{A}}\right)\frac{F(\Lambda_t)}{z_{t,\lambda}-x}\right]\
 ,\nonumber\\
 j(u)
&=&
 \frac{1}{2p'p}\left(X^j+xY^j+x^2Z^j\vphantom{\frac{A}{A}}\right)\frac{F(\Lambda_u)}{z_u(\bar{\kappa})+x}
 \ ,\nonumber\\
 j(s)
&=&
 \left(X^j+xY^j+x^2Z^j\vphantom{\frac{A}{A}}\right)
 \frac{F(\Lambda_s)}{\frac{1}{4}\left(W'+W+\kappa'+\kappa\right)^2-M_B^2}\
 ,\label{pwe32}
\end{eqnarray}
where $j$ is an element of the set $\left(A,B,A',B'\right)$.
Furthermore, there are the relations in the $t$-channel
\begin{eqnarray}
 X_1^j(\pm)
&=&
 -\left[\frac{1}{2}+\frac{\pm\Delta_t^0+\bar{\kappa}}{\pi}\,\int\frac{d\lambda}{f_\lambda(\pm\bar{\kappa})}\right]
 X^j(\pm)\ ,\nonumber\\
 X_2^j(\pm)
&=&
 \frac{\pm\Delta_t^0+\bar{\kappa}}{\pi}\,\int\frac{d\lambda}{f_\lambda(\pm\bar{\kappa})}X^j(\pm)\
 .\label{pwe33}
\end{eqnarray}
The coefficients $X^j$, $Y^j$ and $Z^j$ can easily be extracted
from the invariants and they are given for the various exchange
processes in appendix \ref{kadampinv}.

With the partial wave projection
\begin{equation}
 j_L(i)=\frac{1}{2}\int_{-1}^1dx\,P_L(x)\,j(i)\ ,\label{pwe34}
\end{equation}
where $i=t,u,s$, we find the partial wave projections of the
invariants
\begin{eqnarray}
 j^{\pm}_L(t)
&=&
 \frac{1}{2p'p}\left[\left(X_1^j(\pm)+z_t(\pm\bar{\kappa})Y_1^j(\pm)\vphantom{\frac{A}{A}}\right)U_L(\Lambda_t,z_t(\pm\bar{\kappa}))
 \right.\nonumber\\
&&
 \phantom{\frac{1}{2p'p}\left[\right.}
 +\left(X_2^j(\pm)+z_{t,\lambda}Y_2^j(\pm)\vphantom{\frac{A}{A}}\right)U_L(\Lambda_t,z_{t,\lambda})
 \nonumber\\
&&
 \phantom{\frac{1}{2p'p}\left[\right.}\left.
 -Y_1^j(\pm)R_L(\Lambda_t,z_t(\pm\bar{\kappa}))-Y_2^j(\pm)R_L(\Lambda_t,z_{t,\lambda})\right]\nonumber\\
 j_L(u)
&=&
 \frac{(-1)^L}{2p'p}\left[\left(X^j-z_u(\bar{\kappa})Y^j+z^2_u(\bar{\kappa})Z^j\vphantom{\frac{A}{A}}\right)
 U_L(\Lambda_u,z_u(\bar{\kappa}))\right.\nonumber\\
&&
 \phantom{\frac{(-1)^L}{2p'p}\left[\right.}\left.
 -\left(-Y^j+z_u(\bar{\kappa})Z^j\vphantom{\frac{A}{A}}\right)R_L(\Lambda_u,z_u(\bar{\kappa}))
 -Z^jS_L(\Lambda_u,z_u(\bar{\kappa}))\vphantom{\frac{A}{A}}\right]\nonumber\\
 j_L(s)
&=&
 \left[X^j\,\delta_{L,0}+\frac{1}{3}\,Y^j\,\delta_{L,1}+\frac{1}{3}\left(\frac{2}{5}\,\delta_{L,2}+\delta_{L,0}\right)Z^j\right]
 \nonumber\\
&&
 \times\frac{F(\Lambda_s)}{\frac{1}{4}\left(W'+W+\kappa'+\kappa\right)^2-M_B^2}\
 ,\label{pwe35}
\end{eqnarray}
where
\begin{eqnarray}
 U_L(\Lambda,z)&=&\frac{1}{2}\int_{-1}^1dx\,\frac{P_L(x)F(\Lambda)}{z-x}\ ,\nonumber\\
 R_L(\Lambda,z)&=&\frac{1}{2}\int_{-1}^1dx\,P_L(x)F(\Lambda)\ ,\nonumber\\
 S_L(\Lambda,z)&=&\frac{1}{2}\int_{-1}^1dx\,xP_L(x)F(\Lambda)\ .\label{pwe36}
\end{eqnarray}

\section{Conclusion and Discussion}

In two papers, paper I and this one, we have presented the results
for meson-baryon, or more specifically $\pi N$-scattering in the
Kadyshevsky formalism. In paper I we have presented the results
for meson exchange amplitudes and a second quantization procedure
for the quasi field present in the Kadyshevsky formalism is given.
We studied the frame-dependence, i.e. the $n$-dependence, of the
Kadyshevsky integral equation, which we continued in this paper.

Couplings containing derivatives and higher spin fields may cause
differences and problems as far as the results in the Kadyshevsky
formalism and the Feynman formalism are concerned. This is
discussed in paper I by means of an example. After a second glance
the results in both formalisms are the same, however, they contain
extra frame dependent contact terms. Two methods are shortly
introduced and applied, which discuss a second source extra terms:
the TU and the GJ method. The extra terms coming from this second
source cancel the former ones exactly. Both formalisms yield the
same results. With the use of (one of) these methods the final
results for the S-matrix or amplitude are covariant and frame
independent ($n$-independent). For practical purposes we have
introduced and discussed the $\bar{P}$-method and last but nog
least we have shown that the TU method can be derived from the BMP
theory.

In this paper we have presented the results for baryon exchange.
It also contains a formal introduction and detail discussion of
so-called pair suppression. We have formally implemented
"absolute" pair suppression and applied it to the baryon exchange
processes, although it is in principle possible to also allow for
some pair production. The formalism used is based on the TU
method. For the resulting amplitudes, we have shown, to our
knowledge for the first time, that they are causal, covariant and
$n$-independent. Moreover, the amplitudes are just a factor $1/2$
of the usual Feynman expressions. The amplitudes contain only
positive energy (or if one wishes, only negative energy) initial
and final states. This is particularly convenient for the
Kadyshevsky integral equation. It should be mentioned that
negative energy is present inside an amplitude via the
$\Delta(x-y)$ propagator. This is, however, also the case in the
academic example of the infinite dense anti-neutron star.

The last part of this paper contains the partial wave expansion.
This is used for solving the Kadyshevsky integral equation and to
introduce the phase-shifts.

\begin{appendices}
\section{Kadyshevsky Amplitudes and Invariants}\label{kadampinv}

\subsection{Meson Exchange}

\subsubsection*{Scalar Meson Exchange, diagram (a)}

\begin{eqnarray}
  M^{(a)}_{\kappa',\kappa}
&=&
 g_{PPS}g_{S}\,\left[\bar{u}(p')u(p)\right]D^{(1)}(\Delta_t,n,\bar{\kappa})\ ,
\end{eqnarray}
where
$D^{(1)}(\Delta_t,n,\bar{\kappa})=\frac{1}{2A_t}\cdot\frac{1}{\Delta_t\cdot
n+\bar{\kappa}-A_t+i\varepsilon}$
\begin{eqnarray}
 A_S
&=&
 g_{PPS}g_{S}\,D^{(1)}(\Delta_t,n,\bar{\kappa})\ .
\end{eqnarray}
\begin{eqnarray}
 X_S^A
&=&
 g_{PPS}g_{S}\ .
\end{eqnarray}

\subsubsection*{Scalar Meson Exchange, diagram (b)}

\begin{eqnarray}
 M^{(b)}_{\kappa',\kappa}
&=&
 g_{PPS}g_{S}\,\left[\bar{u}(p's')u(p)\right]
 D^{(1)}(-\Delta_t,n,\bar{\kappa})\ .\qquad
\end{eqnarray}
\begin{eqnarray}
 A_S
&=&
 g_{PPS}g_{S}\,D^{(1)}(-\Delta_t,n,\bar{\kappa})\ .
\end{eqnarray}
\begin{eqnarray}
 X_S^A
&=&
 g_{PPS}g_{S}\ .
\end{eqnarray}

\subsubsection*{Pomeron Exchange}

\begin{eqnarray}
 M_{\kappa'\kappa}&=&\frac{g_{PPP}g_P}{M}\,\left[\bar{u}(p's')u(p)\right]\ .
 \label{pom1}
\end{eqnarray}
\begin{eqnarray}
 A_{P}&=&\frac{g_{PPP}g_P}{M}\ .\label{pom2}
\end{eqnarray}
The partial wave projection is obtained by applying \eqref{pwe34}
straightforward

\subsubsection*{Vector Meson Exchange, diagram (a)}

\begin{eqnarray}
 M^{(a)}_{\kappa',\kappa}
&=&
 -g_{VPP}\ \bar{u}(p's')\left[\vphantom{\frac{A}{A}}2g_V\slQ
 -\frac{g_V}{M_V^2}\left((M_f-M_i)+\kappa'\sln\right)
 \right.\nonumber\\
&&
 \times\left(\frac{1}{4}(s_{p'q'}-s_{pq}+u_{pq'}-u_{p'q})-\left(m_f^2-m_i^2\right)+2\bar{\kappa}n\cdot Q\right)
 \nonumber\\
&&
 +\frac{f_V}{2M_V}\left(2(M_f+M_i)\slQ+\frac{1}{2}\left(u_{pq'}+u_{p'q}\right)
 -\frac{1}{2}\left(s_{p'q'}+s_{pq}\right)\right)\nonumber\\
&&
 -\frac{f_V}{2M_V^3}\left(\frac{1}{2}\left(M_f^2+M_i^2\right)+\frac{1}{2}\left(m_f^2+m_i^2\right)
 \right.\nonumber\\
&&
 \phantom{-\frac{f_V}{2M_V^3}(}
 -\frac{1}{2}\left(\frac{1}{2}(t_{p'p}+t_{q'q})+u_{pq'}+s_{pq}\right)
 \nonumber\\
&&
 \phantom{-\frac{f_V}{2M_V^3}(}\left.
 +\left(M_f+M_i\right)\kappa'\sln +\frac{1}{4}\left(\kappa'-\kappa\right)^2
 -\left(p'+p\right)\cdot n\bar{\kappa}\vphantom{\frac{A}{A}}\right)
 \nonumber\\
&&
 \left.\times
 \left(\frac{1}{4}(s_{p'q'}-s_{pq})+\frac{1}{4}(u_{pq'}-u_{p'q})-\left(m_f^2-m_i^2\right)+2\bar{\kappa}n\cdot Q\right)
 \right]u(ps)\nonumber\\
&&
 \times D^{(1)}(\Delta_t,n,\bar{\kappa})\ .
\end{eqnarray}
\begin{eqnarray}
 A_V
&=&
 -g_{VPP}\left[ -\frac{g_V}{M_V^2}\left(M_f-M_i\right)
 \left(\frac{1}{4}(s_{p'q'}-s_{pq}+u_{pq'}-u_{p'q})-\left(m_f^2-m_i^2\right)
 \right.\right.\nonumber\\*
&&
 \left.\vphantom{\frac{A}{A}}
 +2\bar{\kappa}n\cdot Q\right)
 +\frac{f_V}{4M_V}\left(u_{pq'}+u_{p'q}-s_{p'q'}-s_{pq}\right)
 -\frac{f_V}{2M_V^3}\left(\frac{1}{2}\left(M_f^2+M_i^2\right)\right.\nonumber\\*
&&
  +\frac{1}{2}\left(m_f^2+m_i^2\right)
 -\frac{1}{2}\left(\frac{1}{2}(t_{p'p}+t_{q'q})+u_{pq'}+s_{pq}\right)
 +\frac{1}{4}\left(\kappa'-\kappa\right)^2
 \nonumber\\
&&
 \left.
 -\left(p'+p\right)\cdot n\bar{\kappa}\vphantom{\frac{A}{A}}\right)
 \left(\frac{1}{4}(s_{p'q'}-s_{pq})+\frac{1}{4}(u_{pq'}-u_{p'q})-\left(m_f^2-m_i^2\right)
 \vphantom{\frac{A}{A}}\right.\nonumber\\
&&
 \left.\left.\vphantom{\frac{A}{A}}
 +2\bar{\kappa}n\cdot Q\right)\right]\ D^{(1)}(\Delta_t,n,\bar{\kappa})\ ,\nonumber\\
 B_V
&=&
 -2g_{VPP}\left[g_V+\frac{f_V}{2M_V}\left(M_f+M_i\right)\right]D^{(1)}(\Delta_t,n,\bar{\kappa})\ ,\nonumber\\
 A'_V
&=&
 \frac{g_{VPP}\kappa'}{M_V^2}\left[g_V+\frac{f_V}{2M_V}\left(M_f+M_i\right)\right]
 \left(\frac{1}{4}(s_{p'q'}-s_{pq}+u_{pq'}-u_{p'q})\right.\nonumber\\
&&
 \phantom{\frac{g_{VPP}\kappa'}{M_V^2}[}\left.
 -\left(m_f^2-m_i^2\right)
 +2\bar{\kappa}n\cdot Q\vphantom{\frac{A}{A}}\right)
 D^{(1)}(\Delta_t,n,\bar{\kappa})\ .
\end{eqnarray}
\begin{eqnarray}
 X_V^A
&=&
 -g_{VPP}\left[ -\frac{g_V}{M_V^2}\left(M_f-M_i\right)
 \left(\frac{1}{4}\left(E'+\mathcal{E}'\right)^2-\frac{1}{4}\left(E+\mathcal{E}\right)^2
 -\frac{1}{4}\left(M_f^2-M_i^2\right)\right.\right.\nonumber\\
&&
 \left.-\frac{3}{4}\left(m_f^2-m_i^2\right)+\frac{1}{2}\left(E'\mathcal{E}-E\mathcal{E}'\right)
 +\bar{\kappa}(\mathcal{E}'+\mathcal{E})\right)+\frac{f_V}{4M_V}\left(M_f^2+M_i^2\right.\nonumber\\
&&
 \left.
 +m_f^2+m_i^2-2\left(E'\mathcal{E}+E\mathcal{E}'\right)
 -\left(E'+\mathcal{E}'\right)^2-\left(E+\mathcal{E}\right)^2\right)
 -\frac{f_V}{4M_V^3}\left(\vphantom{\frac{A}{A}}M_f^2\right.\nonumber\\
&&
 +M_i^2+m_f^2+m_i^2+\frac{1}{2}\left(\kappa'-\kappa\right)^2
 -\frac{1}{2}\left(M_f^2+3M_i^2+3m_f^2+m_i^2\right.\nonumber\\
&&
 \left.\left.
 -2E'E-2\mathcal{E}'\mathcal{E}
 -4\mathcal{E}'E+\left(E+\mathcal{E}\right)^2\right)
  -2\left(E'+E\right)\bar{\kappa}\vphantom{\frac{A}{A}}\right)
 \nonumber\\
&&
 \times\left(\frac{1}{4}\left(E'+\mathcal{E}'\right)^2-\frac{1}{4}\left(E+\mathcal{E}\right)^2
 -\frac{1}{4}\left(M_f^2-M_i^2\right)-\frac{3}{4}\left(m_f^2-m_i^2\right)
 \right.\nonumber\\
&&
 \phantom{\times(}\left.\left.
 +\frac{1}{2}\left(E'\mathcal{E}-E\mathcal{E}'\right)
 \vphantom{\frac{A}{A}}+\bar{\kappa}(\mathcal{E}'+\mathcal{E})\right)\right]\ ,\nonumber\\
 Y_V^A
&=&
 -\frac{g_{VPP}\,f_V\,p'p}{M_V}\left[1+\frac{1}{4M_V^2}
 \left(\left(E'+\mathcal{E}'\right)^2-\left(E+\mathcal{E}\right)^2
 -\left(M_f^2-M_i^2\right)\right.\right.\nonumber\\
&&
 \left.\left.\phantom{-\frac{g_{VPP}\,f_V\,p'p}{M_V}[}
 -3\left(m_f^2-m_i^2\right)
 +2\left(E'\mathcal{E}-E\mathcal{E}'\right)+4\bar{\kappa}(\mathcal{E}'+\mathcal{E})\right)\right]\ ,\nonumber\\
 X_V^B
&=&
 -2g_{VPP}\left[g_V+\frac{f_V}{2M_V}\left(M_f+M_i\right)\right]\
 ,\nonumber\\
 X_V^{A'}
&=&
 \frac{g_{VPP}\kappa'}{4M_V^2}\left[g_V+\frac{f_V}{2M_V}\left(M_f+M_i\right)\right]
 \left(\left(E'+\mathcal{E}'\right)^2-\left(E+\mathcal{E}\right)^2
 \right.\nonumber\\*
&&
 \left.\vphantom{\frac{A}{A}}
 -\left(M_f^2-M_i^2\right)
 -3\left(m_f^2-m_i^2\right)+2\left(E'\mathcal{E}-E\mathcal{E}'\right)
 +4\bar{\kappa}(\mathcal{E}'+\mathcal{E})\right)\ .
\end{eqnarray}

\subsubsection*{Vector Meson Exchange, diagram (b)}

\begin{eqnarray}
 M^{(b)}_{\kappa',\kappa}
&=&
 -g_{VPP}\ \bar{u}(p's')\left[\vphantom{\frac{A}{A}}2g_V\slQ
 -\frac{g_V}{M_V^2}\left((M_f-M_i)-\kappa\sln\right)
 \right.\nonumber\\
&&
 \times\left(\frac{1}{4}(s_{p'q'}-s_{pq}+u_{pq'}-u_{p'q})-\left(m_f^2-m_i^2\right)-2\bar{\kappa}n\cdot Q\right)
 \nonumber\\
&&
 +\frac{f_V}{2M_V}\left(2(M_f+M_i)\slQ+\frac{1}{2}\left(u_{pq'}+u_{p'q}\right)
 -\frac{1}{2}\left(s_{p'q'}+s_{pq}\right)\right)\nonumber\\
&&
 -\frac{f_V}{2M_V^3}\left(\frac{1}{2}\left(M_f^2+M_i^2\right)+\frac{1}{2}\left(m_f^2+m_i^2\right)
 \right.\nonumber\\
&&
 \phantom{-\frac{f_V}{2M_V^3}(}
 -\frac{1}{2}\left(\frac{1}{2}(t_{p'p}+t_{q'q})+u_{pq'}+s_{pq}\right)
 \nonumber\\
&&
 \phantom{-\frac{f_V}{2M_V^3}(}\left.
 -\left(M_f+M_i\right)\kappa\sln+\frac{1}{4}\left(\kappa'-\kappa\right)^2
 +\left(p'+p\right)\cdot n\bar{\kappa}\vphantom{\frac{A}{A}}\right)
 \nonumber\\
&&
 \left.\times
 \left(\frac{1}{4}(s_{p'q'}-s_{pq}+u_{pq'}-u_{p'q})-\left(m_f^2-m_i^2\right)-2\bar{\kappa}n\cdot Q\right)
 \right]u(ps)\nonumber\\
&&
 \times D^{(1)}(-\Delta_t,n,\bar{\kappa})\ .
\end{eqnarray}
\begin{eqnarray}
 A_V
&=&
 -g_{VPP}\left[ -\frac{g_V}{M_V^2}\left(M_f-M_i\right)
 \left(\frac{1}{4}(s_{p'q'}-s_{pq}+u_{pq'}-u_{p'q})-\left(m_f^2-m_i^2\right)
 \right.\right.\nonumber\\
&&
 \left.\vphantom{\frac{A}{A}}
 -2\bar{\kappa}n\cdot Q\right)
 +\frac{f_V}{4M_V}\left(u_{pq'}+u_{p'q}-s_{p'q'}-s_{pq}\right)
 -\frac{f_V}{2M_V^3}\left(\frac{1}{2}\left(M_f^2+M_i^2\right)\right.\nonumber\\
&&
 +\frac{1}{2}\left(m_f^2+m_i^2\right)
 -\frac{1}{2}\left(\frac{1}{2}(t_{p'p}+t_{q'q})+u_{pq'}+s_{pq}\right)
 +\frac{1}{4}\left(\kappa'-\kappa\right)^2
 \nonumber\\
&&
 \left.
 +\left(p'+p\right)\cdot n\bar{\kappa}\vphantom{\frac{A}{A}}\right)
 \left(\frac{1}{4}(s_{p'q'}-s_{pq})+\frac{1}{4}(u_{pq'}-u_{p'q})-\left(m_f^2-m_i^2\right)
 \right.\nonumber\\
&&
 \left.\left.\vphantom{\frac{A}{A}}
 -2\bar{\kappa}n\cdot Q\right)\right]\ D^{(1)}(-\Delta_t,n,\bar{\kappa})\ ,\nonumber\\
 B_V
&=&
 -2g_{VPP}\left[g_V+\frac{f_V}{2M_V}\left(M_f+M_i\right)\right]D^{(1)}(-\Delta_t,n,\bar{\kappa})\ ,\nonumber\\
 A'_V
&=&
 -\frac{g_{VPP}\kappa}{M_V^2}\left[g_V+\frac{f_V}{2M_V}\left(M_f+M_i\right)\right]
 \left(\frac{1}{4}(s_{p'q'}-s_{pq}+u_{pq'}-u_{p'q})\right.\nonumber\\
&&
 \phantom{\frac{g_{VPP}\kappa'}{M_V^2}[}\left.
 -\left(m_f^2-m_i^2\right)
 -2\bar{\kappa}n\cdot Q\vphantom{\frac{A}{A}}\right)
 D^{(1)}(-\Delta_t,n,\bar{\kappa})\ .
\end{eqnarray}
\begin{eqnarray}
 X_V^A
&=&
 -g_{VPP}\left[ -\frac{g_V}{M_V^2}\left(M_f-M_i\right)
 \left(\frac{1}{4}\left(E'+\mathcal{E}'\right)^2-\frac{1}{4}\left(E+\mathcal{E}\right)^2
 -\frac{1}{4}\left(M_f^2-M_i^2\right)\right.\right.\nonumber\\
&&
 \left.-\frac{3}{4}\left(m_f^2-m_i^2\right)+\frac{1}{2}\left(E'\mathcal{E}-E\mathcal{E}'\right)
 -\bar{\kappa}(\mathcal{E}'+\mathcal{E})\right)+\frac{f_V}{4M_V}\left(M_f^2+M_i^2\right.\nonumber\\
&&
 \left.
 +m_f^2+m_i^2
 -2\left(E'\mathcal{E}+E\mathcal{E}'\right)
 -\left(E'+\mathcal{E}'\right)^2-\left(E+\mathcal{E}\right)^2\right)
 -\frac{f_V}{4M_V^3}\left(\vphantom{\frac{A}{A}}M_f^2\right.\nonumber\\
&&
 +M_i^2+m_f^2+m_i^2+\frac{1}{2}\left(\kappa'-\kappa\right)^2
 -\frac{1}{2}\left(M_f^2+3M_i^2+3m_f^2+m_i^2\right.\nonumber\\
&&
 \left.\left.
 -2E'E-2\mathcal{E}'\mathcal{E}
 -4\mathcal{E}'E+\left(E+\mathcal{E}\right)^2\right)
 +2\left(E'+E\right)\bar{\kappa}\vphantom{\frac{A}{A}}\right)\nonumber\\
&&
 \times
 \left(\frac{1}{4}\left(E'+\mathcal{E}'\right)^2-\frac{1}{4}\left(E+\mathcal{E}\right)^2
 -\frac{1}{4}\left(M_f^2-M_i^2\right)-\frac{3}{4}\left(m_f^2-m_i^2\right)
 \right.\nonumber\\
&&
 \left.\left.\vphantom{\frac{A}{A}}
 +\frac{1}{2}\left(E'\mathcal{E}-E\mathcal{E}'\right)
 -\bar{\kappa}(\mathcal{E}'+\mathcal{E})\right)\right]\ ,\nonumber\\
 Y_V^A
&=&
 -\frac{g_{VPP}\,f_V\,p'p}{M_V}\left[1+\frac{1}{4M_V^2}
 \left(\left(E'+\mathcal{E}'\right)^2-\left(E+\mathcal{E}\right)^2
 -\left(M_f^2-M_i^2\right)\right.\right.\nonumber\\
&&
 \left.\left.\phantom{-\frac{g_{VPP}\,f_V\,p'p}{M_V}[}
 -3\left(m_f^2-m_i^2\right)
 +2\left(E'\mathcal{E}-E\mathcal{E}'\right)-4\bar{\kappa}(\mathcal{E}'+\mathcal{E})\right)\right]\ ,\nonumber\\
 X_V^B
&=&
 -2g_{VPP}\left[g_V+\frac{f_V}{2M_V}\left(M_f+M_i\right)\right]\
 ,\nonumber\\
 X_V^{A'}
&=&
 -\frac{g_{VPP}\kappa}{4M_V^2}\left[g_V+\frac{f_V}{2M_V}\left(M_f+M_i\right)\right]
 \left(\left(E'+\mathcal{E}'\right)^2-\left(E+\mathcal{E}\right)^2
 \right.\nonumber\\
&&
 \left.\vphantom{\frac{A}{A}}
 -\left(M_f^2-M_i^2\right)
 -3\left(m_f^2-m_i^2\right)+2\left(E'\mathcal{E}-E\mathcal{E}'\right)
 -4\bar{\kappa}(\mathcal{E}'+\mathcal{E})\right)\ .
\end{eqnarray}

\subsection{Baryon Exchange/Resonance}
\subsubsection*{Baryon Exchange, Scalar coupling}

\begin{eqnarray}
 M^S_{\kappa',\kappa}
&=&
 \frac{g_{S}^2}{2}\ \bar{u}(p's')\left[\frac{1}{2}\left(M_f+M_i\right)+M_B-\slQ+\sln\bar{\kappa}\right]u(ps)
 D^{(2)}\left(\Delta_u,n,\bar{\kappa}\right)\ ,\label{kadamplbeps1}
\end{eqnarray}
where the denominator function is
$D^{(2)}\left(\Delta_i,n,\bar{\kappa}\right)
=\left[\vphantom{\frac{A}{A}}\left(\bar{\kappa}+ \Delta_i\cdot
n\right)^2-A_i^2\right]^{-1}$, $i=u,s$.
\begin{eqnarray}
 A_{S}
&=&
 \frac{g_S^2}{2}
 \left[\vphantom{\frac{A}{A}} \frac{1}{2}\left(M_f+M_i\right)+M_B\right]\
 D^{(2)}\left(\Delta_u,n,\bar{\kappa}\right)\ , \nonumber\\
 B_{S}
&=&
 -\frac{g_S^2}{2}\
 D^{(2)}\left(\Delta_u,n,\bar{\kappa}\right)\ ,\nonumber\\
 A'_{S}
&=&
 \frac{g_S^2}{2}\,\bar{\kappa}\,
 D^{(2)}\left(\Delta_u,n,\bar{\kappa}\right)\ .\label{kadamplbeps2}
\end{eqnarray}
\begin{eqnarray}
 X_{S}^A
&=&
 -\frac{g_S^2}{2}
 \left[\vphantom{\frac{A}{A}}
 \frac{1}{2}\left(M_f+M_i\right)+M_B\right]\ ,\nonumber\\
 X_{S}^B
&=&
 \frac{g_S^2}{2}\ ,\nonumber\\
 X_{S}^{A'}
&=&
 -\frac{g_S^2}{2}\,\bar{\kappa}\ .\label{kadamplbeps3}
\end{eqnarray}

\subsubsection*{Baryon Exchange, Pseudo Scalar coupling}

The expressions for baryon exchange with pseudo scalar coupling
are the same as \eqref{kadamplbeps1}-\eqref{kadamplbeps3} with the
substitution $M_B\rightarrow-M_B$.

\subsubsection*{Baryon Resonance, Scalar coupling}

\begin{eqnarray}
 M^S_{\kappa',\kappa}
&=&
 \frac{g_{S}^2}{2}\ \bar{u}(p's')\left[\frac{1}{2}\left(M_f+M_i\right)+M_B+\slQ+\sln\bar{\kappa}\right]u(ps)
 D^{(2)}\left(\Delta_s,n,\bar{\kappa}\right)\ .\label{kadamplbrps1}
\end{eqnarray}
\begin{eqnarray}
 A_{S}
&=&
 \frac{g_S^2}{2}
 \left[\vphantom{\frac{A}{A}} \frac{1}{2}\left(M_f+M_i\right)+M_B\right]\
 D^{(2)}\left(\Delta_s,n,\bar{\kappa}\right)\ , \nonumber\\
 B_{S}
&=&
 \frac{g_S^2}{2}\
 D^{(2)}\left(\Delta_s,n,\bar{\kappa}\right)\ ,\nonumber\\
 A'_{S}
&=&
 \frac{g_S^2}{2}\,\bar{\kappa}\,
 D^{(2)}\left(\Delta_s,n,\bar{\kappa}\right)\ .\label{kadamplbrps2}
\end{eqnarray}
\begin{eqnarray}
 X_{S}^A
&=&
 -\frac{g_S^2}{2}
 \left[\vphantom{\frac{A}{A}}
 \frac{1}{2}\left(M_f+M_i\right)+M_B\right]\ ,\nonumber\\
 X_{S}^B
&=&
 -\frac{g_S^2}{2}\ ,\nonumber\\
 X_{S}^{A'}
&=&
 -\frac{g_S^2}{2}\ \bar{\kappa}\ .\label{kadamplbrps3}
\end{eqnarray}

\subsubsection*{Baryon Resonance, Pseudo Scalar coupling}

The expressions for baryon resonance with pseudo scalar coupling
are the same as \eqref{kadamplbrps1}-\eqref{kadamplbrps3} with the
substitution $M_B\rightarrow-M_B$.

\subsubsection*{Baryon Exchange Vector coupling}

\begin{eqnarray}
 M^V_{\kappa',\kappa}
&=&
 \frac{f_{V}^2}{2m_\pi^2}\bar{u}(p's')\,\left[\vphantom{\frac{A}{A}}
 -\left(\frac{1}{2}\left(M_f+M_i\right)-M_B\right)\right.\nonumber\\*
&&
 \times\left(-\frac{1}{2}\left(M_f^2+M_i^2\right)+\frac{1}{2}\left(u_{p'q}+u_{pq'}\right)
 +\left(M_f+M_i\right)\slQ\right.\nonumber\\*
&&
 \phantom{\times(}
 \left.-\frac{1}{2}\left(\kappa'-\kappa\right)\left(p'-p\right)\cdot n
 +\frac{1}{2}\left(\kappa'-\kappa\right)\left[\sln,\slQ\right]
 -\frac{1}{2}\left(\kappa'-\kappa\right)^2\right)\nonumber\\
&&
 -\frac{1}{2}\left(u_{pq'}-M_i^2\right)\left(\frac{1}{2}\left(M_f-M_i\right)+\slQ
 +\frac{1}{2}\left(\kappa'-\kappa\right)\sln\right)\nonumber\\
&&
 -\frac{1}{2}\left(u_{p'q}-M_f^2\right)\left(-\frac{1}{2}\left(M_f-M_i\right)+\slQ
 -\frac{1}{2}\left(\kappa'-\kappa\right)\sln\right)\nonumber\\
&&
 +\bar{\kappa}\left(-\frac{1}{2}\left(M_f-M_i\right)(p'-p)\cdot n\,-(p'-p)\cdot n\
 \slQ\,+2Q\cdot n\ \slQ\right.\nonumber\\
&&
 \phantom{+\bar{\kappa}(}
 -\frac{1}{2}\left(M_f-M_i\right)\left(\kappa'-\kappa\right)+\frac{1}{2}\left(M_f-M_i\right)[\sln,\slQ]
 \nonumber\\
&&
 \left.\left.\phantom{+\bar{\kappa}(}
 +\frac{1}{2}\left(M_f^2+M_i^2\right)\,\sln
 -\frac{1}{2}\left(u_{p'q}+u_{pq'}\right)\sln\right)\right]u_i(p)
 D^{(2)}\left(\Delta_u,n,\bar{\kappa}\right)\ .\label{kadamplbepv1}
\end{eqnarray}
\begin{eqnarray}
 A_{V}
&=&
 \frac{f^2_V}{2m_\pi^2}\left[
 -\left(\frac{1}{2}\left(M_f+M_i\right)-M_B\right)\left(
 -\frac{1}{2}\left(M_f^2+M_i^2\right)+\frac{1}{2}(u_{p'q}+u_{pq'})
 \right.\right.\nonumber\\
&&
 \left.\phantom{\frac{f^2_V}{2m_\pi^2}[}
 -\frac{1}{2}(\kappa'-\kappa)(p'-p)\cdot n
 -\frac{1}{2}(\kappa'-\kappa)^2\right)-\frac{\bar{\kappa}}{2}\left(M_f-M_i\right)
 \nonumber\\
&&
 \phantom{\frac{f^2_V}{2m_\pi^2}[}
 \times(p'-p)\cdot n
 +\frac{1}{4}\left(u_{p'q}-u_{pq'}-M_f^2+M_i^2\right)\left(M_f-M_i\right)
 \nonumber\\
&&
 \phantom{\frac{f^2_V}{2m_\pi^2}[}\left.
 -\frac{\bar{\kappa}}{2}\left(M_f-M_i\right)(\kappa'-\kappa)\right]
 D^{(2)}\left(\Delta_u,n,\bar{\kappa}\right)\ ,\nonumber\\
 B_{V}
&=&
 \frac{f^2_V}{2m_\pi^2}\left[
 -\left(\frac{1}{2}\left(M_f+M_i\right)-M_B\right)\left(M_f+M_i\right)
 +\frac{1}{2}\left(\vphantom{\frac{A}{A}}M_f^2+M_i^2\right.\right.\nonumber\\
&&
 \left.\left.\phantom{\frac{f^2_V}{2m_\pi^2}[}
 -u_{p'q}-u_{pq'}\right)-\bar{\kappa}(p'-p)\cdot n+2\bar{\kappa}n\cdot Q\right]
 D^{(2)}\left(\Delta_u,n,\bar{\kappa}\right)\ ,\nonumber\\
 A'_{V}
&=&
 \frac{f^2_V}{4m_\pi^2}\left[\vphantom{\frac{A}{A}}
 \left(M_i^2-u_{pq'}\right)\kappa'+\left(M_f^2-u_{p'q}\right)\kappa
 \right]D^{(2)}\left(\Delta_u,n,\bar{\kappa}\right)\ ,\nonumber\\
 B'_{V}
&=&
 -\frac{f^2_V}{4m_\pi^2}\left[\vphantom{\frac{A}{A}}
 \kappa'M_i-\kappa M_f-(\kappa'-\kappa)M_B
 \right]D^{(2)}\left(\Delta_u,n,\bar{\kappa}\right)\ .\label{kadamplbepv2}
\end{eqnarray}
\begin{eqnarray}
  X_{V}^A
&=&
 -\frac{f^2_V}{2m_\pi^2}\left[
 -\left(\frac{1}{2}\left(M_f+M_i\right)-M_B\right)\left(
 \frac{1}{2}\left(m_f^2+m_i^2\right)-E'\mathcal{E}-E\mathcal{E}'
 \right.\right.\nonumber\\
&&
 \left.\phantom{-\frac{f^2_V}{2m_\pi^2}\left[\right.}
 -\frac{1}{2}(\kappa'-\kappa)\left(E'-E\right)
 -\frac{1}{2}(\kappa'-\kappa)^2\right)-\frac{\bar{\kappa}}{2}\left(M_f-M_i\right)
 \nonumber\\
&&
 \phantom{-\frac{f^2_V}{2m_\pi^2}\left[\right.}
 \times\left(E'-E\right)-\frac{1}{4}\left(m_f^2-m_i^2+2E'\mathcal{E}-2E\mathcal{E}'\right)
 \left(M_f-M_i\right)\nonumber\\
&&
 \phantom{-\frac{f^2_V}{2m_\pi^2}\left[\right.}\left.
 -\frac{\bar{\kappa}}{2}\left(M_f-M_i\right)(\kappa'-\kappa)\right]\
 ,\nonumber\\
 Y_{V}^A
&=&
 \frac{f^2_V\,p'p}{m_\pi^2}
 \left[\frac{1}{2}\left(M_f+M_i\right)-M_B\right]\ ,\nonumber\\
 X_{V}^B
&=&
 \frac{f^2_V}{2m_\pi^2}\left[
 \left(\frac{1}{2}\left(M_f+M_i\right)-M_B\right)\left(M_f+M_i\right)
 +\frac{1}{2}\left(m_f^2+m_i^2\right.\right.\nonumber\\*
&&
 \left.\left.\phantom{\frac{f^2_V}{2m_\pi^2}[}
 -2E'\mathcal{E}-2E\mathcal{E}'\right)+\bar{\kappa}\left(E'-E\right)\cdot n
 -\bar{\kappa}\left(\mathcal{E}'+\mathcal{E}\right)\right]
 \ ,\nonumber\\
 Y_{V}^B
&=&
 \frac{f^2_V\,p'p}{m_\pi^2}\ ,\nonumber\\
 X_{V}^{A'}
&=&
 \frac{f^2_V}{4m_\pi^2}\left[
 \kappa'\left(m_f^2-2E\mathcal{E}'\right)+\kappa\left(m_i^2-2E'\mathcal{E}\right)\right]\
 ,\nonumber\\
 Y_{V}^{A'}
&=&
 \frac{f^2_V\bar{\kappa}\,p'p}{m_\pi^2}\
 ,\nonumber\\
 X_{V}^{B'}
&=&
 \frac{f^2_V}{4m_\pi^2}\left[
 \kappa'M_i-\kappa M_f-\left(\kappa'-\kappa\right)M_B\right]\ .\label{kadamplbepv3}
\end{eqnarray}

\subsubsection*{Baryon Exchange, Pseudo Vector coupling}

The expressions for baryon exchange with pseudo vector coupling
are the same as \eqref{kadamplbepv1}-\eqref{kadamplbepv3} with the
substitution $M_B\rightarrow-M_B$.

\subsubsection*{Baryon Resonance, Vector coupling}

\begin{eqnarray}
 M^V_{\kappa',\kappa}
&=&
 \frac{f_{V}^2}{m_\pi^2}\bar{u}(p's')\left[\vphantom{\frac{A}{A}}
 -\left(\frac{1}{2}\left(M_f+M_i\right)-M_B\right)\right.\nonumber\\
&&
 \times\left(-\frac{1}{2}\left(M_f^2+M_i^2\right)+\frac{1}{2}\left(s_{p'q'}+s_{pq}\right)
 -\left(M_f+M_i\right)\slQ\right.\nonumber\\
&&
 \phantom{\times(}\left.
 -\frac{1}{2}\left(\kappa'-\kappa\right)\left(p'-p\right)\cdot n
 -\frac{1}{2}\left(\kappa'-\kappa\right)\left[\sln,\slQ\right]
 -\frac{1}{2}\left(\kappa'-\kappa\right)^2\right)\nonumber\\
&&
 +\frac{1}{2}\left(s_{p'q'}-M_f^2\right)\left(\frac{1}{2}\left(M_f-M_i\right)+\slQ
 +\frac{1}{2}\left(\kappa'-\kappa\right)\sln\right)\nonumber\\
&&
 +\frac{1}{2}\left(s_{pq}-M_i^2\right)\left(-\frac{1}{2}\left(M_f-M_i\right)+\slQ
 -\frac{1}{2}\left(\kappa'-\kappa\right)\sln\right)\nonumber\\
&&
 +\bar{\kappa}\left(-\frac{1}{2}\left(M_f-M_i\right)(p'-p)\cdot n\,+(p'-p)\cdot n\
 \slQ\,+2Q\cdot n\ \slQ\right.\nonumber\\
&&
 \phantom{+\bar{\kappa}(}
 -\frac{1}{2}\left(M_f-M_i\right)\left(\kappa'-\kappa\right)-\frac{1}{2}\left(M_f-M_i\right)[\sln,\slQ]
 \nonumber\\
&&
 \left.\left.\phantom{+\bar{\kappa}(}
 +\frac{1}{2}\left(M_f^2+M_i^2\right)\,\sln
 -\frac{1}{2}\left(s_{p'q'}+s_{pq}\right)\sln\right)\right]u_i(p)
 D^{(2)}\left(\Delta_s,n,\bar{\kappa}\right)\ .\label{kadamplbrpv1}
\end{eqnarray}
\begin{eqnarray}
 A_{V}
&=&
 \frac{f^2_V}{2m_\pi^2}\left[
 -\left(\frac{1}{2}\left(M_f+M_i\right)-M_B\right)\left(\frac{1}{2}(s_{p'q'}+s_{pq})
 -\frac{1}{2}\left(M_f^2+M_i^2\right)\right.\right.\nonumber\\*
&&
 \left.\phantom{ \frac{f^2_V}{2m_\pi^2}[}
 -\frac{1}{2}(\kappa'-\kappa)(p'-p)\cdot n
 -\frac{1}{2}(\kappa'-\kappa)^2\right)-\frac{\bar{\kappa}}{2}\left(M_f-M_i\right)
 (p'-p)\cdot n\nonumber\\*
&&
 \phantom{ \frac{f^2_V}{2m_\pi^2}[}
 +\frac{1}{4}\left(s_{p'q'}-s_{pq}-M_f^2+M_i^2\right)\left(M_f-M_i\right)
 \nonumber\\
&&
 \phantom{\frac{f^2_V}{2m_\pi^2}[}\left.
 -\frac{\bar{\kappa}}{2}\left(M_f-M_i\right)(\kappa'-\kappa)\right]
 D^{(2)}\left(\Delta_s,n,\bar{\kappa}\right)\ ,\nonumber\\
 B_{V}
&=&
 \frac{f^2_V}{2m_\pi^2}\left[
 \left(\frac{1}{2}\left(M_f+M_i\right)-M_B\right)\left(M_f+M_i\right)
 +\frac{1}{2}\left(\vphantom{\frac{A}{A}}s_{p'q'}+s_{pq}-M_f^2-M_i^2\right)\right.\nonumber\\
&&
 \left.\phantom{\frac{f^2_V}{2m_\pi^2}[}
 +\bar{\kappa}(p'-p)\cdot n+2\bar{\kappa}n\cdot Q\right]
 D^{(2)}\left(\Delta_s,n,\bar{\kappa}\right)\ ,\nonumber\\
 A'_{V}
&=&
 \frac{f^2_V}{4m_\pi^2}\left[\vphantom{\frac{A}{A}}
 \left(M_i^2-s_{pq}\right)\kappa'+\left(M_f^2-s_{p'q'}\right)\kappa
 \right]D^{(2)}\left(\Delta_s,n,\bar{\kappa}\right)\ ,\nonumber\\
 B'_{V}
&=&
 \frac{f^2_V}{4m_\pi^2}\left[\vphantom{\frac{A}{A}}
 \kappa'M_i-\kappa M_f-(\kappa'-\kappa)M_B
 \right]D^{(2)}\left(\Delta_s,n,\bar{\kappa}\right)\ .\label{kadamplbrpv2}
\end{eqnarray}
\begin{eqnarray}
 X^A_{V}
&=&
 -\frac{f^2_V}{2m_\pi^2}\left[
 -\frac{1}{2}\left(\frac{1}{2}\left(M_f+M_i\right)-M_B\right)\left(\vphantom{\frac{A}{A}}
 \left(E'+\mathcal{E}'\right)^2+\left(E+\mathcal{E}\right)^2
 \right.\right.\nonumber\\
&&
 \left.\phantom{-\frac{f^2_V}{2m_\pi^2}[}
 -\left(M_f^2+M_i^2\right)
 -\frac{1}{2}(\kappa'-\kappa)\left(E'-E\right)-\frac{1}{2}(\kappa'-\kappa)^2\right)
 \nonumber\\
&&
 \phantom{-\frac{f^2_V}{2m_\pi^2}[}
 -\frac{\bar{\kappa}}{2}\left(M_f-M_i\right)\left(E'-E\right)
 +\frac{1}{4}\left(\left(E'+\mathcal{E}'\right)^2-\left(E+\mathcal{E}\right)^2\right.
 \nonumber\\
&&
 \phantom{-\frac{f^2_V}{2m_\pi^2}[}\left.\left.\vphantom{\frac{a}{a}}
 -M_f^2+M_i^2\right)\left(M_f-M_i\right)
 -\frac{\bar{\kappa}}{2}\left(M_f-M_i\right)(\kappa'-\kappa)\right]\
 ,\nonumber\\
 X^B_{V}
&=&
 -\frac{f^2_V}{2m_\pi^2}\left[
 \left(\frac{1}{2}\left(M_f+M_i\right)-M_B\right)\left(M_f+M_i\right)
 +\frac{1}{2}\left(E'+\mathcal{E}'\right)^2\right.\nonumber\\
&&
 \left.\phantom{-\frac{f^2_V}{2m_\pi^2}[}
 +\frac{1}{2}\left(E+\mathcal{E}\right)^2
 -\frac{1}{2}\left(M_f^2+M_i^2\right)
 +\bar{\kappa}\left(E'-E\right)+\bar{\kappa}\left(\mathcal{E}'+\mathcal{E}\right)\right]\
 ,\nonumber\\
 X^{A'}_{V}
&=&
 -\frac{f^2_V}{4m_\pi^2}\left[
 \left(M_i^2-\left(E+\mathcal{E}\right)^2\right)\kappa'
 +\left(M_f^2-\left(E'+\mathcal{E}'\right)^2\right)\kappa\right]\ ,\nonumber\\
 X^{B'}_{V}
&=&
 -\frac{f^2_V}{4m_\pi^2}\left[\vphantom{\frac{A}{A}}
 \kappa'M_i-\kappa M_f-\left(\kappa'-\kappa\right)M_B\right]\ .\label{kadamplbrpv3}
\end{eqnarray}

\subsubsection*{Baryon Resonance, Pseudo Vector coupling}

The expressions for baryon resonance with pseudo vector coupling
are the same as \eqref{kadamplbrpv1}-\eqref{kadamplbrpv3} with the
substitution $M_B\rightarrow-M_B$.

\subsubsection*{${\frac{3}{2}}^+$ Baryon Exchange, Gauge invariant
coupling}

\begin{eqnarray}
 M_{\kappa',\kappa}
&=&
 -\frac{g_{gi}^2}{2}\,\bar{u}(p's')\left[\vphantom{\frac{A}{A}}\right.\nonumber\\
&&
 \frac{1}{2}\,\bar{P}^2_u
 \left(\frac{1}{2}(M_f+M_i)+M_\Delta-\slQ+\bar{\kappa}\sln\right)
 \left(m_f^2+m_i^2-t_{q'q}\right)\nonumber\\
&&
 -\frac{1}{3}\,\bar{P}^2_u
 \left(\left(\frac{1}{2}\,\left(M_f+M_i\right)+M_\Delta\right)\slq\slq'
 +\frac{1}{2}\,\left(u_{pq'}-M_i^2\right)\slq\right.\nonumber\\
&&
 \phantom{-\frac{1}{3}\,\bar{P}^2_u(}\left.
 +\frac{1}{2}\,\left(s_{pq}+t_{q'q}-M_i^2-m_f^2-3m_i^2\right)\slq'
 +\bar{\kappa}\sln\slq\slq'\right)\nonumber\\
&&
 -\frac{1}{12}\left(\left(
 \bar{P}^2_u+\frac{M_\Delta}{2}\left(M_f-M_i\right)\right)\slq
 +\frac{M_\Delta}{2}\left(s_{pq}-M_i^2-2m_i^2\right)\right.\nonumber\\
&&
 \phantom{-\frac{1}{12}(}\left.
 -\frac{M_\Delta}{2}\,\slq'\slq
 +M_\Delta\bar{\kappa}\sln\slq\right)\left(\bar{P}_u\cdot q'\right)
 \nonumber\\
&&
 +\frac{1}{12}\left(\left(
 \bar{P}^2_u+\frac{M_\Delta}{2}\left(M_f-M_i\right)\right)\slq'
 +\frac{M_\Delta}{2}\left(M_i^2-u_{pq'}\right)\right.\nonumber\\
&&
 \phantom{+\frac{1}{12}(}\left.
 -\frac{M_\Delta}{2}\,\slq\slq'
 +M_\Delta\bar{\kappa}\sln\slq'\right)
 \left(\bar{P}_u\cdot q\right)\nonumber\\
&&
 \left.
 -\frac{1}{24}\left(\frac{1}{2}\,\left(M_f+M_i\right)+M_\Delta-\slQ
 +\bar{\kappa}\sln\right)
 \left(\bar{P}_u\cdot q'\right)\left(\bar{P}_u\cdot q\right)\right]u(ps)\nonumber\\
&&
 \times D^{(2)}\left(\Delta_u,n,\bar{\kappa}\right)\ .
\end{eqnarray}
Here, $\bar{P}^2_u$ is defined in \eqref{smatrixpnD7a}. All the
expressions for the {\it slashed} terms (i.e. $\slq$, $\slq'$,
etc.), can be found in \eqref{ur6}. Furthermore
\begin{eqnarray}
 \bar{P}_u\cdot q'
&=&
 \left(\vphantom{\frac{A}{A}}-M_f^2+M_i^2-3m_f^2-m_i^2
 +s_{p'q'}-u_{pq'}+t_{q'q}-2\bar{\kappa}(p'-p)\cdot n\right.\nonumber\\
&&
 \phantom{\left)(\right.}\left.
 +4\bar{\kappa}n\cdot Q-\left(\kappa'^2-\kappa^2\right)
 \vphantom{\frac{A}{A}}\right)\ ,\nonumber\\
 \bar{P}_u\cdot q'
&=&
 \left(\vphantom{\frac{A}{A}}M_f^2-M_i^2-m_f^2-3m_i^2+s_{pq}-u_{p'q}+t_{q'q}+2\bar{\kappa}(p'-p)\cdot n
 \right.\nonumber\\
&&
 \phantom{\left)(\right.}\left.
 +4\bar{\kappa}n\cdot Q+\left(\kappa'^2-\kappa^2\right)
 \vphantom{\frac{A}{A}}\right)\ .
\end{eqnarray}

\begin{eqnarray}
 A_{\Delta}
&=&
 -\frac{g_{gi}^2}{2}\left\{
 \frac{1}{2}\,\bar{P}^2_u
 \left[\frac{1}{2}\left(M_f+M_i\right)+M_\Delta\right](m_f^2+m_i^2-t_{q'q})\right.\nonumber\\*
&&
 -\frac{1}{3}\,\bar{P}^2_u
 \left[\left(\frac{1}{2}\left(M_f+M_i\right)+M_\Delta\right)
 \left(\frac{1}{2}(u_{p'q}+u_{pq'})-\frac{1}{2}\left(M_f^2+M_i^2\right)
 \right.\right.\nonumber\\*
&&
 \phantom{-\frac{1}{3}\,\bar{P}^2_u}\left.
 -\frac{1}{2}(\kappa'-\kappa)(p'-p)\cdot n-\frac{1}{2}(\kappa'-\kappa)^2\right)
 +\frac{1}{4}\left(u_{pq'}-M_i^2\right)\left(M_f-M_i\right)\nonumber\\*
&&
 \phantom{-\frac{1}{3}\,\bar{P}^2_u}\left.
 -\frac{1}{4}\left(s_{pq}+t_{q'q}-M_i^2-m_f^2-3m_i^2\right)\left(M_f-M_i\right)
 -\bar{\kappa}\left(M_f-M_i\right)n\cdot Q\right]\nonumber\\
&&
 -\frac{1}{12}\left[\frac{1}{2}\left(\bar{P}^2_u+\frac{M_\Delta}{2}\left(M_f-M_i\right)\right)\left(M_f-M_i\right)
 +\frac{1}{2}\,M_\Delta\right.\nonumber\\
&&
 \phantom{-\frac{1}{12}[}
 \times\left(s_{pq}-M_i^2-2m_i^2\right)
 -\frac{1}{2}\,M_\Delta\left(\frac{1}{2}(s_{p'q'}+s_{pq})-\frac{1}{2}\left(M_f^2+M_i^2\right)\right.\nonumber\\
&&
 \phantom{-\frac{1}{12}[}\left.
 -\frac{1}{2}(\kappa'-\kappa)(p'-p)\cdot n
 -\frac{1}{2}(\kappa'-\kappa)^2\right)
 +\bar{\kappa}M_\Delta\left(\vphantom{\frac{A}{A}}n\cdot p'+n\cdot Q\right.\nonumber\\
&&
 \phantom{-\frac{1}{12}[}\left.\left.
 +\frac{1}{2}(\kappa'-\kappa)\right)\right]
 \left(\bar{P}_u\cdot q'\right)
 \nonumber\\
&&
 -\frac{1}{12}\left[\frac{1}{2}\left(\bar{P}_u^2+\frac{M_\Delta}{2}\left(M_f-M_i\right)\right)\left(M_f-M_i\right)
 -\frac{1}{2}\,M_\Delta\left(M_i^2-u_{pq'}\right)\right.\nonumber\\
&&
 \phantom{+\frac{1}{12}[}
 +\frac{1}{2}\,M_\Delta\left(\frac{1}{2}(u_{p'q}+u_{pq'})-\frac{1}{2}\left(M_f^2+M_i^2\right)
 \right.\nonumber\\
&&
 \phantom{+\frac{1}{12}[}\left.
 -\frac{1}{2}(\kappa'-\kappa)(p'-p)\cdot n
 -\frac{1}{2}(\kappa'-\kappa)^2\right)
 -\bar{\kappa}M_\Delta\left(\vphantom{\frac{A}{A}}-n\cdot p'+n\cdot Q\right.\nonumber\\
&&
 \phantom{+\frac{1}{12}[}\left.\left.
 -\frac{1}{2}(\kappa'-\kappa)\right)\right]
 \left(\bar{P}_u\cdot q\right)\nonumber\\
&&
 \left.
 -\frac{1}{24}\left[\frac{1}{2}\left(M_f+M_i\right)+M_\Delta\right]
 \left(\bar{P}_u\cdot q'\right)\left(\bar{P}_u\cdot q\right)
 \right\}D^{(2)}\left(\Delta_u,n,\bar{\kappa}\right)\ .
\end{eqnarray}

\begin{eqnarray}
 B_{\Delta}
&=&
 -\frac{g_{gi}^2}{2}\left\{
 -\frac{1}{2}\,\bar{P}_u^2
 \left(m_f^2+m_i^2-t_{q'q}\right)\right.\nonumber\\
&&
 -\frac{1}{3}\,\bar{P}_u^2
 \left[\left(\frac{1}{2}\left(M_f+M_i\right)+M_\Delta\right)\left(M_f+M_i\right)
 +\frac{1}{2}\left(u_{pq'}-M_i^2\right)\right.\nonumber\\
&&
 \phantom{-\frac{1}{3}\,\bar{P}_u^2}\left.
 +\frac{1}{2}\left(s_{pq}+t_{q'q}-M_i^2-m_f^2-3m_i^2\right)
 +2\bar{\kappa}(p'-p)\cdot n
 +\frac{1}{2}\left(\kappa'^2-\kappa^2\right)\right]\nonumber\\
&&
 -\frac{1}{12}\left(\bar{P}_u^2+M_\Delta M_f\right)\left(\bar{P}_u\cdot q'\right)
 +\frac{1}{12}\left(\bar{P}_u^2-M_\Delta M_i\right)\left(\bar{P}_u\cdot q\right)
 \nonumber\\
&&
 \left.+\frac{1}{24}
 \left(\bar{P}_u\cdot q'\right)\left(\bar{P}_u\cdot q\right)
 \right\}D^{(2)}\left(\Delta_u,n,\bar{\kappa}\right)\ .
\end{eqnarray}

\begin{eqnarray}
 A'_{\Delta}
&=&
 -\frac{g_{gi}^2}{2}\left\{\vphantom{\frac{A}{A}}
 \frac{\bar{\kappa}}{2}\,\bar{P}_u^2\left(m_f^2+m_i^2-t_{q'q}\right)\right.\nonumber\\
&&
 -\frac{1}{3}\,\bar{P}_u^2
 \left[\frac{1}{4}\,(\kappa'-\kappa)\left(u_{pq'}-M_i^2\right)
 -\frac{1}{4}\,(\kappa'-\kappa)\left(s_{pq}+t_{q'q}-M_i^2
 \right.\right.\nonumber\\
&&
 \phantom{-\frac{1}{3}\,\bar{P}_u^2[}\left.
 -m_f^2-3m_i^2\right)
 +\bar{\kappa}\left(-\frac{1}{2}\left(M_f^2+M_i^2\right)
 +\frac{1}{2}(u_{p'q}+u_{pq'})\right.\nonumber\\
&&
 \left.\left.\phantom{-\frac{1}{3}\,\bar{P}_u^2[}
 -\frac{1}{2}(\kappa'-\kappa)(p'-p)\cdot n-(\kappa'-\kappa)Q\cdot n
 -\frac{1}{2}\,(\kappa'-\kappa)^2\right)\right]\nonumber\\
&&
 -\frac{1}{24}\left[(\kappa'-\kappa)\bar{P}_u^2-M_\Delta\left(\kappa M_f+\kappa'M_i\right)
 \vphantom{\frac{A}{A}}\right]
 \left[s_{p'q'}+s_{pq}-u_{p'q}-u_{pq'}\vphantom{\frac{A}{A}}\right.\nonumber\\
&&
 \left.\phantom{-\frac{1}{24}[}
 +2t_{q'q}-4m_f^2-4m_i^2+8\bar{\kappa}n\cdot Q\right]
 \nonumber\\
&&
 \left.
 +\frac{\bar{\kappa}}{24}\
 \left(\bar{P}_u\cdot q'\right)\left(\bar{P}_u\cdot q\right)
 \right\}D^{(2)}\left(\Delta_u,n,\bar{\kappa}\right)\ .
\end{eqnarray}

\begin{eqnarray}
 B'_{\Delta}
&=&
 \frac{g_{gi}^2}{12}\left\{
 \bar{P}_u^2
 \left[M_i\kappa'-M_f\kappa+M_\Delta\left(\kappa'-\kappa\right)\vphantom{\frac{A}{A}}\right]
 +\frac{M_\Delta\kappa'}{4}\left(\bar{P}_u\cdot q'\right)\right.\nonumber\\
&&
 \phantom{\frac{g_{gi}^2}{12}\{}\left.
 -\frac{M_\Delta\kappa}{4}\left(\bar{P}_u\cdot q\right)
 \right\}D^{(2)}\left(\Delta_u,n,\bar{\kappa}\right)\ .
\end{eqnarray}

\begin{eqnarray}
 X_{\Delta}^A
&=&
 \frac{g_{gi}^2}{2}\left\{
 \left(\bar{P}_u^2\right)_{CM}\left[\frac{1}{2}\left(M_f+M_i\right)+M_\Delta\right]
 \mathcal{E}'\mathcal{E}\right.\nonumber\\
&&
 -\frac{1}{3}\left(\bar{P}_u^2\right)_{CM}
 \left[\left(\frac{1}{2}\left(M_f+M_i\right)+M_\Delta\right)
 \left(\frac{1}{2}\left(M_f^2+M_i^2+m_f^2+m_i^2\right.\right.\right.\nonumber\\
&&
 \phantom{-\frac{1}{3}[}\left.
 -2E'\mathcal{E}-2\mathcal{E}'E\right)
 -\frac{1}{2}\left(M_f^2+M_i^2\right)
 -\frac{1}{2}(\kappa'-\kappa)\left(E'-E\right)\nonumber\\
&&
 \phantom{-\frac{1}{3}[}\left.
 -\frac{1}{2}(\kappa'-\kappa)^2\right)
 +\frac{1}{4}\left(m_f^2-2E\mathcal{E}'\right)\left(M_f-M_i\right)
 -\frac{1}{4}\left(\left(E+\mathcal{E}\right)^2\vphantom{\frac{A}{A}}\right.\nonumber\\
&&
 \left.\left.\phantom{-\frac{1}{3}[}
 -2\mathcal{E}'\mathcal{E}-M_i^2-2m_i^2\right)\left(M_f-M_i\right)
 -\frac{1}{2}\,\bar{\kappa}\left(M_f-M_i\right)\left(\mathcal{E}'+\mathcal{E}\right)\right]\nonumber\\
&&
 -\frac{1}{12}\left[\frac{1}{2}\left(\left(\bar{P}_u^2\right)_{CM}
 +\frac{M_\Delta}{2}\left(M_f-M_i\right)\right)\left(M_f-M_i\right)
 +\frac{1}{2}\,M_\Delta\left(\vphantom{\frac{A}{A}}\left(E+\mathcal{E}\right)^2\right.\right.
 \nonumber\\
&&
 \phantom{-\frac{1}{12}[}\left.
 -M_i^2-2m_i^2\right)
 -\frac{1}{2}\,M_\Delta\left(\frac{1}{2}\left(E'+\mathcal{E}'\right)^2+\frac{1}{2}\left(E+\mathcal{E}\right)^2
 -\frac{1}{2}\left(M_f^2+M_i^2\right)\right.\nonumber\\
&&
 \phantom{-\frac{1}{12}[}\left.
 -\frac{1}{2}(\kappa'-\kappa)\left(E'-E\right)
 -\frac{1}{2}(\kappa'-\kappa)^2\right)
 +\bar{\kappa}M_\Delta\left(E'+\frac{1}{2}\left(\mathcal{E}'+\mathcal{E}\right)\right.
 \nonumber\\
&&
 \phantom{-\frac{1}{12}[}\left.\left.
 +\frac{1}{2}(\kappa'-\kappa)\right)\right]
 \left(\bar{P}_u\cdot q'\right)_{CM}
 \nonumber\\
&&
 -\frac{1}{12}\left[\frac{1}{2}\left(\left(\bar{P}_u^2\right)_{CM}
 +\frac{M_\Delta}{2}\left(M_f-M_i\right)\right)\left(M_f-M_i\right)
 -\frac{1}{2}\,M_\Delta\left(m_f^2-2E\mathcal{E}'\right)\right.\nonumber\\
&&
 \phantom{+\frac{1}{12}[}
 +\frac{1}{2}\,M_\Delta\left(\frac{1}{2}\left(m_f^2+m_i^2-2E'\mathcal{E}-2\mathcal{E}'E\right)
 -\frac{1}{2}(\kappa'-\kappa)\left(E'-E\right)
 \right.\nonumber\\
&&
 \phantom{+\frac{1}{12}[}\left.\left.
 -\frac{1}{2}(\kappa'-\kappa)^2\right)
 -\bar{\kappa}M_\Delta\left(-E'+\frac{1}{2}\left(\mathcal{E}'+\mathcal{E}\right)
 -\frac{1}{2}(\kappa'-\kappa)\right)\right]
 \left(\bar{P}_u\cdot q\right)_{CM}\nonumber\\
&&
 \left.
 -\frac{1}{24}\left[\frac{1}{2}\left(M_f+M_i\right)+M_\Delta\right]
 \left(\bar{P}_u\cdot q'\right)_{CM}\left(\bar{P}_u\cdot q\right)_{CM}\right\}\
 ,
\end{eqnarray}
where
\begin{eqnarray}
 \left(\bar{P}_u^2\right)_{CM}
&=&
 \left[\frac{1}{2}\left(M_f^2+M_i^2+m_f^2+m_i^2-2E'\mathcal{E}-2\mathcal{E}'E\right)
 +\kappa'\kappa\right.\nonumber\\
&&
 \phantom{[}\left.\vphantom{\frac{A}{A}}
 +\bar{\kappa}\left(E'+E-\mathcal{E}'-\mathcal{E}\right)\right]\ ,\nonumber\\
 \left(\bar{P}_u\cdot q'\right)_{CM}
&=&
 \left[-M_f^2-3m_f^2+\left(E'+\mathcal{E}'\right)^2+2E\mathcal{E}'-2\mathcal{E}'\mathcal{E}
 -2\bar{\kappa}(E'-E)\vphantom{\frac{A}{A}}\right.\nonumber\\
&&
 \phantom{[}\left.\vphantom{\frac{A}{A}}
 +2\bar{\kappa}\left(\mathcal{E}'+\mathcal{E}\right)-\left(\kappa'^2-\kappa^2\right)\right]
 \ ,\nonumber\\
 \left(\bar{P}_u\cdot q\right)_{CM}
&=&
 \left[-M_i^2-3m_i^2+\left(E+\mathcal{E}\right)^2
 +2E'\mathcal{E}-2\mathcal{E}'\mathcal{E}+2\bar{\kappa}\left(E'-E\right)
 \vphantom{\frac{A}{A}}\right.\nonumber\\
&&
 \phantom{[}\left.\vphantom{\frac{A}{A}}
 +2\bar{\kappa}\left(\mathcal{E}'+\mathcal{E}\right)+\left(\kappa'^2-\kappa^2\right)\right]
 \ .
\end{eqnarray}

\begin{eqnarray}
 Y_{\Delta}^A
&=&
 \frac{g_{gi}^2\,p'p }{2}\left\{
 \left[\frac{1}{2}\left(M_f+M_i\right)+M_\Delta\right]2\mathcal{E}'\mathcal{E}\right.\nonumber\\
&&
 -\frac{5}{3}\left(\bar{P}_u^2\right)_{CM}
 \left[\frac{1}{2}\left(M_f+M_i\right)+M_\Delta\right]\nonumber\\
&&
 -\frac{2}{3}\left[\left(\frac{1}{2}\left(M_f+M_i\right)+M_\Delta\right)
 \left(\frac{1}{2}\left(M_f^2+M_i^2+m_f^2+m_i^2-2E'\mathcal{E}\right.\right.\right.\nonumber\\
&&
 \phantom{-\frac{2}{3}[}\left.\left.
 -2\mathcal{E}'E\right)-\frac{1}{2}\left(M_f^2+M_i^2\right)
 -\frac{1}{2}(\kappa'-\kappa)\left(E'-E\right)-\frac{1}{2}(\kappa'-\kappa)^2\right)\nonumber\\
&&
 \phantom{-\frac{2}{3}[}
 -\frac{1}{4}\left(\vphantom{\frac{A}{A}}\left(E+\mathcal{E}\right)^2-2\mathcal{E}'\mathcal{E}-M_i^2-2m_i^2\right)\left(M_f-M_i\right)
 \nonumber\\
&&
 \phantom{-\frac{2}{3}[}\left.
 +\frac{1}{4}\left(m_f^2-2E\mathcal{E}'\right)\left(M_f-M_i\right)
 -\frac{1}{2}\,\bar{\kappa}\left(M_f-M_i\right)\left(\mathcal{E}'+\mathcal{E}\right)\right]\nonumber\\
&&
 -\frac{1}{12}
 \left[-M_f^2-M_i^2-3m_f^2-3m_i^2+\left(E'+\mathcal{E}'\right)^2+\left(E+\mathcal{E}\right)^2
 \right.\nonumber\\
&&
 \left.\phantom{-\frac{1}{12}[}
 +2E'\mathcal{E}+2E\mathcal{E}'-4\mathcal{E}'\mathcal{E}
 +4\bar{\kappa}\left(\mathcal{E}'+\mathcal{E}\right)\right]\left[M_f-M_i\right]\nonumber\\
&&
 \left.
 -\frac{M_\Delta}{6}
 \left(\bar{P}_u\cdot q\right)_{CM}
 \right\}\ .
\end{eqnarray}

\begin{eqnarray}
 Z^A_{\Delta}
&=&
 -\frac{5g_{gi}^2(p'p)^2}{3}\left[\frac{1}{2}\left(M_f+M_i\right)+M_\Delta\vphantom{\frac{A}{A}}\right]\ .
\end{eqnarray}

\begin{eqnarray}
 X^B_{\Delta}
&=&
 \frac{g_{gi}^2}{2}\left\{
 -\left(\bar{P}_u^2\right)_{CM}\mathcal{E}'\mathcal{E}
 -\frac{1}{3}\left(\bar{P}_u^2\right)_{CM}
 \left[\left(\frac{1}{2}\left(M_f+M_i\right)+M_\Delta\right)\left(M_f+M_i\right)
 \right.\right.\nonumber\\
&&
 \phantom{-\frac{1}{3}[}
 +\frac{1}{2}\left(\left(E+\mathcal{E}\right)^2-2\mathcal{E}'\mathcal{E}-M_i^2-2m_i^2\vphantom{\frac{A}{A}}\right)
 \nonumber\\
&&
 \phantom{-\frac{1}{3}[}\left.
 +\frac{1}{2}\left(m_f^2-2E\mathcal{E}'\right)+2\bar{\kappa}\left(E'-E\right)
 +\frac{1}{2}\left(\kappa'^2-\kappa^2\right)\right]\nonumber\\
&&
 -\frac{1}{12}\left[\left(\bar{P}_u^2\right)_{CM}+M_\Delta M_f\vphantom{\frac{A}{A}}\right]
 \left(\bar{P}_u\cdot q'\right)_{CM}
 +\frac{1}{12}\left[\left(\bar{P}_u^2\right)_{CM}-M_\Delta M_i\vphantom{\frac{A}{A}}\right]
 \left(\bar{P}_u\cdot q\right)_{CM}
 \nonumber\\
&&
 \left.+\frac{1}{24}
 \left(\bar{P}_u\cdot q'\right)_{CM}
 \left(\bar{P}_u\cdot q\right)_{CM}
 \right\}\ .
\end{eqnarray}

\begin{eqnarray}
 Y^B_{\Delta}
&=&
 \frac{g_{gi}^2\,p'p}{2}\left\{\vphantom{\frac{A}{A}}
 -2\mathcal{E}'\mathcal{E}
 +\frac{1}{3}\left(\bar{P}_u^2\right)_{CM}\right.
 \nonumber\\
&&
 -\frac{2}{3}
 \left[\left(\frac{1}{2}\left(M_f+M_i\right)+M_\Delta\right)\left(M_f+M_i\right)
 +\frac{1}{2}\left(\left(E+\mathcal{E}\right)^2-2\mathcal{E}'\mathcal{E}\vphantom{\frac{A}{A}}
 \right.\right.\nonumber\\
&&
 \phantom{-\frac{2}{3}[}\left.\left.
 -M_i^2-2m_i^2\vphantom{\frac{A}{A}}\right)
 +\frac{1}{2}\left(m_f^2-2E\mathcal{E}'\right)+\bar{\kappa}\left((\kappa'-\kappa)+2(E'-E)\right)\right]\nonumber\\
&&
 +\frac{1}{6}
 \left[M_f^2-M_i^2+3m_f^2-3m_i^2-\left(E'+\mathcal{E}'\right)^2+\left(E+\mathcal{E}\right)^2
 \vphantom{\frac{A}{A}}\right.\nonumber\\
&&
 \phantom{+\frac{1}{6}[}\left.\left.\vphantom{\frac{A}{A}}
 -2E\mathcal{E}'+2E'\mathcal{E}
 +4\bar{\kappa}\left(E'-E\right)+2\left(\kappa'^2-\kappa^2\right)\right]
 \right\}\ .
\end{eqnarray}

\begin{eqnarray}
 Z^B_{\Delta}
&=&
 \frac{g_{gi}^2(p'p)^2}{3}\ .
\end{eqnarray}

\begin{eqnarray}
 X^{A'}_{\Delta}
&=&
 \frac{g_{gi}^2}{2}\left\{
 \bar{\kappa}\left(\bar{P}_u^2\right)_{CM}\mathcal{E}'\mathcal{E}
 -\frac{1}{3}\left(\bar{P}_u^2\right)_{CM}
 \left[\frac{1}{4}\,(\kappa'-\kappa)\left(m_f^2-2E\mathcal{E}'
 +2\mathcal{E}'\mathcal{E}
 \right.\right.\right.\nonumber\\
&&
 \phantom{\frac{g_{gi}^2}{2}\{}\left.
 -\left(E+\mathcal{E}\right)^2+M_i^2+2m_i^2\vphantom{\frac{A}{A}}\right)
 +\bar{\kappa}\left(\frac{1}{2}\left(m_f^2+m_i^2\right)-E'\mathcal{E}-\mathcal{E}'E
 \right.\nonumber\\
&&
 \phantom{\frac{g_{gi}^2}{2}\{}\left.\left.
 -\frac{1}{2}(\kappa'-\kappa)\left(E'-E\right)-\frac{1}{2}\,(\kappa'-\kappa)^2
 -\frac{1}{2}(\kappa'-\kappa)\left(\mathcal{E}'+\mathcal{E}\right)
 \right)\right]\nonumber\\
&&
 -\frac{1}{12}\left[(\kappa'-\kappa)\left(\bar{P}_u^2\right)_{CM}
 -M_\Delta \left(\kappa M_f+\kappa'M_i\right)\vphantom{\frac{A}{A}}\right]
 \left[\left(E'+\mathcal{E}'\right)^2\right.\nonumber\\
&&
 \phantom{-\frac{1}{12}[}
 +\left(E+\mathcal{E}\right)^2
 +2E'\mathcal{E}+2E\mathcal{E}'-2\mathcal{E}'\mathcal{E}
 -M_f^2-M_i^2\nonumber\\
&&
 \phantom{-\frac{1}{12}[}\left.
 -3m_f^2-3m_f^2+4\bar{\kappa}\left(\mathcal{E}'+\mathcal{E}\right)
 \vphantom{\frac{a}{a}}\right]
 \nonumber\\*
&&
 \left.+\frac{\bar{\kappa}}{24}
 \left(\bar{P}_u\cdot q'\right)_{CM}
 \left(\bar{P}_u\cdot q\right)_{CM}
 \right\}\ .
\end{eqnarray}

\begin{eqnarray}
 Y^{A'}_{\Delta}
&=&
 \frac{g_{gi}^2\,p'p}{2}\left\{\vphantom{\frac{A}{A}}
 2\bar{\kappa}\mathcal{E}'\mathcal{E}
 -\frac{5\bar{\kappa}}{3}\left(\bar{P}_u^2\right)_{CM}\right.
 \nonumber\\
&&
 -\frac{2}{3}\left[\frac{1}{4}\,(\kappa'-\kappa)\left(m_f^2-2E\mathcal{E}'
 -\left(E+\mathcal{E}\right)^2+2\mathcal{E}'\mathcal{E}+M_i^2+2m_i^2\vphantom{\frac{A}{A}}\right)
 \right.\nonumber\\
&&
 \phantom{-\frac{2}{3}[}
 +\bar{\kappa}\left(\frac{1}{2}\left(m_f^2+m_i^2\right)-E'\mathcal{E}-\mathcal{E}'E
 -\frac{1}{2}(\kappa'-\kappa)\left(E'-E\right)\right.\nonumber\\
&&
 \phantom{-\frac{1}{3}[+\bar{\kappa}(}\left.\left.
 -\frac{1}{2}\,(\kappa'-\kappa)^2
 -\frac{1}{2}(\kappa'-\kappa)\left(\mathcal{E}'+\mathcal{E}\right)
 \right)\right]\nonumber\\
&&
 -\frac{\left(\kappa'-\kappa\right)}{12}
 \left[-M_f^2-M_i^2-3m_f^2-3m_i^2+\left(E'+\mathcal{E}'\right)^2+\left(E+\mathcal{E}\right)^2
 \right.\nonumber\\
&&
 \phantom{-\frac{\left(\kappa'-\kappa\right)}{12}}\left.\left.
 +2E'\mathcal{E}+2E\mathcal{E}'-4\mathcal{E}'\mathcal{E}
 +4\bar{\kappa}\left(\mathcal{E}'+\mathcal{E}\right)\vphantom{\frac{a}{a}}\right]\right\}\ .
\end{eqnarray}

\begin{eqnarray}
 Z^{A'}_{\Delta}
&=&
 -\frac{5g_{gi}^2(p'p)^2\bar{\kappa}}{3}\ .
\end{eqnarray}

\begin{eqnarray}
 X^{B'}_{\Delta}
&=&
 -\frac{g_{gi}^2}{12}\left\{\vphantom{\frac{A}{A}}
 \left(\bar{P}_u^2\right)_{CM}\left[M_i\kappa'-M_f\kappa+M_\Delta\left(\kappa'-\kappa\right)\right]
 \right.\nonumber\\
&&
 \phantom{-\frac{g_{gi}^2}{12}\{}\left.
 +\frac{M_\Delta\,\kappa'}{4}
 \left(\bar{P}_u\cdot q'\right)_{CM}
 -\frac{M_\Delta\,\kappa}{4}
 \left(\bar{P}_u\cdot q\right)_{CM}
 \right\}\ .
\end{eqnarray}

\begin{eqnarray}
 Y^{B'}_{\Delta}
&=&
 -\frac{g_{gi}^2p'p}{6}\left[M_i\kappa'-M_f\kappa+M_\Delta\left(\kappa'-\kappa\right)\vphantom{\frac{A}{A}}\right]\ .
\end{eqnarray}

\subsubsection*{${\frac{3}{2}}^+$ Baryon Resonance, Gauge invariant
coupling}

\begin{eqnarray}
 M_{\kappa',\kappa}
&=&
 -\frac{g_{gi}^2}{2}\,\bar{u}(p's')\left[\vphantom{\frac{A}{A}}\right.\nonumber\\
&&
 \frac{1}{2}\,\bar{P}^2_s
 \left(\frac{1}{2}(M_f+M_i)+M_\Delta+\slQ+\bar{\kappa}\sln\right)
 \left(m_f^2+m_i^2-t_{q'q}\right)\nonumber\\
&&
 -\frac{1}{3}\,\bar{P}^2_s
 \left(\left(\frac{1}{2}\,\left(M_f+M_i\right)+M_\Delta\right)\slq'\slq
 -\frac{1}{2}\,\left(s_{pq}-M_i^2\right)\slq'\right.\nonumber\\
&&
 \phantom{-\frac{1}{3}\,\bar{P}^2_s(}\left.
 -\frac{1}{2}\,\left(u_{pq'}+t_{q'q}-M_i^2-3m_f^2-m_i^2\right)\slq
 +\bar{\kappa}\sln\slq'\slq\right)\nonumber\\
&&
 -\frac{1}{12}\left(\left(\bar{P}^2_s+\frac{M_\Delta}{2}\left(M_f-M_i\right)\right)\slq'
 +\frac{M_\Delta}{2}\left(M_i^2+2m_f^2-u_{pq'}\right)\right.\nonumber\\*
&&
 \phantom{-\frac{1}{12}(}\left.
 +\frac{M_\Delta}{2}\,\slq\slq'
 +M_\Delta\bar{\kappa}\sln\slq'\right)\left(\bar{P}_s\cdot
 q\right)\nonumber\\
&&
 +\frac{1}{12}\left(\left(\bar{P}^2_s+\frac{M_\Delta}{2}\left(M_f-M_i\right)\right)\slq
 +\frac{M_\Delta}{2}\left(s_{pq}-M_i^2\right)\right.\nonumber\\
&&
 \phantom{+\frac{1}{12}(}\left.
 +\frac{M_\Delta}{2}\,\slq'\slq
 +M_\Delta\bar{\kappa}\sln\slq\right)\left(\bar{P}_s\cdot q'\right)\nonumber\\
&&
 \left.-\frac{1}{24}\left(\frac{1}{2}\,\left(M_f+M_i\right)+M_\Delta+\slQ
 +\bar{\kappa}\sln\right)\left(\bar{P}_s\cdot q'\right)\left(\bar{P}_s\cdot q\right)
 \right]u(ps)\nonumber\\
&&
 \times D^{(2)}\left(\Delta_s,n,\bar{\kappa}\right)\ ,\label{m32kappa}
\end{eqnarray}
where $\bar{P}_s^2$ is defined in \eqref{smatrixpnD7a} and the
slashed terms are as, before, defined in \eqref{ur6}. The inner
products in \eqref{m32kappa} are
\begin{eqnarray}
 \bar{P}_s\cdot q'
&=&
 \left(\vphantom{\frac{A}{A}}-M_f^2+M_i^2+3m_f^2+m_i^2
 +s_{p'q'}-u_{pq'}-t_{q'q}-2\bar{\kappa}(p'-p)\cdot n\right.\nonumber\\
&&
 \phantom{(}\left.
 +4\bar{\kappa}n\cdot Q-\left(\kappa'^2-\kappa^2\right)
 \vphantom{\frac{A}{A}}\right)\ ,\nonumber\\
 \bar{P}_s\cdot q
&=&
 \left(\vphantom{\frac{A}{A}}M_f^2-M_i^2+m_f^2+3m_i^2+s_{pq}-u_{p'q}-t_{q'q}+2\bar{\kappa}(p'-p)\cdot n
 \right.\nonumber\\
&&
 \phantom{(}\left.
 +4\bar{\kappa}n\cdot Q+\left(\kappa'^2-\kappa^2\right)
 \vphantom{\frac{A}{A}}\right)\ .
\end{eqnarray}

\begin{eqnarray}
 A_{\Delta}
&=&
 -\frac{g_{gi}^2}{2}\left\{
 \frac{1}{2}\,\bar{P}_s^2
 \left[\frac{1}{2}\left(M_f+M_i\right)+M_\Delta\right](m_f^2+m_i^2-t_{q'q})\right.\nonumber\\
&&
 -\frac{1}{3}\,\bar{P}_s^2
 \left[\left(\frac{1}{2}\left(M_f+M_i\right)+M_\Delta\right)
 \left(\frac{1}{2}(s_{p'q'}+s_{pq})-\frac{1}{2}\left(M_f^2+M_i^2\right)
 \right.\right.\nonumber\\
&&
 \phantom{-\frac{1}{3}\,\bar{P}_s^2[}\left.
 -\frac{1}{2}(\kappa'-\kappa)(p'-p)\cdot n-\frac{1}{2}(\kappa'-\kappa)^2\right)
 +\frac{1}{4}\left(s_{pq}-M_i^2\right)\nonumber\\
&&
 \phantom{-\frac{1}{3}\,\bar{P}_s^2[}\times
 \left(M_f-M_i\right)
 +\frac{1}{4}\left(M_i^2+3m_f^2+m_i^2-u_{pq'}-t_{q'q}\right)\left(M_f-M_i\right)
 \nonumber\\
&&
 \left.\phantom{-\frac{1}{3}\,\bar{P}_s^2[}
 +\bar{\kappa}\left(M_f-M_i\right)n\cdot Q\right]\nonumber\\
&&
 +\frac{1}{12}\left[\frac{1}{2}\left(\bar{P}_s^2+\frac{M_\Delta}{2}\left(M_f-M_i\right)\right)\left(M_f-M_i\right)
 -\frac{1}{2}\,M_\Delta\right.\nonumber\\*
&&
 \phantom{+\frac{1}{12}[}\times
 \left(M_i^2+2m_f^2-u_{pq'}\right)
 -\frac{1}{2}\,M_\Delta\left(\frac{1}{2}(u_{p'q}+u_{pq'})-\frac{1}{2}\left(M_f^2+M_i^2\right)\right.\nonumber\\
&&
 \left.\phantom{+\frac{1}{12}[}
 -\frac{1}{2}(\kappa'-\kappa)(p'-p)\cdot n
 -\frac{1}{2}(\kappa'-\kappa)^2\right)
 -\bar{\kappa}M_\Delta\left(\vphantom{\frac{A}{A}}-n\cdot p'+n\cdot Q\right.\nonumber\\
&&
 \left.\left.\phantom{+\frac{1}{12}[}
 -\frac{1}{2}(\kappa'-\kappa)\right)\right]\left(\bar{P}_s\cdot q\right)
 \nonumber\\
&&
 +\frac{1}{12}\left[\frac{1}{2}\left(\bar{P}_s^2+\frac{M_\Delta}{2}\left(M_f-M_i\right)\right)\left(M_f-M_i\right)
 +\frac{1}{2}\,M_\Delta\left(s_{pq}-M_i^2\right)\right.\nonumber\\
&&
 \phantom{+\frac{1}{12}[}
 +\frac{1}{2}\,M_\Delta\left(\frac{1}{2}(s_{p'q'}+s_{pq})-\frac{1}{2}\left(M_f^2+M_i^2\right)\right.\nonumber\\
&&
 \left.\phantom{+\frac{1}{12}[}
 -\frac{1}{2}(\kappa'-\kappa)(p'-p)\cdot n -\frac{1}{2}(\kappa'-\kappa)^2\right)
 +\bar{\kappa}M_\Delta\left(\vphantom{\frac{A}{A}}n\cdot p'+n\cdot
 Q\right.\nonumber\\
&&
 \left.\left.\phantom{+\frac{1}{12}[}
 +\frac{1}{2}(\kappa'-\kappa)\right)\right]
 \left(\bar{P}_s\cdot q'\right)\nonumber\\
&&
 \left.-\frac{1}{24}\left[\frac{1}{2}\left(M_f+M_i\right)+M_\Delta\right]
 \left(\bar{P}_s\cdot q'\right)\left(\bar{P}_s\cdot q\right)
 \right\}D^{(2)}\left(\Delta_u,n,\bar{\kappa}\right)\ .
\end{eqnarray}

\begin{eqnarray}
 B_{\Delta}
&=&
 -\frac{g_{gi}^2}{2}\left\{
 \frac{1}{2}\,\bar{P}_s^2 \left(m_f^2+m_i^2-t_{q'q}\right)\right.\nonumber\\
&&
 -\frac{1}{3}\,\bar{P}_s^2
 \left[-\left(\frac{1}{2}\left(M_f+M_i\right)+M_\Delta\right)\left(M_f+M_i\right)
 -\frac{1}{2}\left(s_{pq}-M_i^2\right)\right.\nonumber\\
&&
 \left.\phantom{-\frac{1}{3}\,\bar{P}_s^2[}
 +\frac{1}{2}\left(M_i^2+3m_f^2+m_i^2-u_{pq'}-t_{q'q}\right)
 -2\bar{\kappa}(p'-p)\cdot n\right.\nonumber\\
&&
 \left.\phantom{-\frac{1}{3}\,\bar{P}_s^2[}
 -\frac{1}{2}\left(\kappa'^2-\kappa^2\right)\right]\nonumber\\
&&
 -\frac{1}{12}\left(\bar{P}_s^2+M_\Delta M_f\right)
 \left(\bar{P}_s\cdot q\right)
 +\frac{1}{12}\left(\bar{P}_s^2-M_\Delta M_i\right)
 \left(\bar{P}_s\cdot q'\right)\nonumber\\
&&
 \left.+\frac{1}{24}
 \left(\bar{P}_s\cdot q'\right)\left(\bar{P}_s\cdot q\right)
 \right\}D^{(2)}\left(\Delta_s,n,\bar{\kappa}\right)\ .
\end{eqnarray}

\begin{eqnarray}
 A'_{\Delta}
&=&
 -\frac{g_{gi}^2}{2}\left\{
 \frac{\bar{\kappa}}{2}\,\bar{P}_s^2\left(m_f^2+m_i^2-t_{q'q}\right)\right.\nonumber\\
&&
 -\frac{1}{3}\,\bar{P}_s^2
 \left[\frac{1}{4}\,(\kappa'-\kappa)\left(s_{pq}-M_i^2\right)
 +\frac{1}{4}\,(\kappa'-\kappa)\left(M_i^2+3m_f^2+m_i^2\right.
 \right.\nonumber\\
&&
 \phantom{-\frac{1}{3}\,\bar{P}_s^2[}\left.
 -u_{pq'}-t_{q'q}\right)
 +\bar{\kappa}\left(-\frac{1}{2}\left(M_f^2+M_i^2\right)
 +\frac{1}{2}(s_{p'q'}+s_{pq})\right.\nonumber\\
&&
 \left.\left.\phantom{-\frac{1}{3}\,\bar{P}_s^2[}
 -\frac{1}{2}(\kappa'-\kappa)(p'-p)\cdot n+(\kappa'-\kappa)Q\cdot n
 -\frac{1}{2}\,(\kappa'-\kappa)^2\right)\right]\nonumber\\
&&
 +\frac{1}{24}\left[(\kappa'-\kappa)\bar{P}_s^2-M_\Delta\left(\kappa M_f+\kappa'M_i\right)
 \vphantom{\frac{A}{A}}\right]
 \left[s_{p'q'}+s_{pq}-u_{p'q}-u_{pq'}\vphantom{\frac{A}{A}}\right.\nonumber\\
&&
 \phantom{+\frac{1}{24}[}\left.\vphantom{\frac{A}{A}}
 -2t_{q'q}+4m_f^2+4m_i^2+8\bar{\kappa}n\cdot Q\right]
 \nonumber\\
&&
 \left.-\frac{\bar{\kappa}}{24}
 \left(\bar{P}_s\cdot q'\right)\left(\bar{P}_s\cdot q'\right)\right\}
 D^{(2)}\left(\Delta_s,n,\bar{\kappa}\right)\ .
\end{eqnarray}

\begin{eqnarray}
 B'_{\Delta}
&=&
 -\frac{g_{gi}^2}{12}\left\{
 \bar{P}_s^2\left[M_i\kappa'-M_f\kappa+M_\Delta\left(\kappa'-\kappa\right)
 \vphantom{\frac{A}{A}}\right]\right.\nonumber\\
&&
 \phantom{-\frac{g_{gi}^2}{12}\{}\left.
 -\frac{\kappa'M_\Delta}{4}\left(\bar{P}_s\cdot q\right)
 +\frac{\kappa M_\Delta}{4}\left(\bar{P}_s\cdot q'\right)
 \right\}D^{(2)}\left(\Delta_s,n,\bar{\kappa}\right)\ .
\end{eqnarray}

\begin{eqnarray}
 X^A_{\Delta}
&=&
 \frac{g_{gi}^2}{2}\left\{
 \left(\bar{P}_s^2\right)_{CM}
 \left[\frac{1}{2}\left(M_f+M_i\right)+M_\Delta\right]\mathcal{E}'\mathcal{E}\right.\nonumber\\
&&
 -\frac{1}{3}\left(\bar{P}_s^2\right)_{CM}
 \left[\left(\frac{1}{2}\left(M_f+M_i\right)+M_\Delta\right)
 \left(\frac{1}{2}\left(E'+\mathcal{E}'\right)^2
 +\frac{1}{2}\left(E+\mathcal{E}\right)^2\right.\right.\nonumber\\
&&
 \phantom{-\frac{1}{3}[}\left.
 -\frac{1}{2}\left(M_f^2+M_i^2\right)
 -\frac{1}{2}(\kappa'-\kappa)\left(E'-E\right)-\frac{1}{2}(\kappa'-\kappa)^2\right)\nonumber\\
&&
 \phantom{-\frac{1}{3}[}
 +\frac{1}{4}\left(m_f^2+2E\mathcal{E}'+2\mathcal{E}'\mathcal{E}\right)\left(M_f-M_i\right)
 +\frac{1}{4}\left(\left(E+\mathcal{E}\right)^2-M_i^2\right)\nonumber\\
&&
 \left.\phantom{-\frac{1}{3}[}\times
 \left(M_f-M_i\right)
 +\frac{\bar{\kappa}}{2}\left(M_f-M_i\right)\left(\mathcal{E}'+\mathcal{E}\right)\right]\nonumber\\
&&
 +\frac{1}{12}\left[\frac{1}{2}\left(\left(\bar{P}_s^2\right)_{CM}
 +\frac{M_\Delta}{2}\left(M_f-M_i\right)\right)\left(M_f-M_i\right)
 \right.\nonumber\\
&&
 \phantom{+\frac{1}{12}[}
 -\frac{1}{2}\,M_\Delta\left(m_f^2+2E\mathcal{E}'\right)
 -\frac{1}{2}\,M_\Delta\left(\frac{1}{2}\left(m_f^2+m_i^2\right)\right.\nonumber\\
&&
 \left.\phantom{+\frac{1}{12}[}
 -2\left(E'\mathcal{E}+E\mathcal{E}'\right)
 -\frac{1}{2}(\kappa'-\kappa)\left(E'-E\right)
 -\frac{1}{2}(\kappa'-\kappa)^2\right)\nonumber\\
&&
 \left.\phantom{+\frac{1}{12}[}
 -\bar{\kappa}M_\Delta\left(-E'+\frac{1}{2}\left(\mathcal{E}'+\mathcal{E}\right)
 -\frac{1}{2}(\kappa'-\kappa)\right)\right]
 \left(\bar{P}_s\cdot q\right)_{CM}
 \nonumber\\
&&
 +\frac{1}{12}\left[\frac{1}{2}\left(\left(\bar{P}_s^2\right)_{CM}
 +\frac{M_\Delta}{2}\left(M_f-M_i\right)\right)\left(M_f-M_i\right)
 \right.\nonumber\\
&&
 \phantom{+\frac{1}{12}[}
 +\frac{1}{2}\,M_\Delta\left(\left(E+\mathcal{E}\right)^2-M_i^2\right)
 +\frac{1}{2}\,M_\Delta\left(\frac{1}{2}\left(E'+\mathcal{E}'\right)^2+\frac{1}{2}\left(E+\mathcal{E}\right)^2
 \right.\nonumber\\
&&
 \left.\phantom{+\frac{1}{12}[}
 -\frac{1}{2}\left(M_f^2+M_i^2\right)
 -\frac{1}{2}(\kappa'-\kappa)\left(E'-E\right) -\frac{1}{2}(\kappa'-\kappa)^2\right)
 \nonumber\\
&&
 \left.\phantom{+\frac{1}{12}[}
 +\bar{\kappa}M_\Delta\left(E'+\frac{1}{2}\left(\mathcal{E}'+\mathcal{E}\right)
 +\frac{1}{2}(\kappa'-\kappa)\right)\right]
 \left(\bar{P}_s\cdot q'\right)_{CM}\nonumber\\
&&
 \left.-\frac{1}{24}\left[\frac{1}{2}\left(M_f+M_i\right)+M_\Delta\right]
 \left(\bar{P}_s\cdot q'\right)_{CM}\left(\bar{P}_s\cdot q\right)_{CM}
 \right\}\ ,
\end{eqnarray}
where
\begin{eqnarray}
 \left(\bar{P}_s^2\right)_{CM}
&=&
 \left[\frac{1}{2}\left(E'+\mathcal{E}'\right)^2+\frac{1}{2}\left(E+\mathcal{E}\right)^2
 +\kappa'\kappa+\bar{\kappa}\left(E'+E+\mathcal{E}'+\mathcal{E}\right)\right]\
 ,\nonumber\\
 \left(\bar{P}_s\cdot q'\right)_{CM}
&=&
 \left[-M_f^2+m_f^2+\left(E'+\mathcal{E}'\right)^2+2E\mathcal{E}'+2\mathcal{E}'\mathcal{E}
 -2\bar{\kappa}\left(E'-E\right)\vphantom{\frac{A}{A}}\right.\nonumber\\
&&
 \phantom{[}\left.\vphantom{\frac{A}{A}}
 +2\bar{\kappa}\left(\mathcal{E}'+\mathcal{E}\right)
 -\left(\kappa'^2-\kappa^2\right)\right]\ ,\nonumber\\
 \left(\bar{P}_s\cdot q\right)_{CM}
&=&
 \left[-M_i^2+m_i^2+\left(E+\mathcal{E}\right)^2+2E'\mathcal{E}+2\mathcal{E}'\mathcal{E}
 +2\bar{\kappa}\left(E'-E\right)\vphantom{\frac{A}{A}}\right.\nonumber\\
&&
 \phantom{[}\left.\vphantom{\frac{A}{A}}
 +2\bar{\kappa}\left(\mathcal{E}'+\mathcal{E}\right)
 +\left(\kappa'^2-\kappa^2\right)\right]\ .
\end{eqnarray}

\begin{eqnarray}
 Y^A_{\Delta}
&=&
 \frac{g_{gi}^2\,p'p}{2}\left\{
 -\left(\bar{P}_s^2\right)_{CM}\left[\frac{1}{2}\left(M_f+M_i\right)+M_\Delta\right]\right.\nonumber\\
&&
 +\frac{M_\Delta}{6}\left[-\frac{1}{2}\,M_f^2+\frac{1}{2}\,M_i^2+2M_fM_i
 +\frac{1}{2}\left(3m_f^2+m_i^2\right)-\left(E'\mathcal{E}-E\mathcal{E}'\right)
 \right.\nonumber\\
&&
 \left.\phantom{+\frac{M_\Delta}{6}[}
 -\frac{1}{2}\left(E'+\mathcal{E}'\right)^2-\frac{3}{2}\left(E+\mathcal{E}\right)^2
 -4\bar{\kappa}E'-\left(\kappa'^2-\kappa^2\right)\right]\nonumber\\
&&
 +\frac{1}{6}\left[\frac{1}{2}\left(M_f+M_i\right)+M_\Delta\right]
 \left[-M_f^2-M_i^2+m_f^2+m_i^2+\left(E'+\mathcal{E}'\right)^2
 \vphantom{\frac{A}{A}}\right.\nonumber\\
&&
 \phantom{+\frac{1}{6}[}\left.\left.\vphantom{\frac{A}{A}}
 +\left(E+\mathcal{E}\right)^2
 +2E'\mathcal{E}+2E\mathcal{E}'+4\mathcal{E}'\mathcal{E}
 +4\bar{\kappa}\left(\mathcal{E}'+\mathcal{E}\right)\right]
 \right\}\ .
\end{eqnarray}

\begin{eqnarray}
 Z^{A}_{\Delta}
&=&
 -\frac{g_{gi}^2(p'p)^2}{3}\left[\frac{1}{2}\left(M_f+M_i\right)+M_\Delta\right]\ .
\end{eqnarray}

\begin{eqnarray}
 X^B_{\Delta}
&=&
 \frac{g_{gi}^2}{2}\left\{
 \left(\bar{P}_s^2\right)_{CM}\mathcal{E}'\mathcal{E}
 +\frac{1}{3}\left(\bar{P}_s^2\right)_{CM}
 \left[\left(\frac{1}{2}\left(M_f+M_i\right)+M_\Delta\right)\left(M_f+M_i\right)\right.\right.\nonumber\\
&&
 \phantom{+\frac{1}{3}[}
 -\frac{1}{2}\left(m_f^2+2E\mathcal{E}'+2\mathcal{E}'\mathcal{E}\right)
 +\frac{1}{2}\left(\left(E+\mathcal{E}\right)^2-M_i^2\right)
 +2\bar{\kappa}\left(E'-E\right)\nonumber\\
&&
 \phantom{+\frac{1}{3}[}\left.
 +\frac{1}{2}\left(\kappa'^2-\kappa^2\right)\right]\nonumber\\
&&
 -\frac{1}{12}\left[\left(\bar{P}_s^2\right)_{CM}+M_\Delta M_f\vphantom{\frac{A}{A}}\right]
 \left(\bar{P}_s\cdot q\right)_{CM}
 +\frac{1}{12}\left[\left(\bar{P}_s^2\right)_{CM}-M_\Delta M_i\vphantom{\frac{A}{A}}\right]
 \left(\bar{P}_s\cdot q'\right)_{CM}
 \nonumber\\
&&
 \left.+\frac{1}{24}
 \left(\bar{P}_s\cdot q'\right)_{CM}
 \left(\bar{P}_s\cdot q\right)_{CM}
 \right\}\ .
\end{eqnarray}

\begin{eqnarray}
 Y^B_{\Delta}
&=&
 -\frac{g_{gi}^2\,p'p}{6}\left\{\vphantom{\frac{A}{A}}
 \left(\bar{P}_s^2\right)_{CM}
 -M_\Delta\left(M_f+M_i\right)\right.\nonumber\\*
&&
 \phantom{-\frac{g_{gi}^2\,p'p}{6}\{}
 +\frac{1}{2}\left[-M_f^2-M_i^2+m_f^2+m_i^2+\left(E'+\mathcal{E}'\right)^2+\left(E+\mathcal{E}\right)^2
 \vphantom{\frac{a}{a}}\right.\nonumber\\
&&
 \phantom{-\frac{g_{gi}^2\,p'p}{6}\{+\frac{1}{2}[}\left.\left.\vphantom{\frac{a}{a}}
 +2E\mathcal{E}'+2E'\mathcal{E}+4\mathcal{E}'\mathcal{E}
 +4\bar{\kappa}\left(\mathcal{E}'+\mathcal{E}\right)\right]\right\}\ .
\end{eqnarray}

\begin{eqnarray}
 Z^B_{\Delta}
&=&
 \frac{g_{gi}^2(p'p)^2}{3}\ .
\end{eqnarray}

\begin{eqnarray}
 X^{A'}_{\Delta}
&=&
 \frac{g_{gi}^2}{2}\left\{
 \bar{\kappa}\left(\bar{P}_s^2\right)_{CM}\mathcal{E}'\mathcal{E}
 -\frac{1}{3}\left(\bar{P}_s^2\right)_{CM}
 \left[\frac{1}{4}\,(\kappa'-\kappa)\left(\left(E+\mathcal{E}\right)^2-M_i^2\right)\right.\right.
 \nonumber\\
&&
 \phantom{\frac{g_{gi}^2}{2}\{}
 +\frac{1}{4}\,(\kappa'-\kappa)\left(m_f^2+2E\mathcal{E}'+2\mathcal{E}'\mathcal{E}\right)
 +\bar{\kappa}\left(-\frac{1}{2}\left(M_f^2+M_i^2\right)\right.\nonumber\\
&&
 \phantom{\frac{g_{gi}^2}{2}\{}
 +\frac{1}{2}\left(E'+\mathcal{E}'\right)^2+\frac{1}{2}\left(E+\mathcal{E}\right)^2
 -\frac{1}{2}(\kappa'-\kappa)\left(E'-E\right)\nonumber\\
&&
 \left.\left.\phantom{\frac{g_{gi}^2}{2}\{}
 +\frac{1}{2}(\kappa'-\kappa)\left(\mathcal{E}'+\mathcal{E}\right)
 -\frac{1}{2}\,(\kappa'-\kappa)^2\right)\right]\nonumber\\
&&
 +\frac{1}{24}\left[(\kappa'-\kappa)\left(\bar{P}_s^2\right)_{CM}
 -M_\Delta\left(\kappa'M_i+\kappa M_f\right)\vphantom{\frac{A}{A}}\right]
 \left[-M_f^2-M_i^2\vphantom{\frac{A}{A}}\right.\nonumber\\
&&
 \phantom{+\frac{1}{24}[}\left.
 +m_f^2+m_i^2+\left(E'+\mathcal{E}'\right)^2+\left(E+\mathcal{E}\right)^2
 +2E'\mathcal{E}+2E\mathcal{E}'\right.\nonumber\\
&&
 \left.\phantom{+\frac{1}{24}[}
 +4\mathcal{E}'\mathcal{E}+4\bar{\kappa}\left(\mathcal{E}'+\mathcal{E}\right)
 \vphantom{\frac{A}{A}}\right]\nonumber\\
&&
 \left.-\frac{\bar{\kappa}}{24}
 \left(\bar{P}_s\cdot q'\right)_{CM}
 \left(\bar{P}_s\cdot q\right)_{CM}
 \right\}\ .
\end{eqnarray}

\begin{eqnarray}
 Y^{A'}_{\Delta}
&=&
 -\frac{g_{gi}^2\,p'p}{2}\left\{
 \bar{\kappa}\left(\bar{P}_s^2\right)_{CM}
 -\frac{M_\Delta}{3}\left[\kappa'M_i+\kappa M_f\right]\right.
 \nonumber\\
&&
 \phantom{-\frac{g^2\,p'p}{2}\left\{\right.}
 -\frac{\bar{\kappa}}{6}
 \left[-M_f^2-M_i^2+m_f^2+m_i^2+\left(E'+\mathcal{E}'\right)^2+\left(E+\mathcal{E}\right)^2
 \vphantom{\frac{A}{A}}\right.\nonumber\\
&&
 \phantom{-\frac{g^2\,p'p}{2}\left\{\right.-\frac{\bar{\kappa}}{6}[}
 \left.\left.\vphantom{\frac{A}{A}}
 +2E'\mathcal{E}+2E\mathcal{E}'
 +4\mathcal{E}'\mathcal{E}+4\bar{\kappa}\left(\mathcal{E}'+\mathcal{E}\right)\right]
 \right\}\ .
\end{eqnarray}

\begin{eqnarray}
 Z^{A'}_{\Delta}
&=&
 -\frac{g_{gi}^2(p'p)^2}{3}\ .
\end{eqnarray}

\begin{eqnarray}
 X^{B'}_{\Delta}
&=&
 \frac{g_{gi}^2}{12}\left\{
 \left(\bar{P}_s^2\right)_{CM}
 \left[\kappa'M_i-\kappa M_f+\left(\kappa'-\kappa\right)M_\Delta\vphantom{\frac{A}{A}}\right]
 \right.\nonumber\\
&&
 \phantom{\frac{g_{gi}^2}{12}\{}\left.
 -\frac{\kappa'M_\Delta}{4}
 \left(\bar{P}_s\cdot q\right)_{CM}
 +\frac{\kappa M_\Delta}{4}
 \left(\bar{P}_s\cdot q'\right)_{CM}\right\}\ .
\end{eqnarray}

\begin{eqnarray}
 Y^{B'}_{\Delta}
&=&
 \frac{g_{gi}^2M_\Delta\,p'p}{12}\left(\kappa'-\kappa\right)\ .
\end{eqnarray}

\subsection{Useful relations}

\subsubsection{Feynman}\label{URfeyn}

In Feynman formalism the following relations are quit useful
\begin{eqnarray}
 2(q'\cdot q)&=&m_f^2+m_i^2-t\ ,\nonumber\\
 2(p'\cdot p)&=&M_f^2+M_i^2-t\ ,\nonumber\\
 2(p'\cdot q')&=&s-M_f^2-m_f^2\ ,\nonumber\\
 2(p \cdot q)&=&s-M_i^2-m_i^2\ ,\nonumber\\
 2(p\cdot q')&=&M_i^2+m_f^2-u\ ,\nonumber\\
 2(p'\cdot q)&=&M_f^2+m_i^2-u\ .\label{ur1}
\end{eqnarray}
\begin{eqnarray}
 s+u+t=M^2_f+M^2_i+m^2_f+m_i^2\ .\label{ur2}
\end{eqnarray}
\begin{eqnarray}
 \slq &=&\frac{1}{2}\left(M_f-M_i\right)+\slQ\ ,\nonumber\\
 \slq'&=&-\frac{1}{2}\left(M_f-M_i\right)+\slQ\ ,\nonumber\\
 \slq\slq'&=&\left(M_f+M_i\right)\slQ
             -\frac{1}{2}\left(M_f^2+M^2_i\right)+u\ ,\nonumber\\
 \slq'\slq&=&-\left(M_f+M_i\right)\slQ
             -\frac{1}{2}\left(M_f^2+M^2_i\right)+s\ .\label{ur3}
\end{eqnarray}

\subsubsection{Kadyshevsky}\label{URkad}

In Kadyshevsky formalism there are similar relations
\begin{eqnarray}
 2(q'\cdot q)&=&m_f^2+m_i^2-t_{q'q}\ ,\nonumber\\
 2(p'\cdot p)&=&M_f^2+M_i^2-t_{p'p}\ ,\nonumber\\
 2(p'\cdot q')&=&s_{p'q'}-M_f^2-m_f^2\ ,\nonumber\\
 2(p \cdot q)&=&s_{pq}-M_i^2-m_i^2\ ,\nonumber\\
 2(p\cdot q')&=&M_i^2+m_f^2-u_{pq'}\ ,\nonumber\\
 2(p'\cdot q)&=&M_f^2+m_i^2-u_{p'q}\ .\label{ur4}
\end{eqnarray}
\begin{eqnarray}
 s_{p'q'}+s_{pq}+u_{p'q}+u_{pq'}+t_{p'p}+t_{q'q}
&=&
 2\left(M^2_f+M^2_i+m^2_f+m_i^2\right)+\left(\kappa'-\kappa\right)^2\ ,\nonumber\\
 2\sqrt{s_{p'q'}s_{pq}}+u_{p'q}+u_{pq'}+t_{p'p}+t_{q'q}
&=&
 2\left(M^2_f+M^2_i+m^2_f+m_i^2\right)\ .\label{ur5}
\end{eqnarray}
\begin{eqnarray}
 \slq'
&=&
 -\frac{1}{2}\left(M_f-M_i\right)+\slQ-\frac{1}{2}\,\sln(\kappa'-\kappa)\ ,\nonumber\\
 \slq
&=&
 \frac{1}{2}\left(M_f-M_i\right)+\slQ+\frac{1}{2}\,\sln(\kappa'-\kappa)\ ,\nonumber\\
 \slq'\slq
&=&
 -\left(M_f+M_i\right)\slQ+\frac{1}{2}\,\left(s_{p'q'}+s_{pq}\right)
 -\frac{1}{2}\left(M_f^2+M^2_i\right)-\frac{1}{2}\,(\kappa'-\kappa)(p'-p)\cdot n
 \nonumber\\
&&
 -\frac{1}{2}\,(\kappa'-\kappa)\left[\sln,\slQ\right]
 -\frac{1}{2}\,(\kappa'-\kappa)^2\ ,\nonumber\\
 \slq\slq'
&=&
 \left(M_f+M_i\right)\slQ+\frac{1}{2}\,\left(u_{p'q}+u_{pq'}\right)
 -\frac{1}{2}\left(M_f^2+M^2_i\right)-\frac{1}{2}\,(\kappa'-\kappa)(p'-p)\cdot n
 \nonumber\\
&&
 +\frac{1}{2}\,(\kappa'-\kappa)\left[\sln,\slQ\right]
 -\frac{1}{2}\,(\kappa'-\kappa)^2\ ,\nonumber\\
 \sln\slq'
&=&
 \frac{1}{2}\,\left(M_f+M_i\right)\sln-(n\cdot p')
 +\frac{1}{2}\,\left[\sln,\slQ\right]+n\cdot Q
 -\frac{1}{2}\,(\kappa'-\kappa)\ ,\nonumber\\
 \sln\slq
&=&
 -\frac{1}{2}\,\left(M_f+M_i\right)\sln+(n\cdot p')
 +\frac{1}{2}\,\left[\sln,\slQ\right]+n\cdot Q
 +\frac{1}{2}\,(\kappa'-\kappa)\ ,\nonumber\\
 \sln\slq'\slq
&=&
 -\frac{1}{2}\,\left(M_f^2+M_i^2\right)\sln+\frac{1}{2}\,\left(s_{p'q'}+s_{pq}\right)\sln
 +\frac{1}{2}\,\left(M_f-M_i\right)\left[\sln,\slQ\right]+\left(M_f-M_i\right)n\cdot Q
 \nonumber\\
&&
 -\frac{1}{2}\,(\kappa'-\kappa)n\cdot(p'-p)\sln+\left(\kappa'-\kappa\right)\left(n\cdot Q\right)\sln
 -(\kappa'-\kappa)\slQ-2n\cdot(p'-p)\slQ\nonumber\\
&&
 -\frac{1}{2}\,(\kappa'-\kappa)^2\sln\ ,\nonumber\\
 \sln\slq\slq'
&=&
 -\frac{1}{2}\,\left(M_f^2+M_i^2\right)\sln+\frac{1}{2}\,\left(u_{p'q}+u_{pq'}\right)\sln
 -\frac{1}{2}\,\left(M_f-M_i\right)\left[\sln,\slQ\right]-\left(M_f-M_i\right)n\cdot Q
 \nonumber\\
&&
 -\frac{1}{2}\,(\kappa'-\kappa)n\cdot(p'-p)\sln-\left(\kappa'-\kappa\right)\left(n\cdot Q\right)\sln
 +(\kappa'-\kappa)\slQ+2n\cdot(p'-p)\slQ\nonumber\\
&&
 -\frac{1}{2}\,(\kappa'-\kappa)^2\sln
 \ .\label{ur6}
\end{eqnarray}
\end{appendices}


\begin{thebibliography}{99}
\bibitem{JWT1}
J.W.Wagenaar \& T.A.Rijken, "Pion-Nucleon Scattering in
Kadyshevsky Formalism: Meson Exchange Sector", to be published
(2009)
\bibitem{Kad64}
V.G.Kadyshevsky, Sov.\ Phys.\ JETP {\bf 19}, 443 (1964);
V.G.Kadyshevsky, Sov.\ Phys.\ JETP {\bf 19}, 597 (1964)
\bibitem{Kad67}
V.G.Kadyshevsky, Nucl.\ Phys. {\bf B6}, 125 (1968)
\bibitem{Kad68}
V.G.Kadyshevsky and N.D.Mattev, Nuov. Cim. {\bf 55A}, 275 (1968)
\bibitem{Kad70}
C.Itzykson, V.G.Kadyshevsky, I.T.Todorov, Phys.\ Rev.\ {\bf D1},
2823 (1970)
\bibitem{henk1}
 H.Polinder \& T.A.Rijken, Phys.\ Rev. {\bf C72}, 065210 (2005);
 H.Polinder \& T.A.Rijken, Phys.\ Rev. {\bf C72}, 065211 (2005)
\bibitem{Ver76}
 P.A.Verhoeven, "Off-Shell Baryon-Baryon Scattering", Ph.D.
 University of Nijmegen, 1976;
 A.Gersten, P.A.Verhoeven, and J.J.deSwart, Nuovo Cimento {\bf A26}, 375 (1975)
\bibitem{Tak53a}
Y.Takahashi \& H.Umezawa, Prog.\ Theor.\ Phys. {\bf 9}, 14 (1953)
\bibitem{Tak53b}
Y.Takahashi \& H.Umezawa, Prog.\ Theor.\ Phys. {\bf 9}, 501 (1953)
\bibitem{Ume56}
 H.Umezawa, "Quantum Field Theory", North-Holland Publishing
 Company, Amsterdam, 14, 1953 (Chapter X);
 Y.Takahashi,"An Introduction to Field Quantization",
 "International series of monographsin Natural Philosophy", Vol.
 20, Pergamon Press, 1969
\bibitem{Gross69}
D.J.\ Gross and R.\ Jackiw, Nucl.\ Phys.\ {\bf B14}, 269 (1969);
R.\ Jackiw in "lectures on Current algebra and its Applications",
by S.B.\ treiman, R.\ Jackiw,and D.J.\ Gross, Princeton University
Press, Princeton, N.J., 1972
\bibitem{SN78}
J.J. de Swart and M.M. Nagels, Fortschr.\ d.\ Physik {\bf 26}, 215
(1978)
\bibitem{witten}
E.Witten, Nucl. \ Phy. {\bf B160}, 57 (1979)
\bibitem{sakurai69}
J. J. Sakurai, "Currents and Mesons", University of Chicago Press,
Chicago; 1969
\bibitem{sakurai67}
J.J.Sakurai, "Advanced Quantum Mechanics", Addison-Wesley
Publishing Company; 1967
\bibitem{brod}
S.J.\ Brodsky and J.R.\ Primack, Ann.\ Phys.\ (NY) {\bf 52}, 315
(1969)
\bibitem{Bow73}
R.L.\ Bowers and R.L.\ Zimmermann, Phys.\ Rev.\ {\bf D7}, 296
(1973)
\bibitem{Yang50}
C.N.\ Yang and D.\ Feldman, Phys.\ Rev.\ {\bf 79}, 972 (1950)
\bibitem{pasc}
V.Pascalutsa, Phys.\ Rev. {\bf D58}, 096002 (1998)
\bibitem{pasctim}
V.Pascalutsa \& R.Timmermans, Phys.\ Rev. {\bf C60}, 042201 (1999)
\bibitem{brans73}
B.H.Bransden \& R.G.Moorhouse, "The Pion-Nucleon System",
Princeton University Press, Princeton; 1973
\bibitem{Pil67}
H.Pilkuhn, "The Interactions of Hadrons", North-Holland Pub.
Comp., Amsterdam; 1967
\bibitem{Jac59}
M.\ Jacob and G.C.\ Wick, Ann. Phys. (N.Y.) {\bf 7}, 404 (1959)
\end{thebibliography}
\end{document}